\documentclass[manuscript]{aastex}
\usepackage{chngcntr,multirow}

\slugcomment{Not to appear in Nonlearned J., 45.}

\shorttitle{The Young Open Clusters NGC 1624 and NGC 1931}
\shortauthors{Lim et al.}

\begin{document}

\title{Sejong Open Cluster Survey (SOS) - IV. The Young Open Clusters NGC 1624 and NGC 1931}

\author{Beomdu Lim\altaffilmark{1,5}, Hwankyung Sung\altaffilmark{2}, Michael S. Bessell\altaffilmark{3}, Jinyoung S. Kim\altaffilmark{4}, 
Hyeonoh Hur\altaffilmark{2}, and Byeong-Gon Park\altaffilmark{1}}
\email{bdlim1210@kasi.re.kr}

\altaffiltext{1}{Korea Astronomy and Space Science Institute, 776 Daedeokdae-ro, Yuseong-gu, Daejeon 305-348, Korea}
\altaffiltext{2}{Department of Astronomy and Space Science, Sejong University, 209 Neungdong-ro, Gwangjin-gu, Seoul 143-747, Korea}
\altaffiltext{3}{Research School of Astronomy and Astrophysics, Australian National University, MSO, Cotter Road, Weston, ACT 2611, Australia} 
\altaffiltext{4}{Steward Observatory, University of Arizona, 933 N. Cherry Ave. Tucson, AZ 85721-0065, USA}
\altaffiltext{5}{Corresponding author, Korea Research Council of Fundamental Science \& Technology Research Fellow} 

\begin{abstract}
Young open clusters located in the outer Galaxy provide us with an opportunity to study star formation activity in a
different environment from the solar neighborhood. We present a $UBVI$ and H$\alpha$ photometric study 
of the young open clusters NGC 1624 and NGC 1931 that are situated toward the Galactic anticenter. Various 
photometric diagrams are used to select the members of the clusters and to determine the fundamental 
parameters. NGC 1624 and NGC 1931 are, on average, reddened by $\langle E(B-V) \rangle = 0.92 \pm 0.05$ and 
$0.74 \pm 0.17$ mag, respectively. The properties of the reddening toward NGC 1931 indicate an abnormal reddening 
law ($R_{V,\mathrm{cl}} = 5.2 \pm 0.3$). Using the zero-age main sequence fitting method we confirm that 
NGC 1624 is $6.0 \pm 0.6$ kpc away from the Sun, whereas NGC 1931 is at a distance of $2.3 \pm 
0.2$ kpc. The results from isochrone fitting in the Hertzsprung-Russell diagram indicate the ages of NGC 1624 
and NGC 1931 to be less than 4 Myr and 1.5 -- 2.0 Myr, respectively. We derived the initial mass function (IMF) 
of the clusters. The slope of the IMF ($\Gamma _{\mathrm{NGC \ 1624}} = -2.0 \pm 0.2$ and $\Gamma _{\mathrm{NGC 
\ 1931}} = -2.0 \pm 0.1$) appears to be steeper than that of the Salpeter/Kroupa IMF. We discuss the implication of 
the derived IMF based on simple Monte-Carlo simulations and conclude that the property of star formation in the 
clusters seems not to be far different from that in the solar neighborhood.
\end{abstract}


\section{INTRODUCTION}
Young open clusters are useful objects to study the star formation process because about 80 percent of the 
stars in star forming regions (SFRs) are found in clusters with at least 100 members \citep{LL03,
PCA03}. The relation between star formation activity and environmental conditions is one of the most interesting 
issues in star formation studies \citep{CMP08,CMP12}. A basic diagnostic tool for understanding 
star formation processes is the stellar initial mass function (IMF). The concept was first introduced by \citet{Sp55}. 
If the stellar IMF has a universal shape in all SFRs, there must be a fundamental 
factor controlling the star formation process. A number of efforts have been devoted to confirming the universality 
(see review of \citealt{BCM10}), however, there is as yet no firm conclusion concerning the universality or diversity of 
the shape. In this context, the IMF of young open clusters formed in various star-forming environments may 
give a clue to the dependence of star formation processes on environmental conditions. As  
part of an attempt to study this issue, we investigated the young open clusters NGC 1624 and NGC 1931 
in the outer Galaxy, which is probably a low-metallicity environment, according to the gradient of metal 
abundance in the Galactic disk \citep{YCF12}.

The young open cluster NGC 1624 is surrounded by the H {\scriptsize \textsc{II}} region Sh2-212 
with a shell like structure. The cluster comprises several high-mass main sequence (MS) stars, as well as 
a large number of pre-main sequence (PMS) stars \citep{JPO11}. The most luminous star is NGC 1624-2 
(O7f?p -- \citealt{WSA10,SAW11}) located near the cluster center. This star with strong X-ray emission is known 
to be a slowly rotating star with a very strong magnetic field \citep{WAM12}. Out of all the probable PMS 
members one-fifth have circumstellar disks \citep{JPO11}. From their radio observations, \citet{DLK08} found 
an ultra-compact H {\scriptsize \textsc{II}} (UCHII) region at the border of Sh2-212.  A high-mass 
PMS star (No. 228 of \citealt{DLK08}) seems to be associated with the UCHII. A few photometric studies of NGC 1624 
provided the fundamental parameters \citep{MFJ79,CW84,SuBa06,DLK08,JPO11}. The reddening of 
the cluster is about $E(B-V) = 0.7$ -- 0.9 mag. The previously determined distance to the 
cluster was in the range 6.0 -- 6.5 kpc \citep{MFJ79,SuBa06,DLK08,JPO11}, while \citet{CW84} obtained a
far different value of 10.3 kpc. The IMF of the cluster was investigated by \citet{SuBa06} 
and \citet{JPO11}, and its slope was reasonably consistent with the Salpeter/Kroupa IMF 
\citep{Sp55,K01,K02}. It implies that although NGC 1624 is located in the outer Galaxy, 
the nature of the star formation activity is similar to that found in the solar neighbourhood. 

The nearer and younger open cluster NGC 1931 is associated with the glowing nebula Sh2-237. A few bright 
MS stars and many PMS stars constitute the cluster \citep{PES13}. The main ionizing source of the 
bright nebula is thought to be the B0.5 star \citep{GDK75} or two B2 stars \citep{PES13}. A number of PMS star 
candidates were identified by \citet{PES13} from the Two Micron All Sky Survey (2MASS; \citealt{2mass}) 
catalog and {\it Spitzer} InfraRed Array Camera (IRAC) photometry. The spatial 
distribution of members seems to be divided into two stellar groups, a northern and a southern group 
\citep{BB09,PES13}. The fundamental parameters of the cluster were obtained from photometric 
studies \citep{MFJ79,PM86,BPMP94,BB09,PES13}. According to these studies, the reddening and distance 
of the cluster are $E(B-V) = 0.5$ -- 1.0 mag and $d = 1.8$ -- 2.4 kpc, respectively. A recent 
study \citep{PES13} investigated the properties of dust toward the cluster using 
polarimetric and photometric data. The reddening law toward the cluster estimated from 
the Serkowski law \citep{SMF75} and color excess ratios indicated that the size distribution of dust grains 
may be different from that of the general diffuse interstellar medium (ISM). They also derived the IMF of the northern,  southern, and 
the entire cluster regions, respectively. The slope of the IMF appears to be shallower than the 
Salpeter/Kroupa IMF.

The present work on NGC 1624 and NGC 1931 is the fifth paper of the Sejong Open cluster Survey (SOS) 
project which was initiated to provide homogeneous photometric data for many open clusters. 
The overview of the SOS project can be found in \citet{SLB13} (hereafter 
Paper 0). Comprehensive studies of several open clusters NGC 2353, IC 1848, and NGC 1893 
were published as part of the project \citep{LSKI11, LSKBK14,LSKBP14}. In this work, 
we revisit the reddening law as well as the fundamental parameters of these clusters in 
a homogeneous manner. The IMF of the clusters is also studied in order to investigate the property of 
star formation activity in the outer Galaxy. The observation and reliability of our photometry 
are described in Section 2. In Section 3, we present several fundamental parameters of 
the clusters obtained from photometric diagrams. The reddening law toward two clusters is 
also discussed in this section. We construct the Hertzsprung-Russell diagram (HRD) in Section 4 
and derive the IMF of the clusters in Section 5. Several discussions on the age spread of PMS stars 
are made in Section 6. Finally, the comprehensive results from this study are summarized 
in Section 7.

\section{OBSERVATION}

\subsection{AZT-22 1.5m Telescope of Maidanak Astronomical Observatory}

The observations of NGC 1624 was made on 2006 November 24, using the 
AZT-22 1.5m telescope (f/7.74) at Maidanak Astronomical Observatory (MAO) in Uzbekistan. 
All imaging data were acquired using Fairchild 486 CCD camera (SNUCam; \citealt{IKC10}) with the 
standard Bessell $UBVI$ filters \citep{B90}. The field of view (FOV) is about $18\farcm1 \times 18\farcm1$. The characteristics of the 
CCD chip have been described by \citet{LSKI08} in detail. The observations comprised
8 frames that were taken in two sets of exposure times for each band 
-- 3 and 60s in $I$, 3 and 180s in $V$, 5 and 300s in $B$, and 15 and 600s in $U$. To transform 
instrumental magnitudes to standard magnitudes and colors we made observations of several 
equatorial standard stars \citep{MMLCE91} over a wide air mass range of $X \sim$ 1.2 -- 2.0 on 
the same night. The observations are summarized in Table~\ref{tab1}, and the left-hand panel 
in Figure~\ref{fig1} displays the finder chart (dashed line) for the observed stars brighter than $V = 18$ mag. 

All pre-processing to remove instrumental artifacts was carried out using the \textsc{IRAF}\footnote{Image 
Reduction and Analysis Facility is developed and distributed by the National Optical Astronomy 
Observatories, which is operated by the Association of Universities for Research in Astronomy under 
operative agreement with the National Science Foundation.}/\textsc{CCDRED} packages as described in 
\citet{LSKI08}. Simple aperture photometry was performed for the standard stars with an aperture 
size of 14.0 arcsec (26.3 pixels). Atmospheric extinction coefficients were determined from the photometric 
data of the standard stars with a weighted least-square method. The coefficients are presented with photometric 
zero points in Table~\ref{tab2}. Using \textsc{IRAF}/\textsc{DAOPHOT} we conducted point spread function (PSF) photometry 
on stars in the target images using a small fitting radius of 1  full width at half-maximum (FWHM; $\leq$ 1.0 arcsec), 
and then aperture correction was done using the aperture photometry of bright, isolated stars with a photometric 
error smaller than 0.01 mag in individual target images. The instrumental magnitude was 
transformed to the standard magnitude and colors using the transformation equations below 
(Paper 0):

\begin{equation}
M_{\lambda} = m_{\lambda} - (k_{1\lambda} - k_{2\lambda}C)\cdot X + \eta _{\lambda} \cdot C 
+ \alpha _{\lambda} \cdot \hat{UT} + \zeta _{\lambda}
\end{equation}

\noindent where $M_{\lambda}$, $m_{\lambda}$, $k_{1\lambda}$, $k_{2\lambda}$, $\eta _{\lambda}$, 
$C$, $X$, $\alpha _{\lambda}$, $\hat{UT}$, and $\zeta _{\lambda}$ are the standard magnitude, 
instrumental magnitude, the primary extinction coefficient, the secondary extinction coefficient, 
transformation coefficient, relevant color, air mass, time-variation coefficient, time difference relative 
to midnight, and photometric zero point, respectively. We adopted the SNUCam transformation coefficients 
($\eta _{\lambda}$), which were recently refined from those in \citet{LSBKI09}. 

\subsection{Kuiper 61" Telescope of Steward Observatory}

The observations of NGC 1624 and NGC 1931 were made on 2011 October 29, using the Kuiper 61" 
telescope (f/13.5) of Steward Observatory (SO) on Mt. Bigelow in Arizona, USA. We obtained images 
with the Mont4k CCD camera and 5 filters (Bessell $U$, Harris $BV$, Arizona $I$, and H$\alpha$). The 
FOV is about $9\farcm7 \times 9\farcm7$. Later, additional images of NGC 1931 were taken
on 2013 February 5, without the H$\alpha$ filter. Since typical seeing was $1\farcs0$ -- 
$2\farcs0$ (7 -- 14 pixels), the $3\times3$ binning mode ($0\farcs42$ per pixel) allowed us not only to obtain 
an appropriate FWHM ($\sim$ 3 pixels) for PSF photometry but also to decrease the readout time 
($\sim$ 10s). In order to obtain reliable transformation relations for this photometric system across 
a wide color range, we observed several extreme red and blue standard stars \citep{KWRMC98} as 
well as a large number of equatorial standard stars \citep{MMLCE91} at air masses of 1.2 -- 2 during 
the run. In addition, a few Landolt standard star fields (Rubin 149, 152, and PG 2213 -- \citealt{L92}), 
which contain both extreme red and blue stars within a small region, were also observed to obtain the secondary extinction 
coefficients. The observations are summarized in Table~\ref{tab1}, and the finder chart (solid line) for the stars brighter than 
$V = 18$ mag is shown in Figure~\ref{fig1}. 

The Mont4k photometric system of the Kuiper 61" telescope was used in this survey project for the first time. Understanding 
the characteristics of the photometric system is essential for obtaining reliable photometric data. 
We address the characteristics and transformation relations for the Mont4k CCD photometric system 
in the Appendix. Atmospheric extinction coefficients and photometric zero points obtained from the 
aperture photometry of the standard stars are presented in Table~\ref{tab2}. 
PSF photometry and aperture correction for the images of NGC 1624 and NGC 1931 were 
carried out using the same procedures as described in the previous section. The instrumental magnitude 
was transformed to the standard magnitude and colors using the transformation equations addressed 
in the Appendix.

\subsection{The Consistency and Completeness of the Photometric Data}

We have confirmed from a series of studies \citep{LSKI11,LSKBK14,LSKBP14} that the 
photometric data obtained from the photometric system of the AZT-22 1.5m telescope at MAO 
are well-tied to the Johnson-Cousins standard system. But we need to check the homogeneity 
of the photometric data obtained with the Mont4k CCD camera of the Kuiper 61" telescope. 
Since NGC 1624 was observed with the two different photometric systems, it was 
possible for us to compare our photometric data directly. Figure~\ref{fig2} shows 
the comparison between a few sets of modern CCD photometry. Bold dots (black) in the upper panels 
represent the differences between the photometric data obtained at MAO and SO. 
We confirmed that our photometry is in good agreement within 0.01 mag 
(see also Table~\ref{tab3}). The imaging data for NGC 1931 were taken with the Mont4k CCD 
camera of the Kuiper 61" telescope at two different epochs. The internal consistency of the two data sets 
was examined as shown by bold dots in the lower panels of Figure~\ref{fig2}. The photometry is well 
consistent with each other within 0.01 mag. We averaged the data sets for each cluster 
using a weighted average scheme \citep{SL95}.

A couple of photometric studies with a modern CCD camera have been made 
for NGC 1624 (e.g. \citealt{SuBa06,JPO11}). The combined photometric data for the 
cluster were compared with those of the previous studies as marked by squares and 
triangles in the upper panels of Figure~\ref{fig2} (see also Table~\ref{tab3}). The 
photometric data of \citet{SuBa06} are far different from ours, particularly 
the $V$ magnitudes. All their colors are systematically redder than the others. 
There is also a color-dependent trend for $U-B \leq 0.7$. 

The difference between the photometry of \citet{JPO11} and ours is acceptable in $V$ 
and $B-V$, whereas the other colors show systematic differences. Although their $U-B$ color 
is in a good agreement with ours for $U-B \leq 0.4$, for redder stars their $U-B$ colors are
systematically redder than ours. Their $V-I$ color appears bluer than that of the others. 
We note that the previous sets of CCD data for NGC 1624 are likely to 
involve problems in the photometric standardization. On the other hand, the photometry of NGC 1931 by 
\citet{PES13} is in good agreement with ours. We present a summary of the comparison 
in Table~\ref{tab3}. 

Statistical analysis based on photometric data requires the completeness 
of the photometry to be at least 90 percent. In the case of sparse open clusters, the 
completeness limit is not very different from that obtained in nearby field region,
and that relates to the turnover magnitude in the luminosity function of the
observed stars. We estimated the completeness of our photometry from the turnover in the luminosity 
function of all the observed stars, by assuming a linear slope across the entire 
magnitude range. Our photometry is complete down to $V = 19$ mag ($\sim 2 \ M_{\sun}$) for 
NGC 1624 and 20 mag ($\sim 1 \ M_{\sun}$) for NGC 1931. However, the completeness limit may 
be an upper limit because the bright nebulae surrounding the young open clusters may prevent the 
detection of faint stars. The photometric data from this work are available 
in the electronic tables (Table~\ref{tab4} and \ref{tab5}) or from the authors (BL or HS).

\section{PHOTOMETRIC DIAGRAMS}

Photometric studies for young open clusters are based on the two-color diagram (TCD) 
and color-magnitude diagrams (CMDs). With well-calibrated empirical 
relations and stellar evolution models, fundamental parameters, such as reddening, 
distance, and age, can be determined from these photometric diagrams. The fundamental 
parameters allow us to investigate the local spiral arm structure in the Galaxy (Paper 0) as well as the 
properties and evolution of the dust grain through the reddening law \citep[Paper 0, etc]
{PUNO03,LSKI11,LSKBK14,LSKBP14,SB14}. In addition, the age distribution of stars can be used to 
study the star formation history within a star-forming region \citep[etc]{PaSt99,SHBMC04,
SB10,LSKBP14} and the pattern speed of the Galactic spiral arms \citep{DL05}. These studies 
should be based on reliable membership selection. In the following sections, we present 
membership selection criteria, the reddening law, and fundamental parameters of 
NGC 1624 and NGC 1931 with TCDs and CMDs.

\subsection{The Extent of NGC 1624 and NGC 1931}
As most open clusters are unbound stellar systems, i.e. expanding systems 
\citep{LL03}, it is believed from theoretical approaches, that gas expulsion 
affects the dynamical evolution of the remaining clusters significantly \citep{T78,
GB06}. In some SFRs, new generation stars are still forming in the natal 
clouds swept away by the stellar wind and the UV radiation from high-mass stars 
(e.g. \citealt{SHBMC04,KAG08,LSKBP14}). The observed structure is basically a
projection onto the sky. Hence, it is difficult to determine the physically meaningful 
extent of young open clusters. Alternatively, radial surface density profile 
(RSDP) gives an area encompassing almost all the members of a cluster. This is a 
fundamental method to define the size of clusters. 

We determined the radius of NGC 1624 assuming a circular shape. The center of the 
cluster was set to the position of the brightest star NGC 1624-2. The stars 
observed in the $V$ band were counted within concentric rings increased 
by $0\farcm5$. The RSDP of NGC 1624 is plotted in Figure~\ref{fig3}. The error in the 
surface density was assumed to follow Poisson statistics. The surface density decreases 
with the increasing distance from the center until it indistinguishably converges to the 
surface density of field stars. The distance from the center to the convergent point 
($\sim 2\farcm5$) was assigned as the radius of the cluster. \citet{JPO11} also obtained 
a consistent value ($2\farcm0$) with ours using their optical and 2MASS near-infrared (NIR) 
data. 

The extent of NGC 1931 could not be determined from the RSDP. \citet{PES13} 
presented the spatial distribution of PMS members in their figure 25. The spatial distribution 
could not be confined to a small specific region within our small FOV ($\sim 10'  \  \times  \ 10' $) because 
the PMS stars scattered across the emission nebula (Sh2-237). Given that no PMS member has been identified 
northwards from $\delta \sim 34^{\circ} \ 18'$ \citep{PES13}, we assigned the southern region 
covered by our observation as likely to encompass all the 
members of the cluster. The area of these clusters is used to normalize the IMF as well as to 
isolate members.

\subsection{Membership Selection}
The early-type stars (O -- B-type) can be unambiguously identified in the ($U-B$, $B-V$) TCD as they have 
very blue $U-B$ colors (see figure 11 of Paper 0). In addition, the intrinsic colors and absolute 
magnitude of these stars have been well calibrated in the optical passbands, so that the reddening and 
distance of the stars can be reliably determined by comparing the observed colors and magnitude with 
the intrinsic relations. These procedures are described in detail in the following sections. We identified early-type 
members within specific reddening and distance ranges based on the ($U-B$, $B-V$) TCD (Figure~\ref{fig4}) 
and CMDs (Figure~\ref{fig5}). The membership selection criteria for the early-type 
main sequence (MS) stars of NGC 1624 are (1) $V \leq 18$ mag, $0.5 \leq B-V \leq 0.9$, $-1.0 \leq U-B \leq 0.5$, 
and $Q^{\prime} \leq -0.4$, where $Q^{\prime} \equiv (U-B) - 0.72(B-V) - 0.025E(B-V)^2$ (Paper 0), 
(2) $E(B-V) > 0.8$ mag, (3) for late-B type members ($-0.4 < Q^{\prime} \leq -0.2$) the individual distance 
modulus should be between ($V_0-M_V)_{\mathrm{cl}} - 0.75 - 2 \sigma _{V_0-M_V}$ and ($V_0-M_V)_{\mathrm{cl}} 
+ 2 \sigma _{V_0-M_V}$, where ($V_0-M_V)_{\mathrm{cl}}$ and $\sigma _{V_0-M_V}$ are the distance modulus 
of NGC 1624 and the standard deviation of the distance modulus, respectively. The factor -0.75 is introduced 
to take into account the effect of equal mass binaries. The FOV of SNUCam ($18\farcm1 \times 18\farcm1$) 
is wide enough to include the distributed population, most of which may not be associated with NGC 1624. In order 
to prevent the inclusion of such stars we isolated only stars within a radius of $2\farcm5$ from the brightest 
star NGC 1624-2. The star ID 1609 ($V = 14.94$, $V-I = 1.05$, $B-V = 0.87$, $U-B = 0.24$ ) was rejected 
from the member list because the star is too bright to be the member of the cluster at a given color in the CMDs. 
These processes of the membership selection for early-type MS stars were carried out iteratively. A total 
of 14 stars were selected as the early-type MS members of NGC 1624. 

Similar criteria were used to select the early-type MS members of NGC 1931. The criteria are 
(1) $V \leq 17$ mag, $0.2 \leq B-V \leq 0.9$, $-1.0 \leq U-B \leq 0.4$, and $Q^{\prime} 
\leq -0.4$, (2) $E(B-V) > 0.5$ mag, (3) 10.8 mag $\leq (V_0 - M_V)_{\star} \leq$ 12.3 mag for late-B type 
members. Several field interlopers were identified from these criteria. We checked the color excess ratio 
of the most probable members ($U-B \leq -0.1$) and excluded outliers with significantly different color 
excess ratios from those of the most probable members. The abnormal color excess ratios of these outliers 
are caused by the fact that less reddened foreground F or G-type stars would be regarded as highly reddened 
early-type stars from the color-cuts. The star ID 314 ($V = 13.11$, $V-I =  0.53$, $B-V = 0.36$, $U-B = -0.31$) was originally selected 
as an early-type member, however, its spatial position ($\Delta \alpha \sim -3\farcm23$, $\Delta \delta 
\sim +5\farcm11$) is outside the cluster boundary. We regarded the star as a mid-B-type field interloper 
at a similar distance, and therefore the star was excluded from the member list. A total 
of 14 stars were assigned as early-type MS members of NGC 1931. It is worth noting that 
BD+34 1074, which is one of the two brightest stars in NGC 1931, was resolved into four stars, 
of which three (ID 857, 865, and 872) are also the early-type members identified here.

Most low-mass PMS stars have a circumstellar disk, and some of these disks exhibit significant accretion activity 
\citep{LSKBK14,LSKBP14}. According to the standard accretion model for PMS stars \citep{US85,BBB88,K91}, 
material channeled from the circumstellar disk falls onto the surface of the central star along its magnetosphere. 
Various emission lines and hot continuum excess emission arise from accretion columns, preshock infall region, 
and the heated photosphere \citep{CG98,H99}. The generated energy is mainly released at ultraviolet (UV) wavelengths, so that 
the accretion luminosity is related to the $U$ magnitude \citep{GHBC98,CG98}. Hence, a few 
PMS stars with strong UV excess emission can be identified in the $U$ band photometry. 
There is a dotted line parallel to the reddening vector in the ($U-B, B-V$) TCD (the left-hand panels 
of Figure~\ref{fig4}). The stars with a bluer $U-B$ color than the dotted line 
at a given $B-V$ color are either early-type stars or UV excess stars. The criteria for the PMS stars 
with UV excess emission are (1) $\epsilon (U-B) \leq 0.1$, (2) $B-V > 1$ and $U-B < 0.72[(B-V)+0.5)] 
- 0.9$, (3) stars within the empirical PMS locus \citep{SBCKI08} in the $(V, V-I)$ CMD (see Figure~\ref{fig5}). 
The last criterion prevents the inclusion of background early-type stars. We found 3 and 4 PMS stars 
with UV excess emission in NGC 1624 and NGC 1931, respectively.

H$\alpha$ photometry provides a good criterion to find PMS stars in young open clusters 
($\leq$ 3 Myr). Since \citet{SBL97} achieved the successful detection of many PMS stars in 
NGC 2264, this efficient technique has been used to search for PMS members in a series 
of studies (e.g. \citealt{SBL98,SCB00,PSBK00,PS02,SBC04,SB04,SBCKI08,SSB13,
LSKBK14,LSKBP14}). In order to identify H$\alpha$ emission stars photometrically we have defined the 
H$\alpha$ index as H$\alpha$ - ($V + I$)/2 \citep{SCB00}. As shown in the right-hand panels of 
Figure~\ref{fig4}, stars with -0.2 (dashed line) or -0.1 mag (dotted line) smaller H$\alpha$ index than 
the empirical photospheric level (solid line) of normal MS stars are selected as 
H$\alpha$ emission stars or candidates. The H$\alpha$ emission stars and candidates ID 1897 
($V = 18.68$, $B-V = 1.17$, $U-B = 0.41$), ID 2326 ($V = 18.97$, $B-V = 1.62$, $U-B = 1.42$), ID 2356 
($V = 19.11$, $B-V = 1.36$, $U-B = 1.14$), and ID 2588 ($V = 19.12$, $B-V = 1.17$, $U-B = 0.44$) 
in the FOV of NGC 1624 were regarded as foreground stars because their colors are similar 
to those of less reddened foreground stars. In addition, the radius ($2\farcm5$) of the 
cluster was also used to isolate the members. We found 9 H$\alpha$ emission stars and 
candidates associated with NGC 1624. In the same way, we found 28 H$\alpha$ 
emission stars and candidates in NGC 1931, one of which (ID 596; $V = 19.30, 
B-V =1.36, U-B = 1.00$) is likely a foreground late-type star. 

The dust emission from circumstellar disks of PMS stars is prominent at infrared (IR) wavelength, particularly 
the mid-infrared (MIR). A number of young stellar objects in various star-forming regions have been identified 
through extensive imaging surveys with the {\it Spitzer} space telescope \citep[etc]{CMP08,GMM08,KAG08,
SSB09}. We used the {\it Spitzer} Galactic Legacy Infrared Mid-Plane Survey Extraordinaire 
360-degree catalog (GLIMPSE360; \citealt{WAB08,WBM11}) to identify PMS members with MIR excess 
emission. The GLIMPSE360 survey is a ``Warm Mission", so that only the
3.6 and 4.5 $\micron$ bands were available. Unfortunately, NGC 1624 ($l = 155\fdg356$, $b = 2\fdg616$) 
was not covered in the survey program because of its high Galactic latitude. The catalog allows us to select 
PMS members with a circumstellar disk in NGC 1931. A total of 1249 
optical counterparts in the catalog were found within a matching radius of $1\farcs0$. We attempted 
to identify the PMS members in the ($[3.6] - [4.5], V-I$) TCD as shown in Figure~\ref{fig6}. The majority 
of the H$\alpha$ emission stars exhibited MIR excess emission. Stars with a $[3.6] - [4.5]$ color larger 
than 0.2 mag (dashed line) were selected as PMS member candidates, and then the empirical PMS 
locus \citep{SBCKI08} was used to isolate the PMS members among the candidates. From this procedure, 
a total of 54 PMS stars with MIR excess emission were selected as members of NGC 1931. Although it is difficult to 
make a detailed classification due to the absence of information in other IRAC bands (5.8 and 8.0 $\micron$), 
most of them may be Class II objects (see also Figure 18 of \citealt{PES13}). 

It is a well known observational fact that PMS stars, particularly classical T-Tauri (CTTS) stars, are 
variable objects with an amplitude of 0.1 -- 2 mag in the form of irregular variation \citep{GMBHS07}. 
The observations of NGC 1624 and NGC 1931 were made at two different epochs. The time difference 
between the first and the second observation was about $\sim$ 4.9 and 1.2 years for NGC 1624 and 
NGC 1931, respectively. Since the photometric errors from the averaging process used in section 2.3 (originally 
from equation (2) of \citealt{SL95}) represent the consistency of magnitude and colors from several observations, 
a photometric error larger than that expected from the distribution of errors with magnitude may be attributed 
to a genuine variation in brightness. We investigated the photometric error of the stars observed more than 
twice as shown in Figure~\ref{fig7}. The photometric errors ($\sigma _V$) in a given magnitude bin 
($\Delta V = 1$ mag) were averaged, and then its standard deviation ($\sigma _{S.D.}$) was used as a criterion 
for the variability of individual stars. If stars have variations larger than 0.03 mag in brightness and $\sigma _V > \langle 
\sigma _V \rangle + 3 \sigma_{S.D.}$ in the range of $V = 10$ -- 21 mag, we assigned the stars as variables. The star 
ID 727 in NGC 1931 showed a very large variation ($V = 16.34$ mag on 2011 Oct. 29 and $V = 14.86$ mag on 2013 Feb. 5). 

We found 4 variable stars within the radius of NGC 1624. The variable star 
ID 1733 has a red $U-B$ color similar to that of foreground stars. The $U-B$ color of the 
star in 2011 was bluer by 0.24 mag than that observed in 2006. The star may be an 
active late-type star in front of the cluster. The $V$ magnitude and $V-I$ color of the other 
variable stars appear to be commensurate with those of the H$\alpha$ emission stars 
in the PMS locus. A total of 16 stars were identified as variable stars in NGC 1931. The majority of the 
variable stars were crowded into the cluster center. Their $V$ magnitudes and colors 
are similar to those of other PMS members with H$\alpha$ and MIR excess emission, 
some of which ($\sim 44$ percent) are indeed UV excess, H$\alpha$ emission, or 
MIR excess emission stars. Thus, these variable stars were assigned as PMS members of 
NGC 1931. 

A total of 28 stars (14 early-type and 14 PMS stars) in NGC 1624 and 85 stars (14 early-type and 
71 PMS stars) in NGC 1931 were selected as members. The star ID 1773 (No. 228 of \citealt{DLK08}) 
in NGC 1624 is the only star selected using 2 membership selection criteria (UV excess and H$\alpha$ 
emission), and we found 24 out of 71 PMS members in NGC 1931 satisfied more than 2 selection criteria. 
The membership selection for early-type MS members is likely complete, however only a small fraction of 
the PMS stars may have been selected as members of the clusters. According to a study of the young open cluster 
NGC 1893 \citep{LSKBP14}, the detection efficiencies of H$\alpha$ photometry, {\it Spitzer} MIR, 
and {\it Chandra} X-ray data turned out to be about 10, 24, and 85 percent for PMS stars ($> 1 \ M_{\sun}$), 
respectively. This implies that X-ray emission from PMS stars is the most efficient criterion to identify the 
remaining PMS stars. Although NGC 1624 has been observed using the {\it Chandra} X-ray Observatory 
(ObsID 7473, PI Garmire), the exposure times were not long enough to detect PMS members. Using the 
published X-ray source list of \citet{EPG10}, the optical counterparts of X-ray emission sources 
and candidates were searched for with matching radii of 1.0 and 1.5 arcsec, respectively. We found only 
2 X-ray emission sources and 1 candidate with optical counterparts. The brightest early-type member 
NGC 1624-2 is known to be an X-ray emitter \citep{EPG10,WAM12}, and the others turned out to be 
field interlopers given their colors. On the other hand, X-ray observations for NGC 1931 have not yet been 
made. Extensive X-ray imaging observations are clearly required in order to study the nature of PMS stars 
in detail based on complete membership lists. As of now, we merely anticipate that a few hundreds of 
members may exist in each cluster. A discussion on the approximate number is addressed in a later section.

\subsection{Structure of the Clusters}

The structure of young open clusters gives us clues to dynamical evolution in the 
early stages of cluster formation as well as the formation process of stellar clusters  
\citep{EEPZ00}. Many efforts have been made to study the structure of open clusters 
using the techniques of surface density distribution and minimum spanning 
tree \citep[etc]{SBLKL99,SPO07,KAG08,GMM09,KSB10,JPO11,SSB13,LCS13,PES13}. 
In this section, we describe the apparent structure of NGC 1624 and NGC 1931 using the 
surface density distribution, with additional information from previous studies.

The appearance of the HII region Sh2-212 associated with NGC 1624 is close 
to a symmetric shell structure. An inner region filled with hot ionized gas and a 
shocked outer region constitute the apparent gas structure (see figure 1 of 
\citealt{JPO11}). A molecular filament with different velocity components 
in the range of -32 to -37 km s$^{-1}$ surrounds the southern part of the cluster, 
stretching out to the north-west \citep{DLK08}. The filament incubates at least five 
clumps, one of which is likely to be associated with an UCHII in the 
western part of the cluster. The brightest star NGC 1624-2 (O7f?p -- \citealt{WSA10,SAW11}) is 
located near the center of the cluster. Other members are distributed in the vicinity 
of the star as shown in the left-hand panel of Figure~\ref{fig8}. We 
obtained the surface density distribution of NGC 1624 using stars observed 
in the $V$ band (contour). The concentration of stars appears 
high in the center of the cluster, and the spatial distribution approximates to a circular shape. 

On the other hand, NGC 1931 is enclosed within a 
dusty molecular cloud (see figure 1 of \citealt{PES13}). The natal cloud also 
exhibits a hierarchical structure. The inner region is filled with hot ionized gas 
produced by the radiation from a few bright stars while polycyclic aromatic 
hydrocarbon molecules are glowing in the outer region. The right-hand 
panel in Figure~\ref{fig8} shows the spatial distribution of cluster 
members. The early-type MS members appear to be divided into two groups, 
a northern and a southern group. A high stellar density region is seen between 
the groups, implying that mass segregation among the members may not yet be established. 
While many variable PMS members are located in the dense region, the majority of 
H$\alpha$ emission stars are found in the vicinity of the southern group. The PMS 
members with MIR excess emission are spread out across the whole region, however the stars 
in the northern part of the cluster show a weak concentration. The surface 
density distribution (contour) reflects such an elongated 
shape. The apparent shape is in a good agreement with the result of \citet{PES13}. 
These observational properties may be related to the star formation history 
within NGC 1931.

\subsection{Reddening and the Reddening Law}

As most open clusters are formed in the Galactic plane, a region where interstellar matter is 
unevenly distributed, reddening corrections are crucial to obtaining reliable 
physical quantities. The interstellar reddening is basically determined by comparing an observed 
color with the intrinsic one. The ($U-B, B-V$) TCD is a very useful tool because the reddening 
vector has been well established in the diagram [$E(U-B)/E(B-V) = 0.72 + 0.025E(B-V)$ -- Paper 0]. The intrinsic 
color relations of Paper 0  were adopted to obtain the reddening of the individual early-type 
members (see table 1 in Paper 0). 

The reddening of NGC 1624 determined from 14 early-type members is in the range of 
$E(B-V) = 0.83$ -- 1.01 mag, and the mean value is $\langle E(B-V) \rangle = 0.92 \pm 
0.05$ (s.d.) mag. The result is in close agreement with that of previous studies, e.g. 
$E(B-V) = 0.88$ -- 0.94 mag \citep{MFJ79}, 0.84 -- 0.87 mag \citep{CW84}, 0.70 -- 0.90 
mag \citep{SuBa06}, and 0.76 -- 1.00 mag \citep{JPO11}. The reddening of NGC 1931 
obtained from 14 early-type members ranges from $E(B-V) = 0.51$ -- 1.01 mag. 
The mean reddening is $\langle E(B-V) \rangle = 0.74 \pm 0.17$ (s.d.) mag. This result is 
also commensurate with that of previous studies, e.g. $E(B-V) = 0.49$ -- 0.93 mag 
\citep{MFJ79}, 0.47 -- 1.00 mag \citep{PM86}, 0.55 -- 1.00 mag \citep{BPMP94}, 
0.52 -- 0.72 mag \citep{BB09}, and 0.50 -- 0.90 mag \citep{PES13}. The dispersion in 
the reddening indicates that there is differential reddening across each cluster. The 
differential reddening in NGC 1624 appears to be less significant than that found in NGC 1931. 

Unlike early-type MS stars, it is difficult to determine the reddening for the PMS stars 
because the intrinsic colors can be altered by hot and cold spots on the surface, accretion 
activities, and the obscuration by a circumstellar disk \citep{GMBHS07}. Simultaneous imaging 
(color) and spectroscopic (spectral type) observations are required to determine the reddening 
of individual PMS stars accurately. In the absence of such data, we estimated the reddening of 
PMS stars by using a weighted-mean reddening value at a given position from the reddening distribution of 
the early-type members, where the weight was given by an exponential 
function with respect to the distance from the individual early-type members. 

The reddening law toward young open clusters provides an opportunity to study the spatial 
distribution of interstellar matter in the Galaxy as well as the
dust evolution in SFRs. The ratio of total-to-selective extinction ($R_V$) is a basic tool 
to investigate the reddening law, being also a crucial parameter to correct the total 
extinction in the visual band. The parameter is closely related to the size distribution 
of dust grains. The $R_V$ found in several extremely young SFRs present deviations 
from the normal reddening law \citep{G10,HSB12}. Previous studies on NGC 1624 
have adopted the normal reddening law ($R_V = 3.1$, \citealt{CW84,
JPO11}), while \citet{PES13} showed, from photometric and polarimetric data, that the 
reddening law toward NGC 1931 is slightly different from that found in the 
general ISM. In order to check the reddening law toward 
these clusters we investigated various color excess ratios in the optical and 
IR passbands as performed in the series of studies \citep{KSB10,LSKI11,HSB12,
SSB13,LSKBK14,LSKBP14}. 

\citet{LSKBK14} have noted that several mid -- late-B-type MS stars in the young 
open cluster IC 1848 have a bluer color than that of normal B-type MS stars. Since 
the bluer color may not reflect the genuine photospheric color of 
the stars, the early-type members with $U-B \leq 0.3$ for NGC 1624 and with $U-B \leq -0.2$ 
for NGC 1931 were chosen to avoid such a color anomaly. We used 2MASS NIR data (tagged 
as a photometric quality flag `AAA') as well as our optical data for the stars. For MIR data, the {\it Spitzer} GLIMPSE360 
catalog \citep{WAB08,WBM11} was used to obtain the $V$-MIR color excess. The various 
color excesses $E(V-\lambda)$ in the optical and IR passbands were computed by comparing 
the observed colors with the intrinsic relations from Paper 0 and Sung et al. (in preparation). 
The $R_V$ is expressed as a linear relation for each color excess ratio $E(V-\lambda)/E(B-V)$ 
\citep{GV89,SSB13}. We present the color excess ratios of the cluster members in Figure~\ref{fig9}. 
The thick solid line (blue) in the figure represents the slope corresponding to the normal 
reddening law ($R_V = 3.1$).

There is a foreground early-type star in the direction of NGC 1624 which was helpful 
to check the difference between the general ISM and the intracluster medium (ICM). As shown 
in the upper panels of Figure~\ref{fig9}, all the color excess ratios consistently indicate 
that the ICM of NGC 1624 is indistinguishable from the general ISM. The mean 
$R_V$ obtained from different color excess ratios was $3.12 \pm 0.01$ (s. d.). 
This result is in good agreement with that examined by \citet{JPO11}, implying that 
dust evolution in the cluster had already progressed or no grain growth had occurred 
in the natal cloud. We adopted $R_V = 3.1$ and obtained a mean extinction of $\langle 
A_V \rangle = 2.85 \pm 0.17 $ mag. On the other hand, there was no available foreground early-type star 
toward NGC 1931 for checking the foreground reddening law. Given that the reddening 
law toward the neighbouring clusters NGC 1893 and Stock 8 turned out to be normal \citep{JPO08,EPM11,LSKBP14}, 
we assumed that the foreground reddening law in the direction of NGC 1931 was likely to be normal. In the lower 
panels of Figure~\ref{fig9}, the color excess ratios of the members  show a significant deviation 
from the normal reddening law. It implies that the size distribution of dust grain in the ICM of NGC 1931 
is far different from that found the general ISM. Thus, the total extinction in $V$ band toward 
the cluster should be divided into two components as below: 

\begin{equation}
A_V = R_{V,\mathrm{fg}}\times E(B-V)_{\mathrm{fg}} + R_{V,\mathrm{cl}}\times [E(B-V) - E(B-V)_{\mathrm{fg}}]
\end{equation} 

\noindent where $A_V$, $R_{V,\mathrm{fg}}$, $R_{V,\mathrm{cl}}$, $E(B-V)_{\mathrm{fg}}$, and $E(B-V)$ 
represent the total $V$ band extinction, the foreground $R_V$, intracluster $R_V$, 
foreground reddening, and total reddening, respectively.  As mentioned above, 
$R_{V,\mathrm{fg}} = 3.1$ was adopted for the foreground component. We simultaneously 
determined $E(B-V)_{\mathrm{fg}}$ and $R_{V,\mathrm{cl}}$ from the $V$-IR color excess ratios 
using a $\chi ^2$ fitting method. The star ID 800 shows significant excess 
emission at all the wavelengths. Another star ID 872 with a photometric quality flag 
`EEE' in the 2MASS catalog also exhibits excess emission in the MIR passbands. 
We did not include these stars in the examination of the reddening law. 
The mean foreground reddening was found to be $E(B-V)_{\mathrm{fg}} = 0.417 \pm 0.005$ (s. d.) mag, 
corresponding to a point of contact between foreground (thick solid line) and intracluster 
(thin solid line) components in the lower panels of Figure~\ref{fig9}. This value is in 
good agreement with the smallest reddening of the neighbouring cluster NGC 1893 
[$E(B-V) = 0.418$ mag, \citealt{LSKBP14}], which is likely the amount of reddening in 
front of the clusters. The mean intracluster $R_{V,\mathrm{cl}}$ determined from $V$-IR 
color excess ratios was found to be $5.2 \pm 0.3$ (s. d.). The slope corresponding to the 
$R_{V,\mathrm{cl}}$ is shown by thin solid lines in the lower panels of Figure~\ref{fig9}. 
  
The $E(V-I)/E(B-V)$ color excess ratio gives an intracluster $R_{V,\mathrm{cl}}$ of 
4.3 (dashed line). The value is significantly different from the mean $R_{V,\mathrm{cl}}$ 
obtained from the $V$-IR colors excess ratios. We attempted to check our $V-I$ 
colors. As discussed in Section 2.3, there is no considerable systematic difference 
between the photometry of \citet{PES13} and ours. The observations of NGC 1624 
was made on the same night, and the multiwavelength study on the reddening 
law showed a consistent result in the optical -- NIR passbands. Hence, our 
photometry and the adopted empirical relations of Paper 0 do not suggest 
any serious systematic problem. A similar aspect was found in the reddening law 
toward the starburst cluster Westerlund 2 \citep{HPS14}. The authors attributed this 
discrepancy to the unknown behaviour of the $V-I$ color in heavily reddened situations. 
It needs to be confirmed whether or not the linear relation between $R_V$ and $E(V-I)/E(B-V)$ 
is applicable to extremely young and dusty SFRs. We leave this issue for future work in our survey project.

Using the IR extinction curve of \citet{FM07} (equation 4) we tested the reddening 
law once more as shown in Figure~\ref{fig10}. The extinction curve of NGC 1624 (open circles 
in the upper panel) is close to the mean Galactic extinction curve (dashed line -- $k_{\mathrm{IR}} = 
1.06$ and $R_V = 3.00$ -- \citealt{FM07}), while that of NGC 1931 still shows a conspicuous 
deviation from the mean curve with a somewhat large scatter. The $R_V$ was determined 
by a $\chi ^2$ fitting method, and the result was $R_V = 3.8 \pm 0.1$ ($1/\lambda 
< 1 \micron ^{-1}$). However, we cannot accept that a small specific column throughout the 
sky can be filled with significantly large dust grains, although a global variation of $R_V$ along the 
Galactic longitude was found \citep{W77,SB14}. We derived the ICM component 
in the extinction curve after subtracting a foreground reddening of $E(B-V)_\mathrm{{fg}} = 0.42$ mag. 
The lower panel of Figure~\ref{fig10} shows the extinction curve of the ICM component. We found
$R_{V,\mathrm{cl}}$ of $5.5 \pm 0.3$ using the same procedure as above. The result is reasonably 
consistent with that obtained from Figure~\ref{fig9} within the uncertainties. We note 
that the $E(V-I)/E(B-V)$ color excess ratio  seems to be ambiguous in this 
analysis because it is commensurate with the mean Galactic extinction curve as well as 
the IR extinction curve of the cluster ($R_{V,\mathrm{cl}} = 5.5$). 

We adopted the segmented reddening law toward NGC 1931 as presented in  
Figure~\ref{fig9} [$R_{V,\mathrm{fg}} = 3.1$, $E(B-V)_{\mathrm{fg}} = 0.42$ mag, and 
$R_{V,\mathrm{cl}} = 5.2$] and obtained a mean extinction of $\langle A_V 
\rangle = 2.97 \pm 0.88$ mag. This result is somewhat different from that obtained through a polarimetric 
and photometric method ($R_V = 3.2$ -- 3.3 -- \citealt{PES13}). A reason is that the Serkowski 
law \citep{SMF75} may not be applicable in the extremely young and dusty SFRs because of the 
efficiency and time scale for grain alignment, the probable complex structure of the magnetic field 
in the ICM, as well as the depolarization effect by foreground dust grains. The other reason is that 
the authors did not consider the foreground and intracluster components separately, and therefore 
their $R_V$ could be underestimated. Nevertheless, their results also suggest that the reddening 
law in the direction of NGC 1931 is not normal. We conclude that the evolution of dust grains within 
the cluster is still in progress. 

 \subsection{Distance}
Young open clusters ($<$ 10 Myr) are a useful tracer of the local spiral arm structure 
in the Galaxy \citep{DL05}. We have been using the zero-age main sequence (ZAMS) 
fitting method to determine distances to open clusters in a series of studies. The canonical 
ZAMS fitting method should be made after correction for interstellar reddening. In order 
to reduce uncertainties arising from the reddening correction, Paper 0 introduced 
the reddening-independent indices as below: 

\begin{equation}
 Q_{VI} \equiv V - 2.45(V - I) 
\end{equation}
\begin{equation}
 Q_{VJ} \equiv V -1.33(V -J) 
\end{equation}
\begin{equation}
 Q_{VH} \equiv V - 1.17(V -H) 
\end{equation}
\begin{equation}
 Q_{VK_{S}} \equiv V - 1.10(V - K_{S})
\end{equation}

\noindent The modified Johnson $Q$ [$Q^{\prime} = (U-B) - 0.72(B-V)-0.025E(B-V)^2$] 
is used as the abscissa of new CMDs. These reddening-independent indices 
give a few advantages in ZAMS fitting 
procedure. The indices are based on $UBVIJHK_S$ multicolor photometry, and 
thus the ZAMS fitting to the CMDs of young open clusters can be made consistently 
with respect to different colors. The colors of early-type stars are less affected by 
metallicity effects because few strong metallic lines apart from a few light 
elements are, in general, found in their spectra. We have determined the distance of 
young open clusters using these indices \citep{LSKBK14,LSKBP14}.

A large fraction of high-mass stars constitute binary systems 
\citep{SMK12}. The luminosity of such stars appears higher than that expected from ZAMS stars at 
a given color or effective temperature. In addition, physical properties, such as stellar 
rotation and overshooting, cause the MS band to be wider in the high-mass regime \citep{MP13}. Therefore, 
the faintest MS stars at a given color in the CMDs have been regarded as ZAMS 
stars \citep{JH56}. It is worth noting that some of the early-type members may possibly 
reveal a color anomaly as mentioned in the previous section. Allowing for this point we 
carefully fitted the ZAMS relations of Paper 0 to the lower ridge line of the cluster sequence 
in Figure~\ref{fig11} after adjusting the relations above and below it. The lower ridge line 
could be confined between the ZAMS relations (dashed lines) shifted by $\pm$ 0.2 -- 0.3 mag 
for each cluster. The fitted distance modulus of NGC 1624 and NGC 1931 was $13.9 \pm 0.2$ 
mag ($6.0 \pm 0.6$ kpc) and $11.8\pm0.3$ mag ($2.3 \pm 0.3$ kpc), respectively. This result 
places NGC 1624 ($l = 155\fdg356$, $b = 2\fdg616$) in the outer arm and NGC 1931 
($l = 173\fdg898$, $b = 0\fdg281$) in the Perseus arm. 

Previous studies used either ZAMS or isochrone fitting in the determination of 
distance. The distance to NGC 1624 obtained in this work is in good agreement with the results of previous studies, 
e.g. 6.0 kpc \citep{MFJ79,SuBa06,JPO11} and 6.1 -- 6.5 \citep{DLK08}. It is interesting 
that \citet{SuBa06} obtained a distance consistent with that of other studies although 
their photometry shows a serious systematic difference in $V$ (Figure~\ref{fig2}). 
\citet{CW84} obtained a distance of 10.3 kpc using spectroscopic parallax for 2 bright stars. 
In their study, the stars were classified as O6I and B1III, respectively. Since their spectral 
classification is somewhat different from that of a recent study for the brightest star NGC 1624-2 (O7f?p -- 
\citealt{WSA10,SAW11}), their distance may be shifted systematically. A similar discussion 
has been made in \citet{JPO11}. In the case of NGC 1931, \citet{PES13} refined their previous 
results \citep{PM86,BPMP94} and published a new distance of $2.3\pm0.3$ kpc. \citet{BB09} 
obtained a distance of $2.4\pm0.3$ kpc from the 2MASS NIR photometry. Although \citet{MFJ79} reported a rather 
smaller value (1.8 kpc), our result is in good agreement with that of more recent studies. 
 
\section{HERTZSPRUNG-RUSSELL DIAGRAM}

The HRD is a basic tool to understand the evolution of 
stars. The effective temperature ($T_{eff}$) of stars is a crucial parameter for constructing 
the HRD. We used the relations published in Paper 0 (table 5) to obtain the $T_{eff}$ 
and bolometric correction for individual stars. The $T_{eff}$ of the 
star NGC 1624-2 was determined from the spectral type-$T_{eff}$ relation. Although 
the spectral type and luminosity class of the star seem to be uncertain - O5.5V 
\citep{MFJ79}, O6I \citep{CW84}, O7f?p \citep{WSA10,SAW11}, and O6.5f?cp--O8f?cp 
(a variation in the spectral type has been reported by \citealt{WAM12}), we assumed the star 
to be an O7 MS star according to the recent classification \citep{SAW11,WAM12}. It is worth 
noting that the absolute magnitude of NGC 1624-2 ($M_V =$ -4.66 mag) is reasonably 
consistent with that of O7 MS stars ($M_V =$ -4.90 mag) rather than that of O7 supergiant 
stars ($M_V =$ -6.05 -- -6.95 mag, see table 4 of Paper 0). The difference between an O7 MS and giant 
star is only about 165 K in $T_{eff}$, and the stars have the same bolometric correction value (Paper 0). 
If NGC 1624-2 is a supergiant star, the difference between MS and supergiant stars increases 
up to 800 K in $T_{eff}$ and 0.1 mag in bolometric correction. The $T_{eff}$ of other early-type MS 
stars was inferred from the color-$T_{eff}$ relations. We averaged the $T_{eff}$ estimated from each 
color-$T_{eff}$ relation with weights. 

We set the weights of $T_{eff}$ estimated from $U-B$ vs. $T_{eff}$ 
relation to be 0.5, 1.0, 1.5, 2.0, and 2.5 in the $(U-B)_0$ color range of 0.07 to -0.1 mag, -0.1 to -0.3 mag, 
-0.3 to -0.5 mag, -0.5 to -0.7 mag, and -0.7 to -1.2 mag, respectively, set it to be 0 for late-type stars. 
The temperature sensitivity of the $V-I$ color is not high enough to estimate the $T_{eff}$ of early-type stars. We assigned 
each weight of $T_{eff}$ inferred from $B-V$ vs. $T_{eff}$ relation and $V-I$ vs. $T_{eff}$ relation to 1.0 and 
0.5 for stars with $(B-V)_0 < -0.24$ mag. For MS stars with $(B-V)_0 < 0.18$ mag, the weights of $T_{eff}$ 
estimated from $B-V$ vs. $T_{eff}$ relation and $V-I$ vs. $T_{eff}$ relations were set to be 1.0 and 0.7, respectively. 
Since the $V-I$ color is a good 
temperature indicator for cool stars, we only used $V-I$ vs. $T_{eff}$ relations \citep{B95,BCP98} for 
PMS stars. The bolometric correction values for all the members were inferred from their $T_{eff}$ 
using table 5 in Paper 0. We present the HRD of NGC 1624 and NGC 1931 in Figure~\ref{fig12}. 

The brightest star in NGC 1624 is an O7 star \citep{WSA10,SAW11}. If the star is a MS 
star with solar metallicity, its lifetime is smaller than 10 Myr according to \citet{M13}. 
On the other hand, the spectral type of the most luminous star in NGC 1931 is 
somewhat uncertain. \citet{PES13} argued that the main ionizing sources of Sh2-237 may be 
two B2 stars. The MS lifetime of such stars is tens of millions of years. If the stars are in the MS stage, 
the lifetime is the upper age limit of the cluster. We superimposed a few isochrones (solid lines) 
constructed from the stellar evolution models of \citet{EGE12} and \citet{SDF00} 
on the HRD of NGC 1624 with several evolutionary tracks (dashed lines). The position of NGC 1624-2 in the HRD 
is well matched to the 4 Myr isochrone, and thus the turn-off age is about 
4 Myr. We also used the isochrones interpolated from the stellar evolution models of 
\citet{BMC11} for the initial chemical composition of the Galaxy and Large 
Magellanic Cloud (LMC), where the evolutionary tracks with the similar initial rotation 
velocity to that of \citet{EGE12} were selected. The isochrone for the chemical composition of the 
Galaxy gives an age of 3.3 Myr, while the 3.8 Myr isochrone for the slightly lower LMC 
metallicity appears to well predict the $T_{eff}$ and $M_{bol}$ of NGC 1624-2. 
A systematic uncertainty of the turn-off age arising from the difference in metallicity may 
be about 0.5 Myr.

In the case of NGC 1931, it is impossible to infer the turn-off age because 
the most luminous star ID 872 in NGC 1931 is below the ZAMS line. The $U-B$ 
color of the star became bluer by $\sim$ 0.1 mag from 2005 to 2013 (the data from \citealt{PES13}). 
Another early-type stars ID 629 and 854 also showed a similar variation 
in the $U-B$ color. The mysterious young stellar object Walker 90 found in the young 
open cluster NGC 2264 has also shows such a variation in its spectral type as well as colors 
for the last $\sim$ 50 years (from A3 to B4, \citealt{PMAJ08} and references therein). 
The observational properties of the star, such as an abnormal reddening law of $R_V = 3.6$ -- 7.0, 
the inverse P-Cygni profile of Balmer lines, IR excess emission, and non-photospheric 
UV continuum, indicate that the star is likely an intermediate-mass PMS star with an 
accretion disk \citep{PMAJ08}. The star ID 872 in NGC 1931 also showed similar 
photometric properties (the variation in $U-B$ color, an abnormal reddening law, and 
IR excess emission) to those of Walker 90 although its spectral features were not 
confirmed. If the star is at the analogous evolutionary stage to Walker 90, the age of NGC 1931 can 
be conjectured from the age of the S Mon group within NGC 2264 (1.6 -- 3.0 Myr, \citealt{SB10}) 

Most of the PMS members in the two clusters have masses smaller than 3 $M_{\sun}$. 
The number of PMS members identified in NGC 1624 is insufficient to 
investigate the age distribution. The star ID 1773, which is 
known as the main ionizing source of UCHII at the border of NGC 1624, is likely a 
PMS star at a very early evolutionary stage. The age of the star seems to be younger 
than 1 Myr, indicating that star formation is currently taking place in the cluster. 
We estimated the mass accretion rate of the star using its UV excess emission as 
performed in previous studies \citep{RHS00,LSKBK14,LSKBP14}. The mass accretion 
rate of the young PMS star is $\dot{M} = 3.98 \times 10^{-6} M_{\sun} \ \mathrm{yr}^{-1}$. 
On the other hand, more than half of the PMS members are likely to be younger than 4 Myr. 
There are three PMS members (2 H$\alpha$ emission and 
1 UV excess stars) near the ZAMS. These stars seem to be older than 
other members. However, if the stars have a nearly edge-on disk, their luminosities would be 
underestimated. In addition, accretion activities can lead the colors of the stars 
to be bluer, i.e. hotter $T_{eff}$. 

In the HRD of NGC 1931 (the right panel of Figure~\ref{fig12}), most of the PMS 
members are younger than 10 Myr, which was suggested by the open cluster 
data base WEBDA\footnote{http://www.univie.ac.at/webda/}. It is worth noting that WEBDA 
provided incorrect ages for some open clusters. The number of PMS members is 
relatively larger than that identified in NGC 1624 because of its proximity as well as the 
variety of the available data. We investigated the age distribution of the PMS members 
(thick solid line) in Figure~\ref{fig13}, where the bin size is 0.5 Myr. In order to avoid binning 
effects, another histogram (thin solid line) was plotted by shifting the bins by half the bin size. 
A peak in the age distribution appears at 1.5 Myr. The median age is about 2.0 Myr with a 
spread of 4.5 Myr, where the spread was defined as the age difference between the 10 and 
90 percentiles in the cumulative age distribution of the PMS members \citep{SB10}. Since 
all the PMS members were identified using membership selection criteria based on the observational 
properties of warm circumstellar disks and accretion activities, the age distribution could 
be biased to young active stars.

Previous studies estimated the turn-off age of NGC 1624 to be 4 Myr 
\citep{SuBa06,JPO11}. Our estimate (4 Myr) is in good agreement with those.
Other evolution models by \citet{BMC11} for Milky Way and LMC metallicities  
give a slightly younger age (3.3 and 3.8 Myr). The upper value of the ages of PMS members 
($ \leq 4$ Myr) is also consistent with that of \citet{JPO11}. The turn-off age of NGC 
1931 remains uncertain because the evolutionary stage of the brightest star is 
ambiguous. For this reason, \citet{BB09} obtained a somewhat larger turn-off age ($10 \pm 3$ Myr), 
while \citet{PES13} only suggested an upper value of 25 Myr. The latter reported a 
mean age of $2 \pm 1$ Myr for the PMS stars. This is in good agreement with the median age 
from this work. They also found a similar age spread ($\sim 5$ Myr) from their 
CMD. We have further discussions on the age spread of PMS stars in Section 6.

\section{INITIAL MASS FUNCTION}

\subsection{Derivation}

The membership selection of PMS stars is likely incomplete because the membership selection 
criteria used above are more likely to identify young active PMS stars rather than those in a quiescent 
phase. A statistical method was therefore used to fill in the incomplete mass bins of the IMF. 

We included all the stars within the PMS locus in the ($V, V-I$) CMD, that is, bona-fide members 
plus any field interlopers. These 
stars were then placed in the HRD following the same procedure as delineated in the previous section 
(see Figure~\ref{fig12}). The mass of the individual stars was estimated in the HRD by comparing 
their $T_{eff}$ and $M_{bol}$ with those from the evolutionary tracks. For the MS members, 
the stellar evolution model of \citet{EGE12} was used to estimate the mass. The mass of 
the PMS members and the field stars was inferred from the PMS evolution models of 
\citet{SDF00}. In order to minimize the contribution of field interlopers, appropriate 
control fields were chosen within the observed regions (the shaded area in Figure~\ref{fig1}). The stars 
within the PMS locus in the control fields were assumed to be evenly distributed in 
the direction of each cluster. The apparent mass of these stars was estimated by using the same 
procedure as above, where the mean reddening and distance of the two clusters applied to the stars. 
We note that the apparent mass does not mean real stellar mass.

The IMF is, in general, expressed as $\xi \equiv$ $N \over \Delta \log m  \cdot S$, 
where $N$, $\Delta \log m$, and $S$ represent the number of stars within a given 
mass bin, the size of the logarithmic mass bin, and the area of the cluster, 
respectively. We applied a slightly large bin size of $\Delta \log m = 0.4$ to include 
as many stars as possible within a given mass bin. An even larger bin size of 1.0 was adopted for 
the highest mass bin [$10 < M/M_{\sun} \leq 100$ -- we assumed the upper limit of stellar 
mass ($M_{\mathrm{up}}$) to be 100 $M_{\sun}$] because of the small number of high-mass stars. 
The number of stars was counted within each given logarithmic mass bin, and then that was 
normalized by the size of the mass bin and the area of the clusters. The apparent IMF 
of the field interlopers was obtained in the same way. In the case of NGC 1931, the 
extent of the control field was smaller than that of the cluster, and therefore the number 
of field interlopers was corrected by multiplying the areal ratio of the cluster to the control 
field. Finally, the IMF of the two clusters was obtained by subtracting the contribution of field 
interlopers. Figure~\ref{fig14} shows the IMF of the two clusters. In order to avoid 
binning effects we shifted the mass bin by 0.2 and re-derived the IMF (open circle) in the 
same way. The error of the IMF was assumed to follow Poisson statistics. The arrows 
represent the lower limit of the IMF below the completeness limit of our photometry.

Using a least-square method we determined the slope of the IMF in the mass range of 
3 -- 27 $M_{\sun}$ for NGC 1624 and 1 -- 18 $M_{\sun}$ for NGC 1931. The slope 
is $\Gamma = -2.0 \pm 0.2$ for the former and $\Gamma = -2.0 \pm 0.1$ for the latter. 
The IMF of these clusters appears far steeper than the Salpeter/Kroupa IMF 
\citep{Sp55,K01,K02} as well as the results of previous studies, e.g. $\Gamma = -1.18$ to $-1.31$ 
for NGC 1624 \citep{JPO11} and $\Gamma = -1.15$ for NGC 1931 \citep{PES13}. 
However, the small number of high-mass stars ($> 10 \ M_{\sun}$) and the upper limit of the stellar mass 
which was required for the normalization of the IMF may be responsible for the 
steep slope. The IMF of the well-studied young open cluster 
NGC 2264 (dashed line, \citealt{SB10}) is shown in Figure~\ref{fig14} for comparison. 
The IMF of the two clusters seems to resemble that of NGC 2264 within the uncertainties. 
According to a recent review on the IMF \citep{OCH14} the IMF of 
NGC 2264 is in good agreement with that of nearby SFRs. This may imply that the
star formation in NGC 1624 and NGC 1931 is not much different from that in 
the solar neighbourhood. More discussion on the slope of the IMF is addressed in the next section.

We estimated the total number of members and the masses of the two clusters by 
integrating their IMFs. The IMF in the low-mass regime (from the completeness limit 
down to 0.25 $M_{\sun}$) was assumed to be that of NGC 2264. NGC 1624 hosts 
$661^{+81}_{-81}$ members, and a total of $656^{+143}_{-10}$ stars constitute NGC 
1931. The total mass of NGC 1624 and NGC 1931 is about $510^{+111}_{-111} \ M_{\sun}$ 
and $510^{+103}_{-75} \ M_{\sun}$, respectively. The uncertainty in the total masses was 
propagated from the error of the IMF in given mass bins. The total mass is comparable to that 
of a few small open clusters, e.g. $576 \ M_{\sun}$ for NGC 2264 (which is the sum 
of masses of all members from \citealt{SB10}), $550^{+40}_{-40} \ M_{\sun}$ for 
Praesepe \citep{KH07}, and 800 $M_{\sun}$ for the Pleiades \citep{ASMSB01}. These 
clusters are two orders of magnitude lighter than the most massive starburst cluster 
Westerlund 1 in the Galaxy ($> 50,000 \ M_{\sun}$, \citealt{CNCG05,GBSH11,LCS13}). 

\subsection{Implications of the Slope of the IMF}

The IMF of NGC 1624 and NGC 1931 shows a steep slope for stars 
with mass larger than 1 $M_{\sun}$. This could imply that low-mass 
star formation is dominant in the clusters. According to the metallicity variation in the 
Galactic disk \citep{YCF12} the chemical composition at the Galactocentric distance 
of NGC 1624 ($R_{\mathrm{GC}} = 13.7$ kpc) and NGC 1931 ($R_{\mathrm{GC}} = 10.3$ kpc) is 
expected to be 0.1 -- 0.3 dex lower than the solar metallicity. The temperature of 
a molecular cloud may therefore be higher in the low metallicity environment because radiative cooling 
by metallic ions will be lower \citep{CMP12}. Furthermore, there is a relation 
between the mass accretion rate of a star and the temperature of a cloud ($\dot{M} 
\propto T^{3/2}$, \citealt{SHT14}). As a result, high-mass stars may be formed advantageously in 
such a low metallicity environment, rather than active low-mass star formation. However, 
the IMF of the two clusters is inconsistent with this expectation. 

In order to properly interpret the IMF of the clusters we will first discuss a few issues concerning
the subtraction of field stars, dynamical evolution and stochastic effects. 

\begin{figure}
\epsscale{1.0}
\plottwo{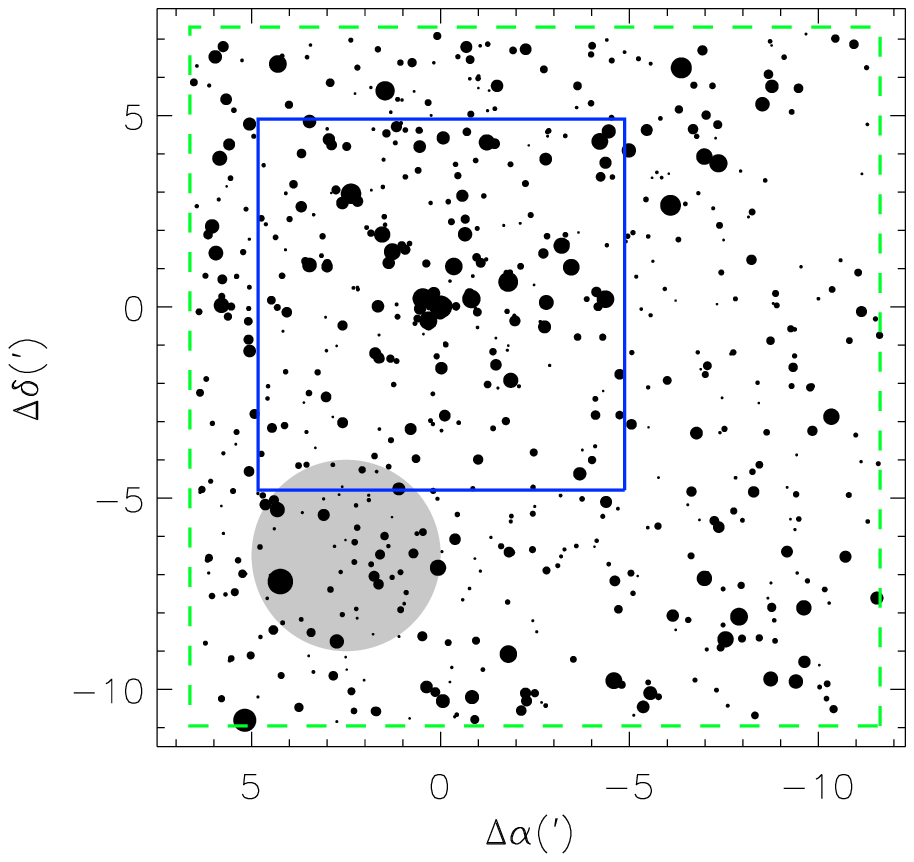}{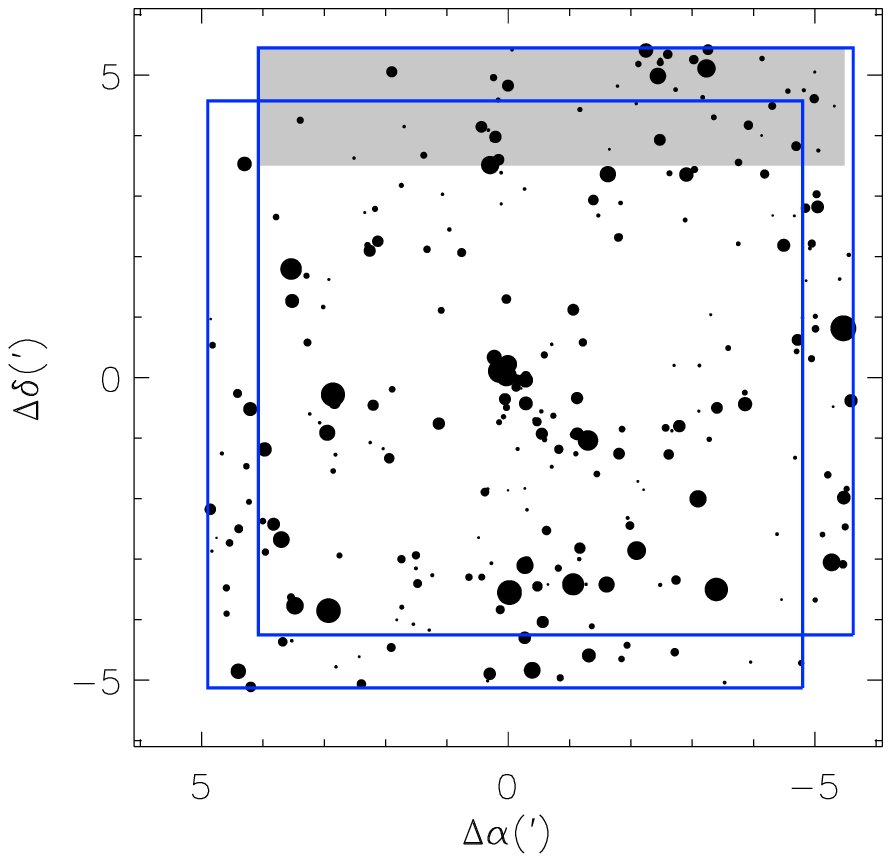}
\caption{Finder charts of NGC 1624 (left) and NGC 1931 (right). Stars brighter than 
$V = 18$ mag are plotted, and the size of the circles is proportional to the brightness of 
individual stars. The positions of stars are relative to the O-type star NGC 1624-2 
($\alpha = 04^h \ 40^m \ 37^{s}.3, \ \delta = +50^{\circ} \ 27' \ 41''.1$) for NGC 1624 and to 
BD+34 1074 ($\alpha = 05^h \ 31^m \ 26^{s}.4, \ \delta = +34^{\circ} \ 14' \ 43''.0$) for 
NGC 1931, respectively. Squares outlined by blue solid lines represent the region observed 
by the Mont4k CCD camera, and the other square (green dashed line) shows the field of 
view of SNUCam. The shaded regions outline the control fields, which are used to estimate 
the density of field interlopers. }
\label{fig1}
\end{figure}

\begin{figure}
\epsscale{1.0}
\plotone{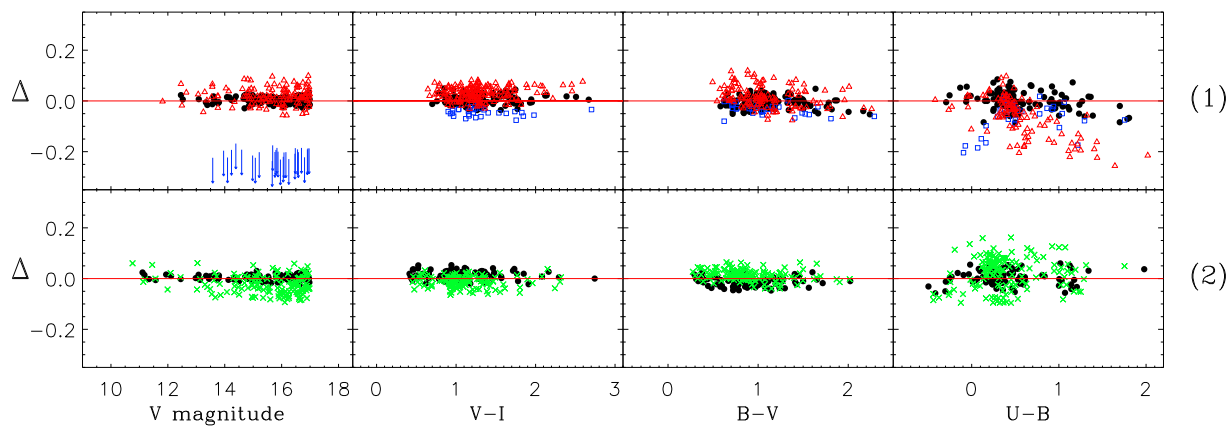}
\caption{Comparison of our photometry with a few previous sets of CCD photometry. The 
upper and lower panels represent the difference between photometric datasets with respect to the
$V$ magnitude, and the $V-I$, $B-V$, and $U-B$ colors for NGC 1624 (1) and NGC 1931 (2). Bold dots (black) 
displays the consistency between our data obtained with two different observational setups. 
Squares (blue), triangles (red), and crosses (green) denote the photometry of \citet{SuBa06}, 
\citet{JPO11}, and \citet{PES13}, respectively. Stars brighter than $V = 17$ mag are used in the comparison 
to avoid the large photometric errors of faint stars. }
\label{fig2}
\end{figure}

\begin{figure}
\epsscale{0.8}
\plotone{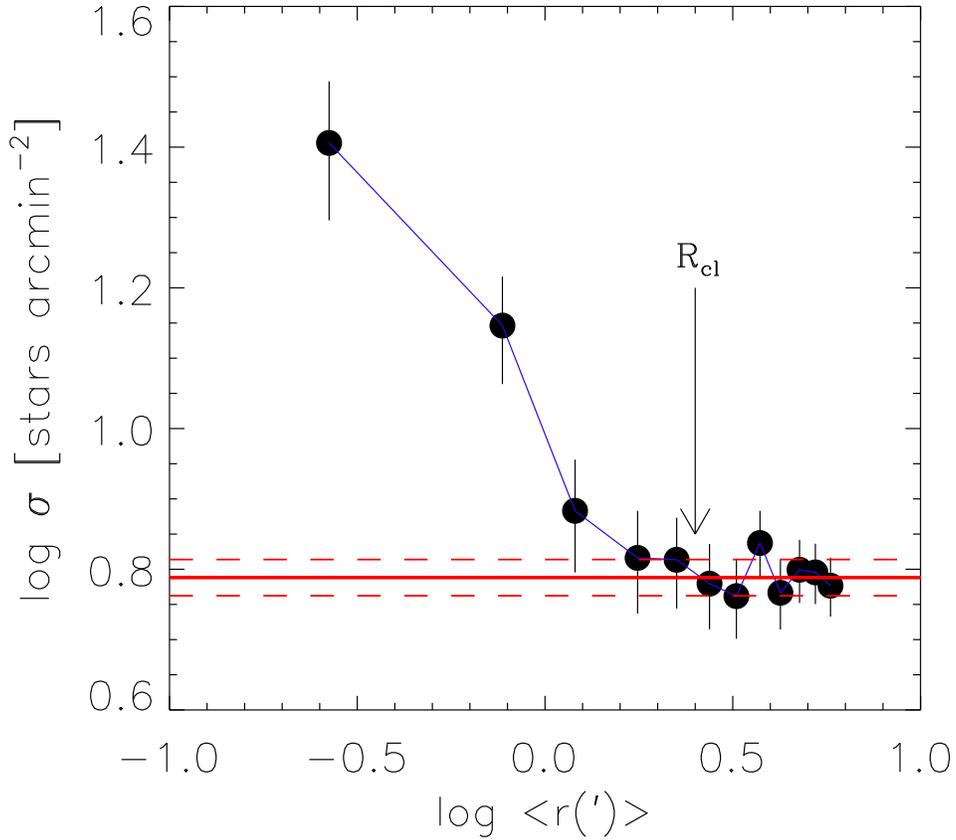}
\caption{The radial surface density profile of NGC 1624. The error at a given surface density 
is assumed to follow Poisson statistics. Solid and dashed lines represent the mean 
surface density of field stars and its standard deviation, respectively. The surface density 
continues to decrease until it reaches the mean surface density of field stars. The radius of 
the cluster is about $2\farcm5$ (arrow). }
\label{fig3}
\end{figure}

\begin{figure}
\epsscale{0.8}
\plotone{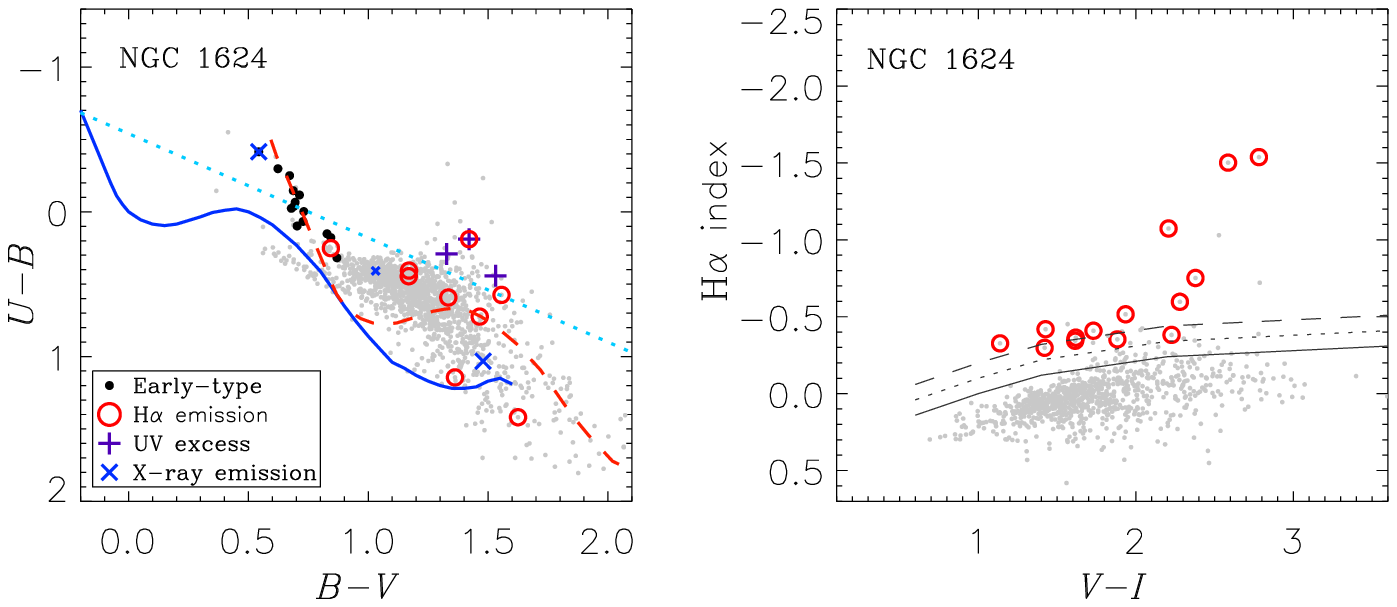}
\plotone{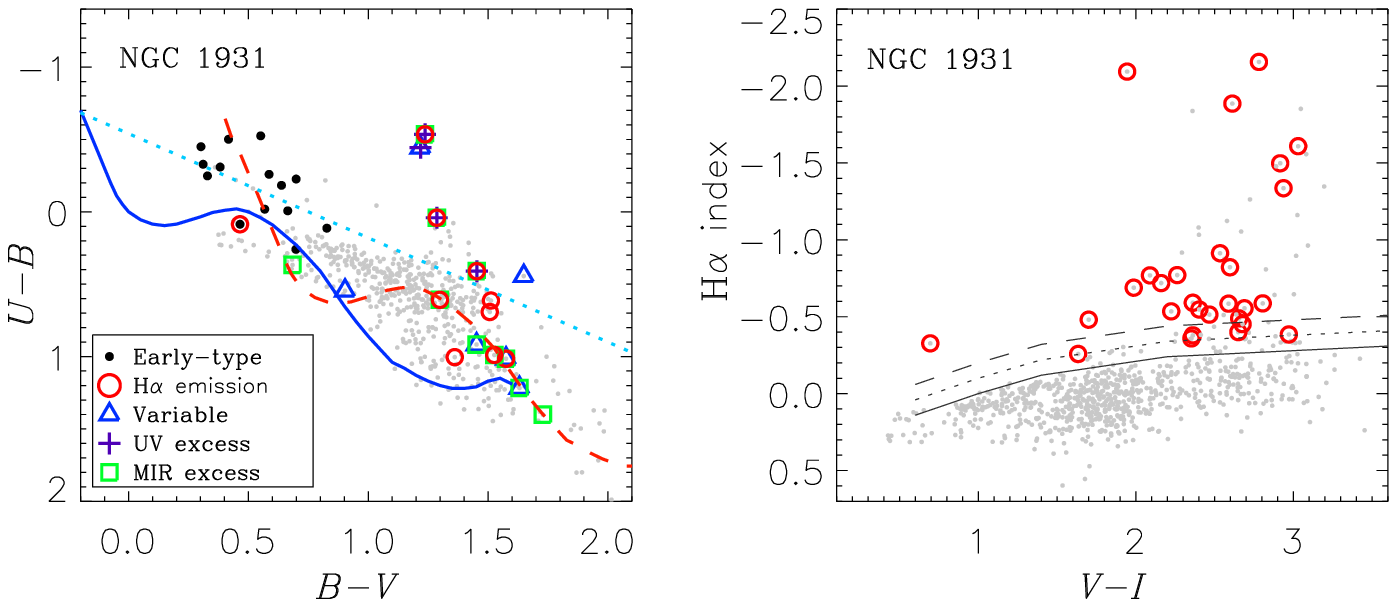}
\caption{Color-color diagrams of NGC 1624 (upper) and NGC 1931 (lower). The small dots represent all 
the observed stars. Different symbols identify early-type main sequence members (black bold dots), 
H$\alpha$ emission stars (red circles), variables (blue triangles), ultraviolet excess stars (violet pluses), X-ray 
emission objects (large blue crosses), an X-ray emission candidate (a small blue cross), and mid-infrared excess 
stars (green squares). Since most pre-main sequence stars are too faint to be detected in the $U$ band, they do not 
appear in the ($U-B$, $B-V$) diagram. The intrinsic and reddened color-color 
relations are overplotted by a solid line (blue) and dashed line (red) in the left-hand side panels. The mean 
reddening of $\langle E(B-V) \rangle = 0.92$ mag for NGC 1624 and 0.74 mag for NGC 1931 is adopted for 
the reddened relations. The dotted line (cyan) denotes a fiducial line to identify UV excess stars. In the right-hand 
side panels, the solid line represents the empirical photospheric level of normal main sequence stars, while the 
dashed and dotted lines are the lower limit of H$\alpha$ emission stars and H$\alpha$ emission candidates. }
\label{fig4}
\end{figure}

\begin{figure}
\epsscale{0.9}
\plotone{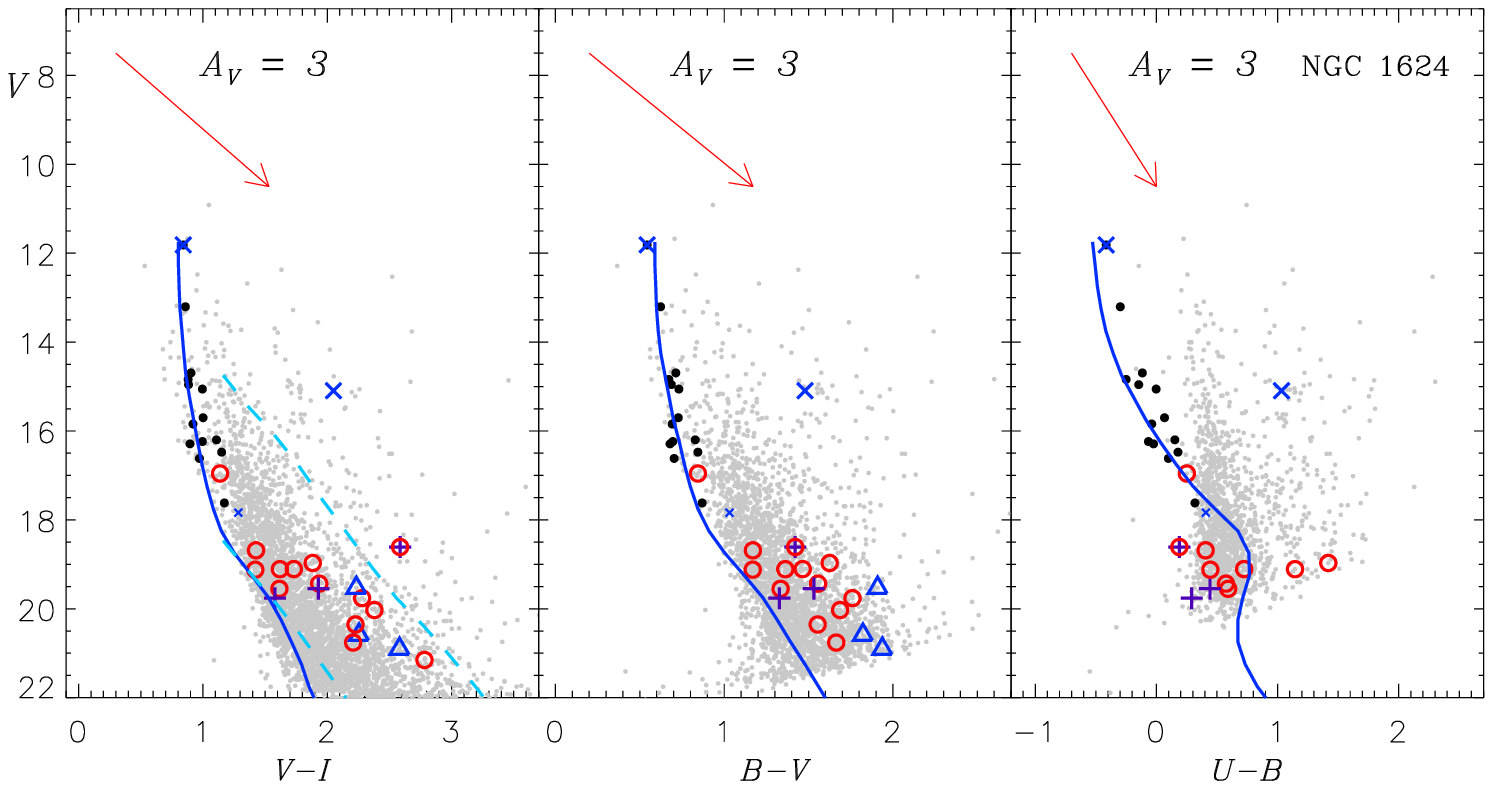}
\plotone{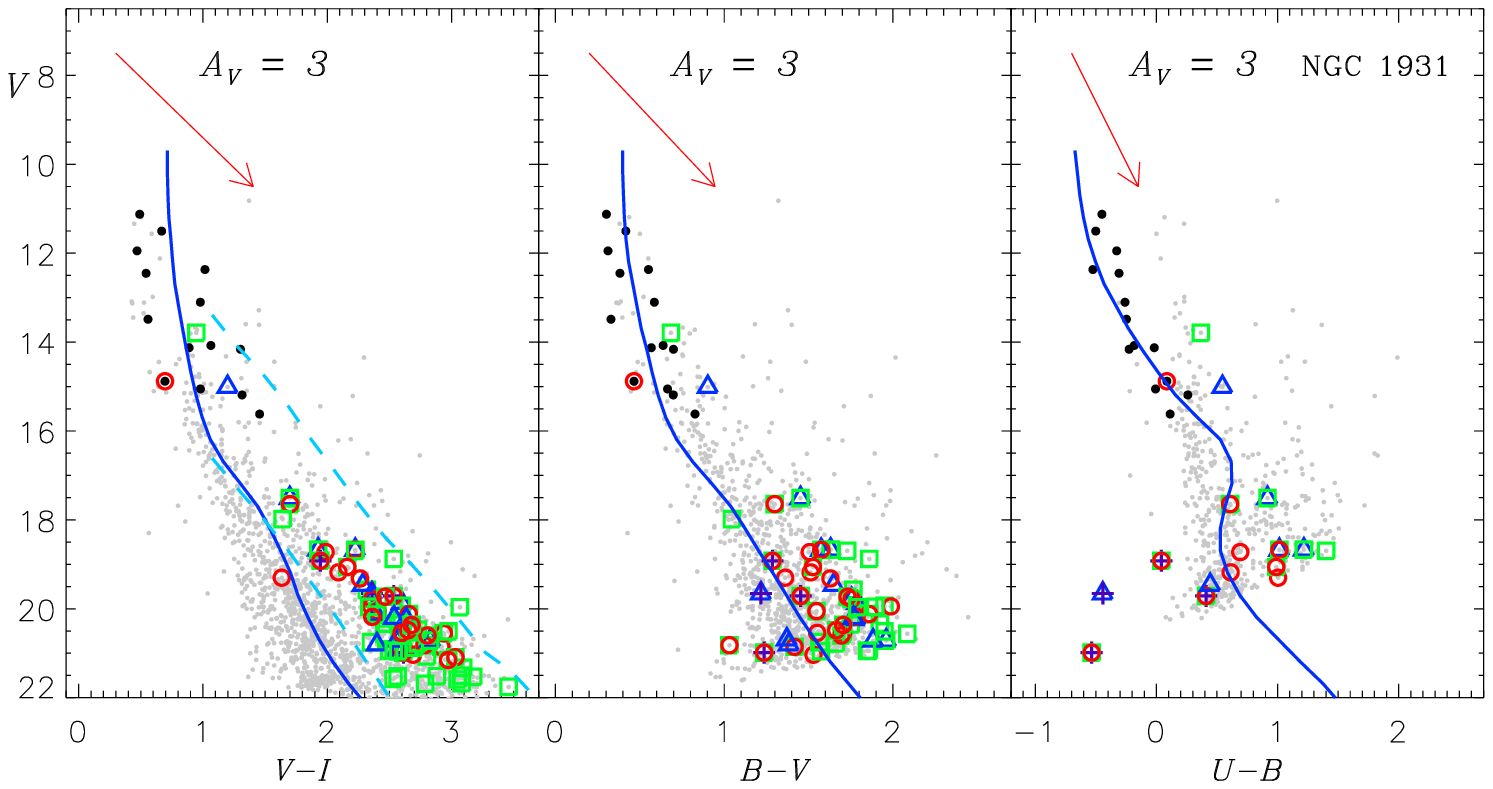}
\caption{Color-magnitude diagrams for NGC 1624 (upper) and NGC 1931 (lower). Left-hand 
side panels: $V-I$ vs. $V$ diagram. Dashed lines (cyan) represent the empirical pre-main sequence 
locus. Middle panels: $B-V$ vs. $V$ diagram. Right-hand side panel: $U-B$ vs. $V$ diagram. 
The solid lines are the reddened zero-age main sequence relations of Paper 0. The arrow 
displays a reddening vector corresponding to $A_V = 3$ mag. The other symbols are the 
same as in Figure~\ref{fig4}.}
\label{fig5}
\end{figure}

\begin{figure}
\epsscale{1.}
\plotone{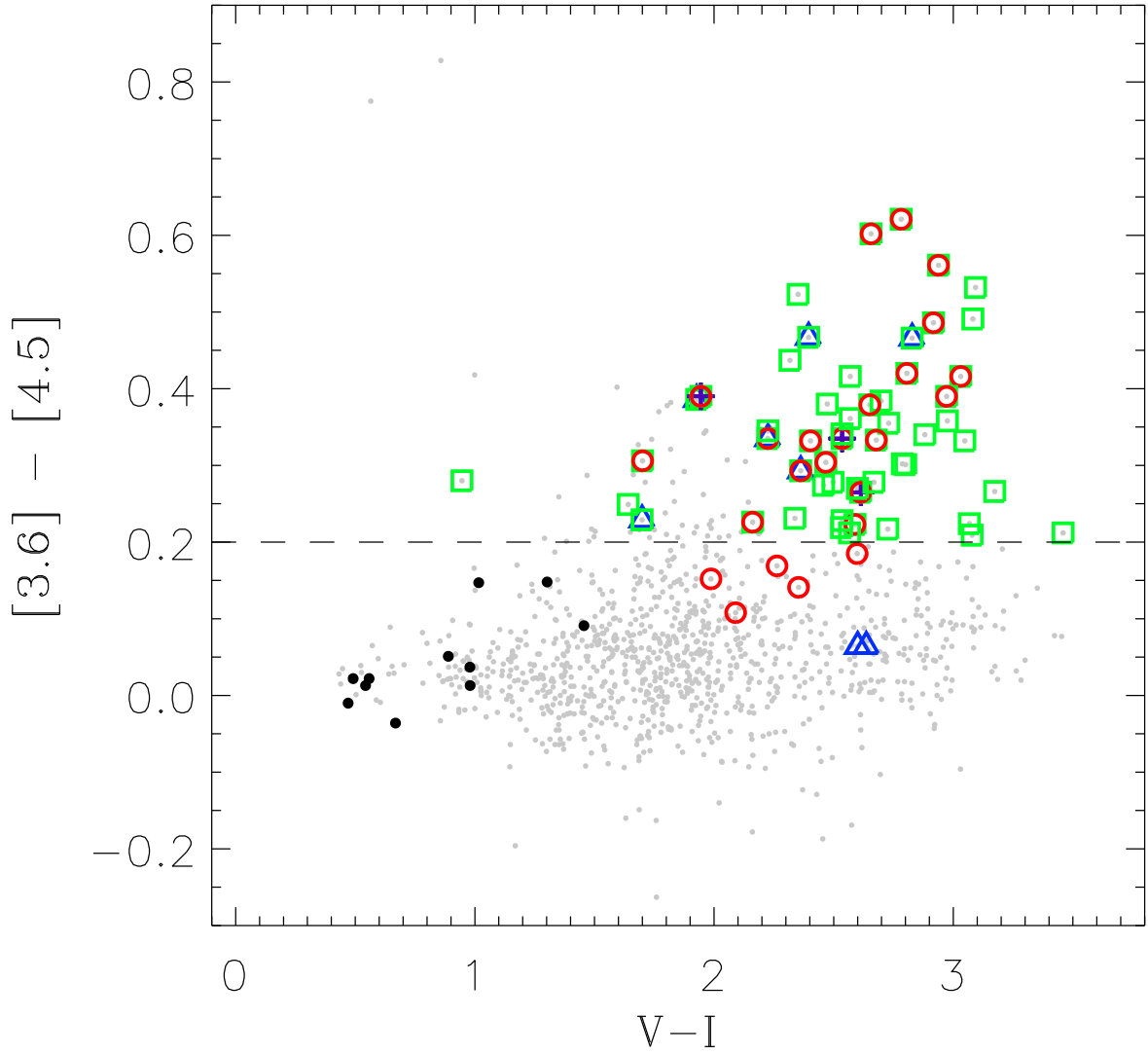}
\caption{The ($[3.6] - [4.5], V-I$) color-color diagram for NGC 1931. Dashed line represents 
the lower limit of the mid-infrared excess stars. The other symbols are the same as in Figure~\ref{fig4}.}
\label{fig6}
\end{figure}

\begin{figure}
\epsscale{0.9}
\plotone{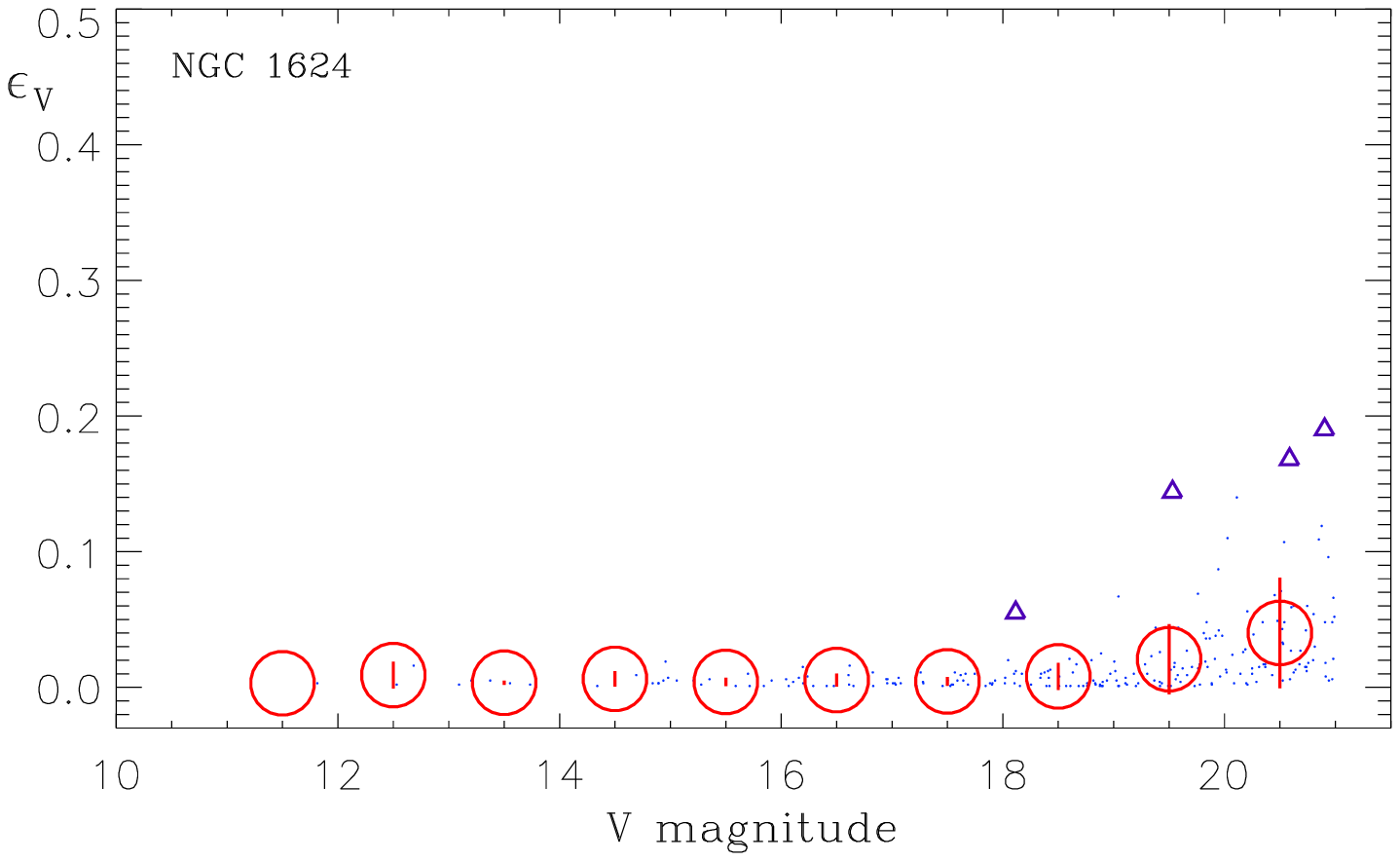}
\plotone{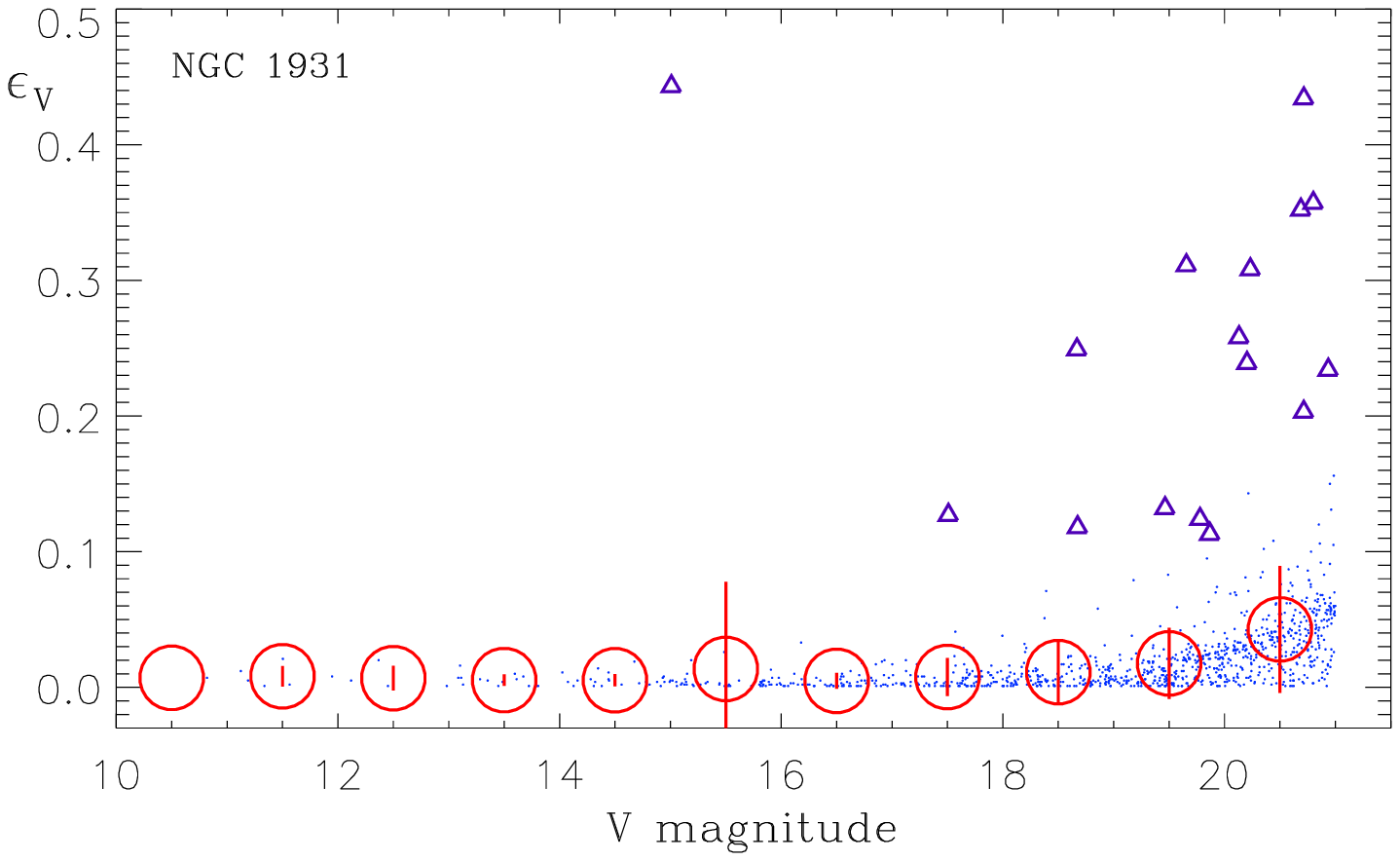}
\caption{Variability test in the $V$ band. The upper and lower panels show the photometric 
errors of stars observed in NGC 1624 (upper) and NGC 1931 (lower). The red circles and 
vertical bars indicate the mean and standard deviation of the photometric errors for a given magnitude bin. 
Triangles denotes variable candidates.}
\label{fig7}
\end{figure}

\begin{figure}
\epsscale{1.}
\plottwo{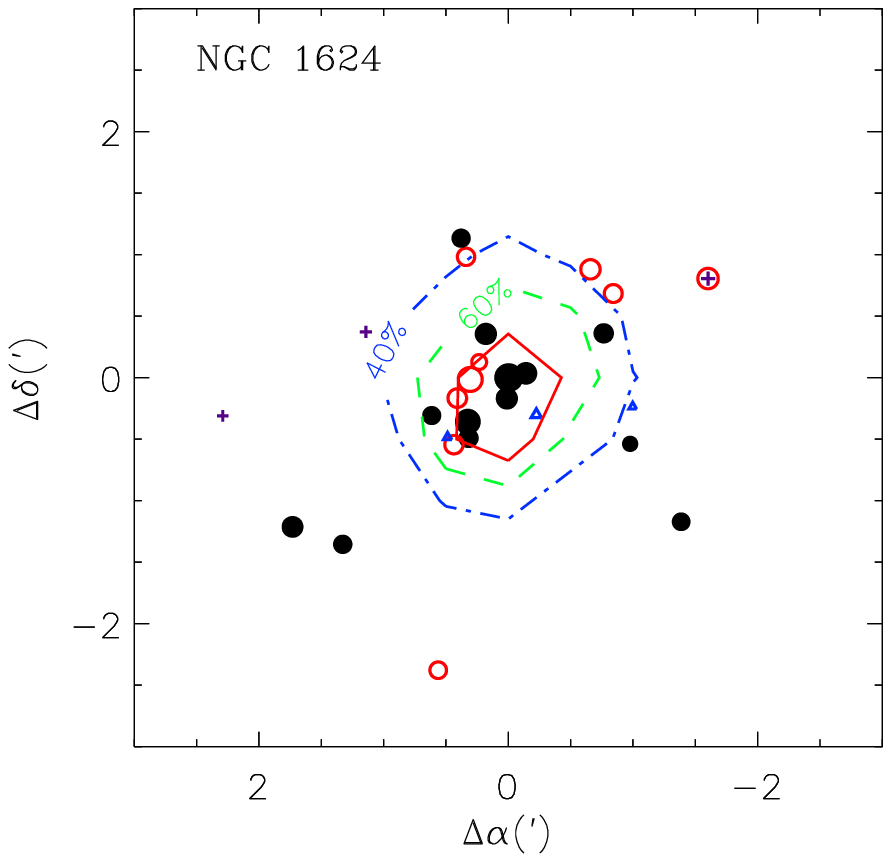}{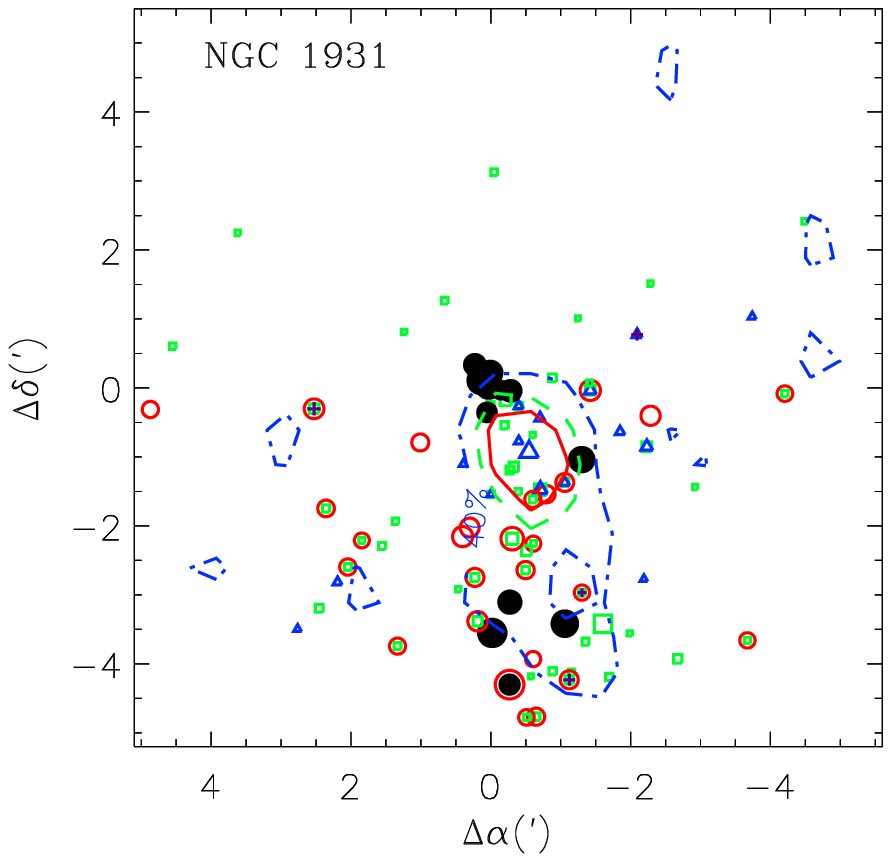}
\caption{The spatial distribution of members of NGC 1624 (left) and NGC 1931 (right). The solid, dashed, 
and dot-dashed lines show contours corresponding to 80, 60, and 40 percent level of the maximum 
surface density. The size of each symbol is proportional to the brightness of individual stars. The other 
symbols are the same as in Figure~\ref{fig4}. }
\label{fig8}
\end{figure}

\begin{figure}
\epsscale{1.}
\plotone{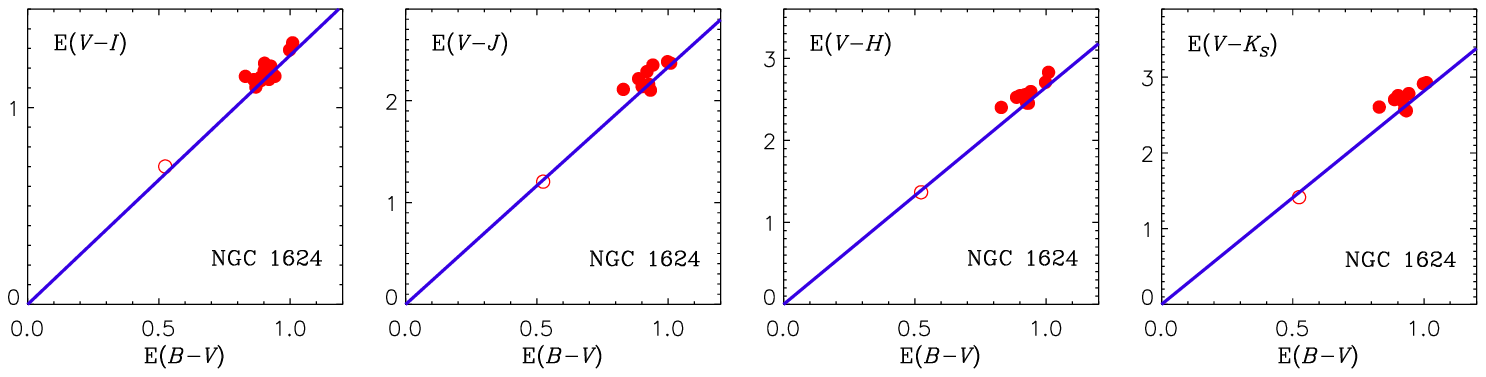}
\plotone{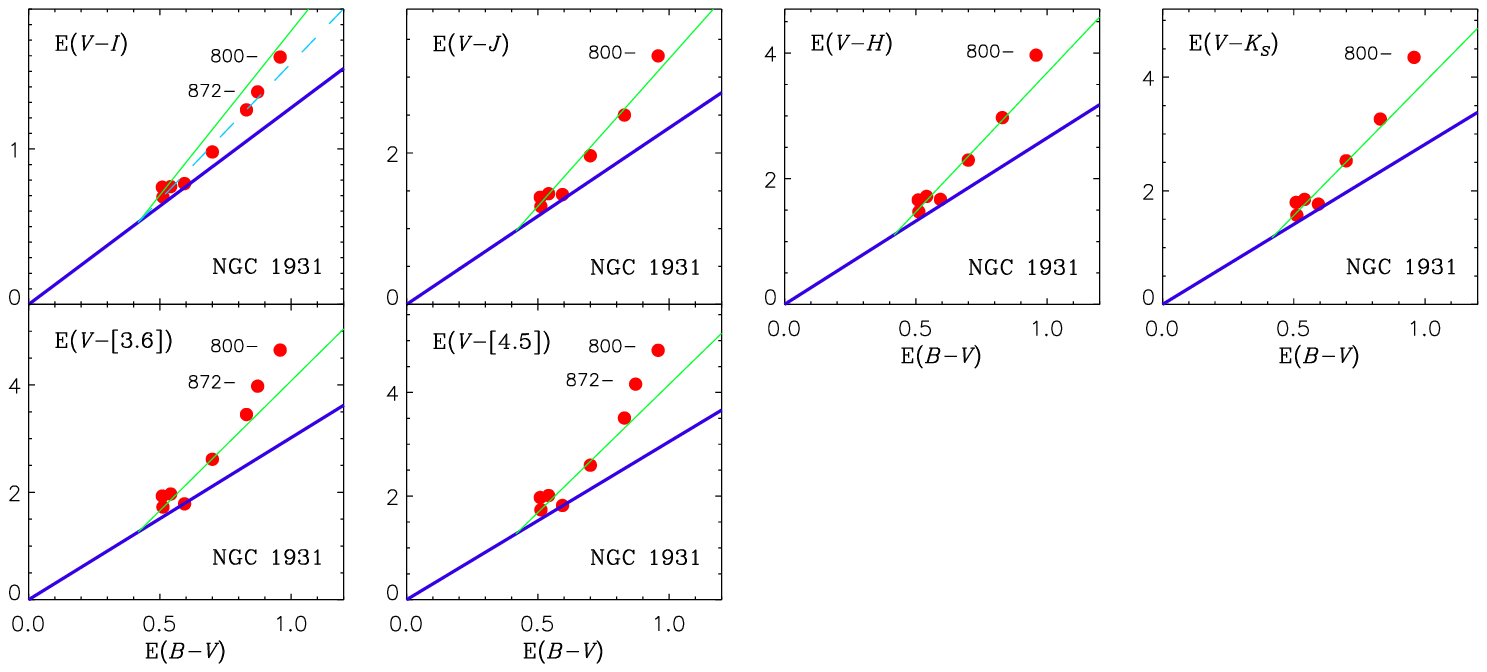}
\caption{Color excess ratios obtained from the early-type star toward NGC 1624 (upper row) 
and NGC 1931 (middle and lower panels). The dots and the open circle denote the cluster members and a foreground 
early-type star. Thick solid (blue), thin solid (green), and dashed (cyan) lines corresponds 
to the normal reddening law ($R_V = 3.1$) and the intracluster reddening relations 
($R_{V,\mathrm{cl}} = 5.2$ and 4.3). The extinction toward NGC 1624 follows the normal 
reddening law, while the color excess ratios toward NGC 1931 exhibit a deviation from the 
normal reddening law. See the main text for details. }
\label{fig9}
\end{figure}

\begin{figure}
\epsscale{0.8}
\plotone{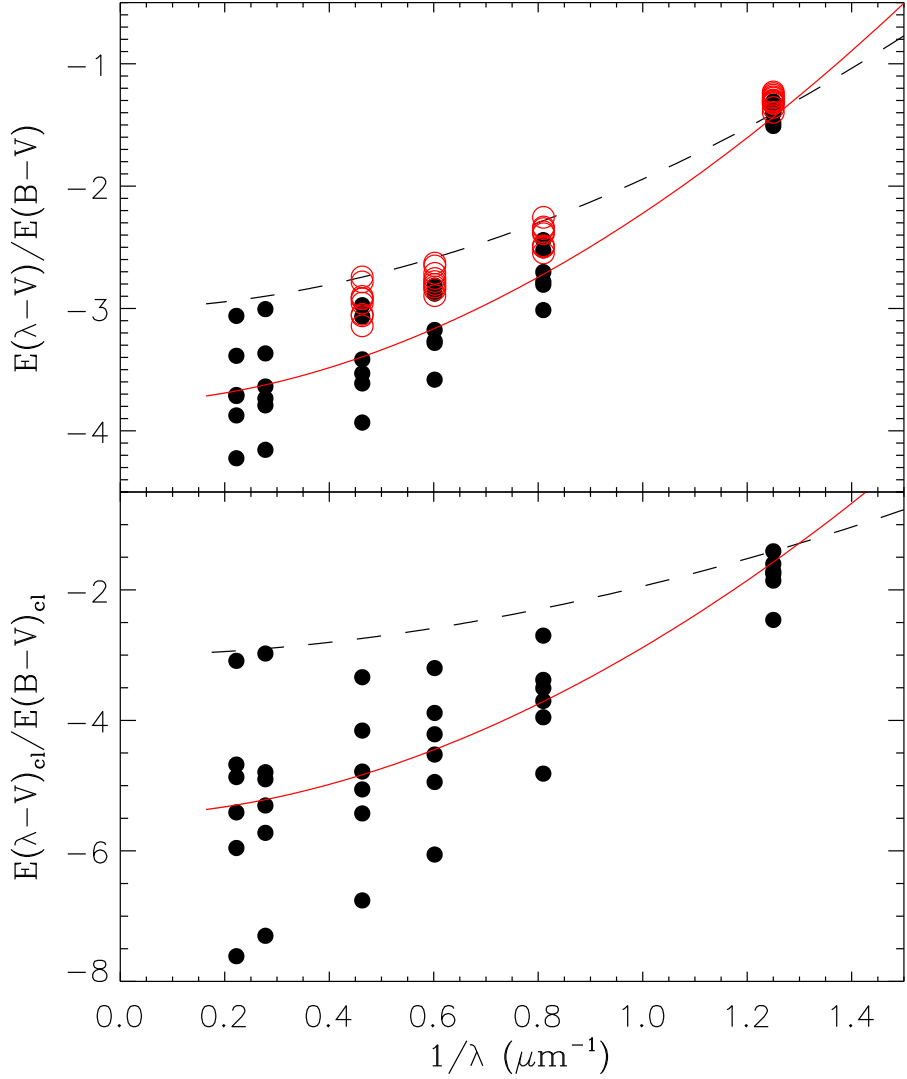}
\caption{Infrared extinction curves of NGC 1624 and NGC 1931. Upper panel: color 
excess ratios in the direction of NGC 1624 (open circle) and NGC 1931 (bold dot). Lower panel: color 
excess ratios for the intracluster medium of NGC 1931. The dashed lines represent the mean Galactic 
extinction curve of \cite{FM07} ($R_V = 3.00$ and $k_{IR} = 1.06$). Fitting solutions are drawn by solid 
lines (upper -- $R_V = 3.8$ and $k_{IR} =1.5$, lower -- $R_{V,\mathrm{cl}} = 5.5$ and $k_{IR} = 2.6$).}
\label{fig10}
\end{figure}

\begin{figure}
\epsscale{1.}
\plotone{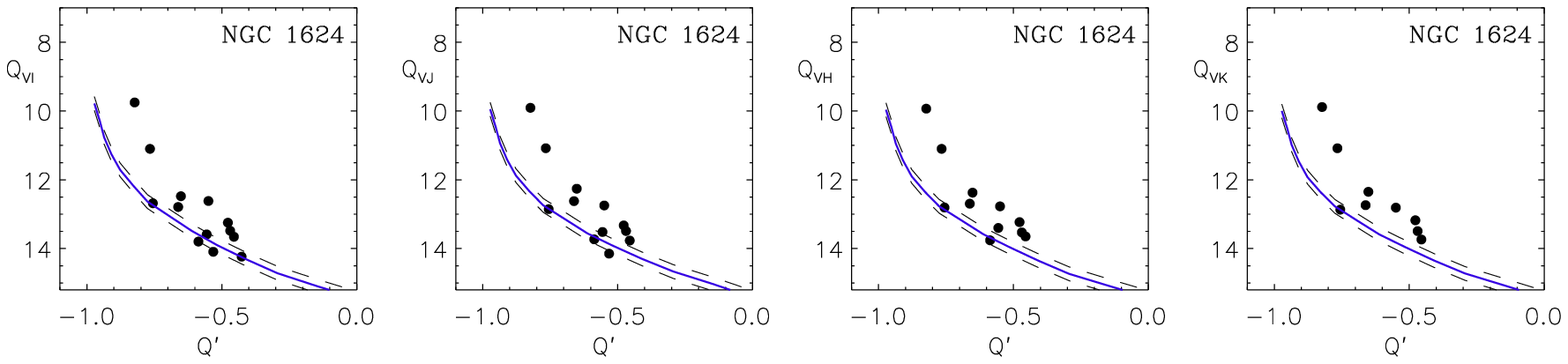}
\plotone{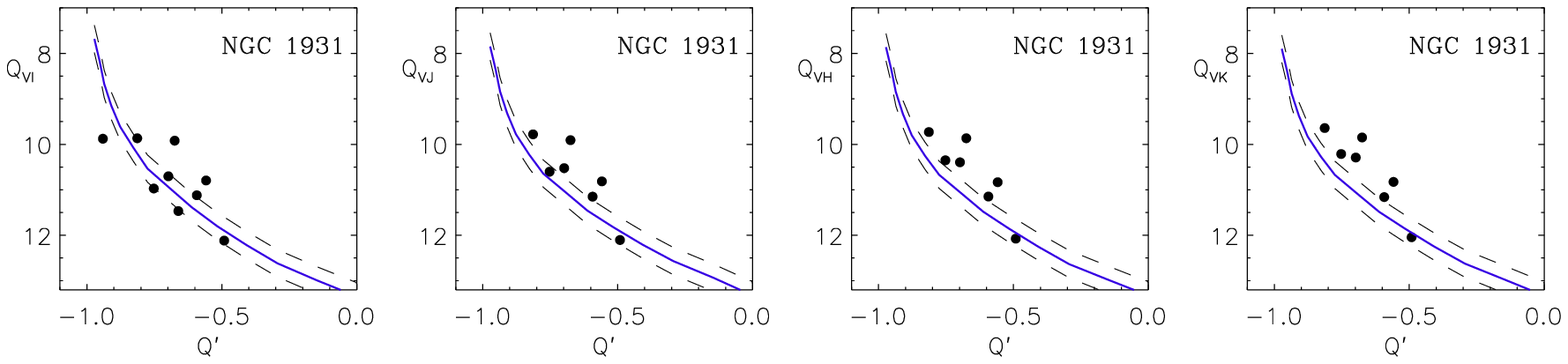}
\caption{Zero-age main sequence fitting to the early-type MS members of NGC 1624 (upper) and 
NGC 1931 (lower). The zero-age main sequence relations of \citet{SLB13} were fitted to the 
lower ridge line of the members. The solid lines (blue) represent the adopted distance moduli of 
$13.9 \pm 0.2$ ($6.0\pm0.6$ kpc) and $11.8\pm0.3$ mag ($2.3 \pm 0.3$ kpc) for NGC 1624 
and NGC 1931, respectively. The dashed lines are the ZAMS relations adjusted by the fitting 
errors.}
\label{fig11}
\end{figure}

\begin{figure}
\epsscale{1.1}
\plottwo{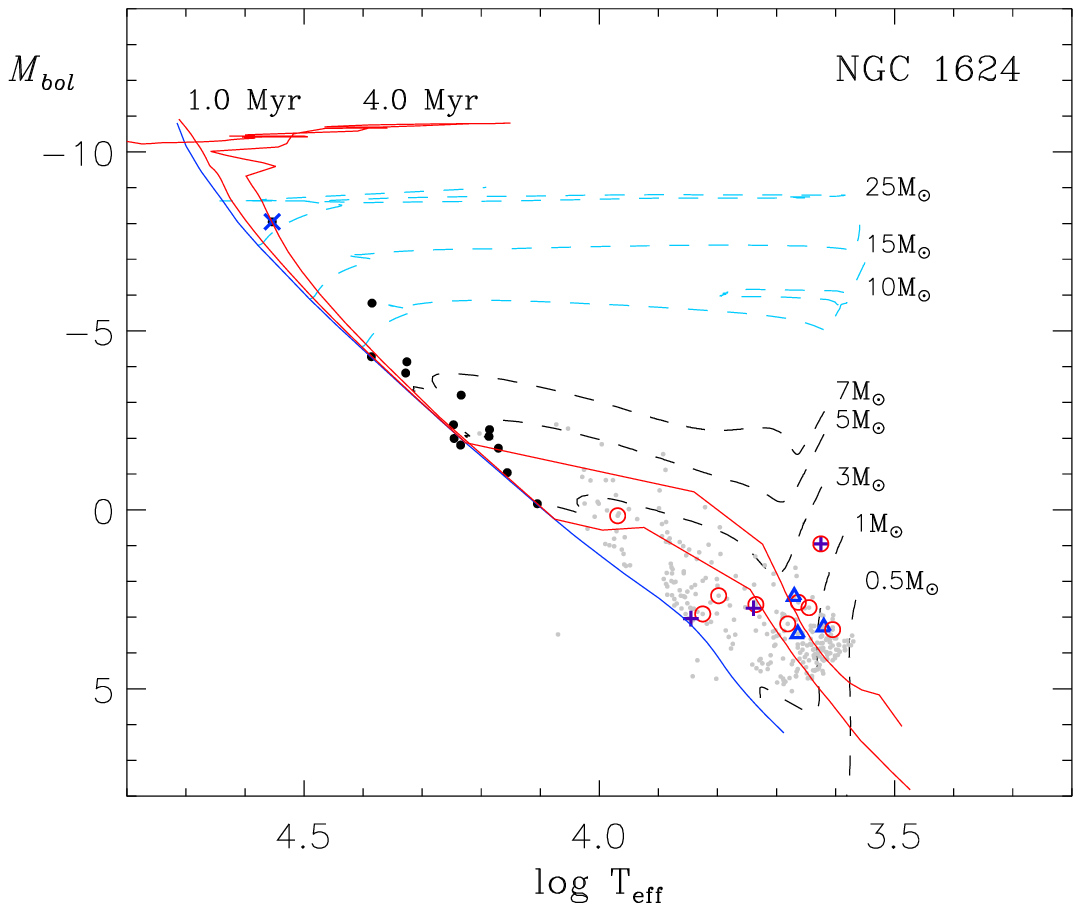}{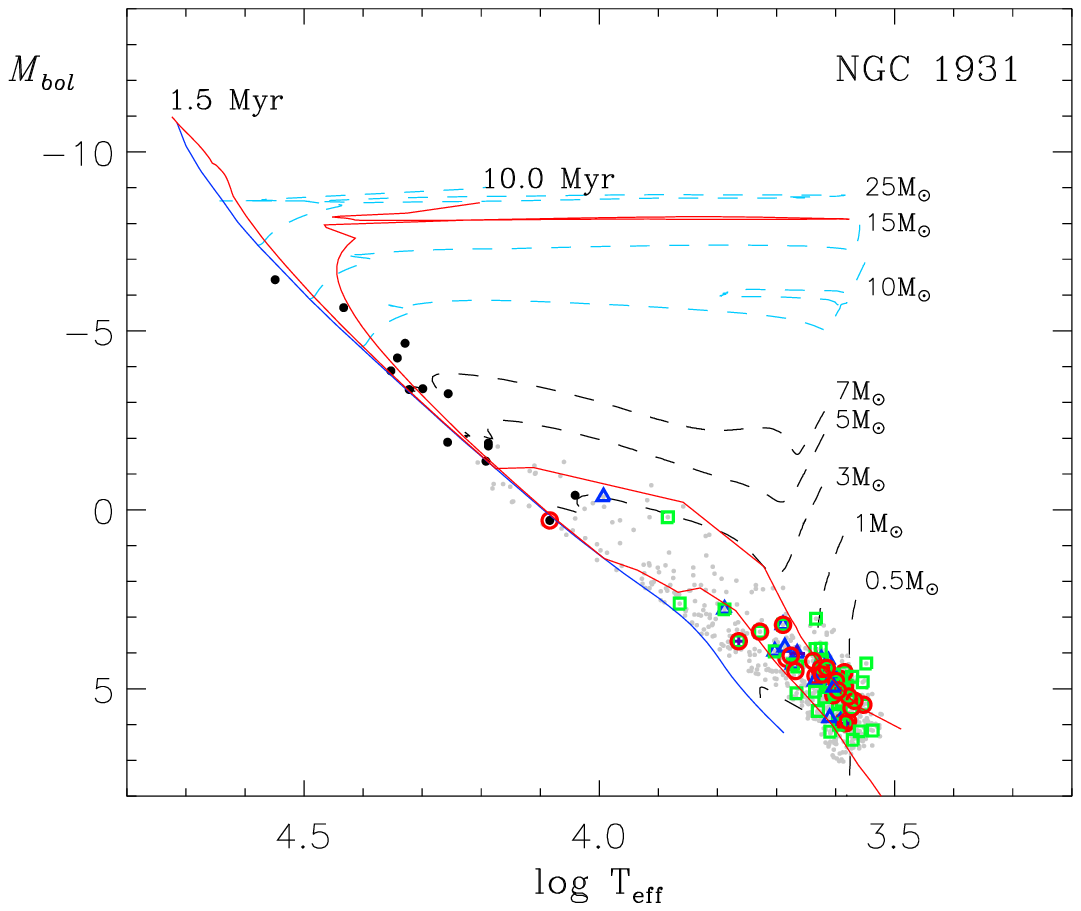}
\caption{The Hertzsprung-Russell diagram of NGC 1624 (left) and NGC 1931 (right). A few 
isochrones (solid line) with different age (0., 1, 1.5, 4, and 10 Myr) are superimposed on the 
diagram with several evolutionary tracks (dashed line, \citealt{SDF00,EGE12}). The other symbols 
are the same as in Figure~\ref{fig4}. }
\label{fig12}
\end{figure}

\begin{figure}
\epsscale{0.5}
\plotone{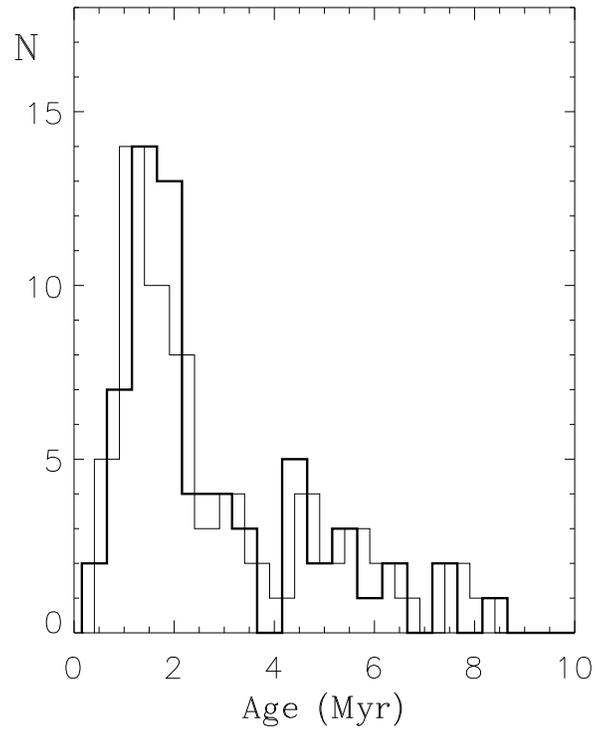}
\caption{The age distribution of PMS members identified in NGC 1931. The histograms are based on different 
binning of the same stars. A peak age appears at 1.5 Myr, and the median age is about 2 Myr with 
a spread of 4.5 Myr.}
\label{fig13}
\end{figure}

\begin{figure}
\epsscale{1.0}
\plottwo{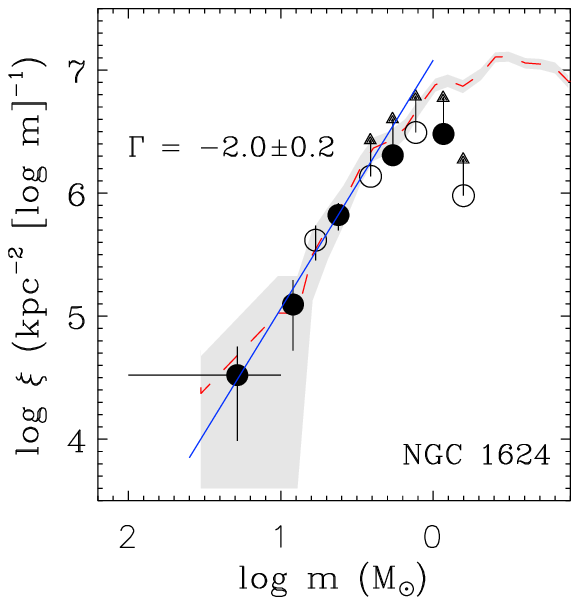}{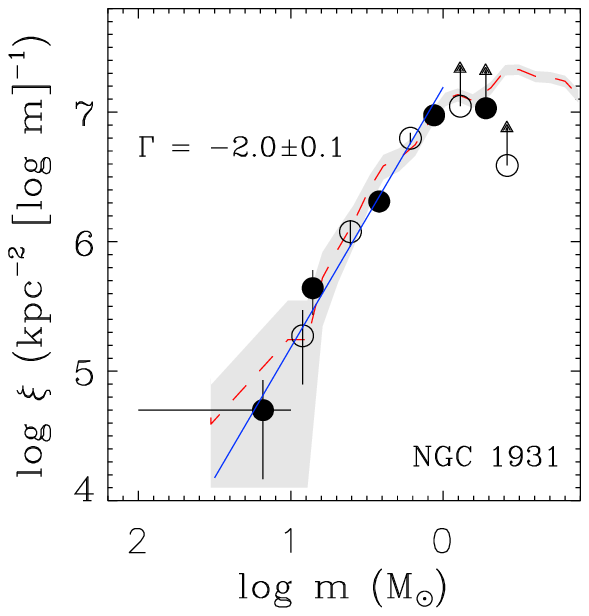}
\caption{The initial mass function of NGC 1624 (left) and NGC 1931 (right). To avoid the binning effect we 
shifted the mass bin by 0.2 and re-derived the IMF (open circle) using the same procedure. The 
IMF of NGC 2264 \citep{SB10} is shown as a dashed line with its uncertainty (shaded area) for comparison. 
Arrows denote the IMF below the completeness limit of our photometry. See the main text for details. }
\label{fig14}
\end{figure}

We carried out checks to see whether the subtraction of field interpolators
had been adequately carried out because had the contribution of field interlopers been improperly 
subtracted within a given mass bin, the resultant IMF would be misleading. We 
examined the $V$ band luminosity function (LF) of stars in each control field using 
the synthetic stellar population model for the Galaxy \citep{RRDP03}. The synthetic LF 
was scaled to have the same area as that of each control field. The observed 
LF was compared with the synthetic one. The LF of stars in the control field for NGC 1931 
($l = 173\fdg90$, $b = 0\fdg28$) is in good agreement with the synthetic one, while 
the stellar population model overestimated the number of stars toward NGC 1624 at 
$V \sim 19$ mag. Because of the Galactic warp found at $l = 90^{\circ}$ \citep{RRDP03}, 
the synthetic stellar population in the direction of NGC 1624 ($l = 155\fdg356$, $b = 2\fdg616$) 
which is linearly interpolated from those between $l = 90^{\circ}$ and $180^{\circ}$ may 
not warrant consistency with the observational result. Alternatively, we obtained the IMF of NGC 1624 
selecting different control fields (the west region $\Delta \alpha = -6\farcm0$, 
$\Delta \delta = +3\farcm5$ or the south-west region $\Delta \alpha = -7\farcm5$, 
$\Delta \delta = -6\farcm5$ in Figure~\ref{fig1}). The slope of the IMF was about $\Gamma = 
-2.2 \pm 0.3$ and $-2.0 \pm 0.3$, respectively. Hence, the uncertainty 
arising from the subtraction of field interlopers has a negligible contribution 
to the slope.

\citet{K01} has presented a segmented IMF based on compiled data. In 
his alpha-plot, a non-negligible scatter was found in the mass range of 1 -- 100 $M_{\sun}$. 
The author attributed the cause of the scatter to dynamical effects and 
Poisson noise. It used to be believed that young open clusters were too young 
to have undergone dynamical mass segregation. However, a few young open clusters show clear signs of mass 
segregation, e.g. Arches cluster \citep{HSBHM13}, NGC 3603 \citep{PGA13}, 
NGC 6231 \citep{RM98,SSB13}, ONC \citep{HH98}, Westerlund 1 \citep{GBSH11,
LCS13}, etc. Several studies, through multiple sets of numerical simulation, suggested  
that the mass segregation found in young clusters has a dynamical origin \citep{MVZ07,MB09,
A09,A10}. If dynamical evolution had efficiently operated within young clusters, 
a number of low-mass members would have evaporated into the outskirts of the cluster 
region. Consequently, one would expect the IMF to be biased toward bright stars, and therefore 
show a shallow slope. The steep slope of the IMF in this work is inconsistent with this 
expectation of the effects. Thus, the contribution of the dynamical effect may be less significant for the clusters.

\begin{figure}
\epsscale{1.}
\plotone{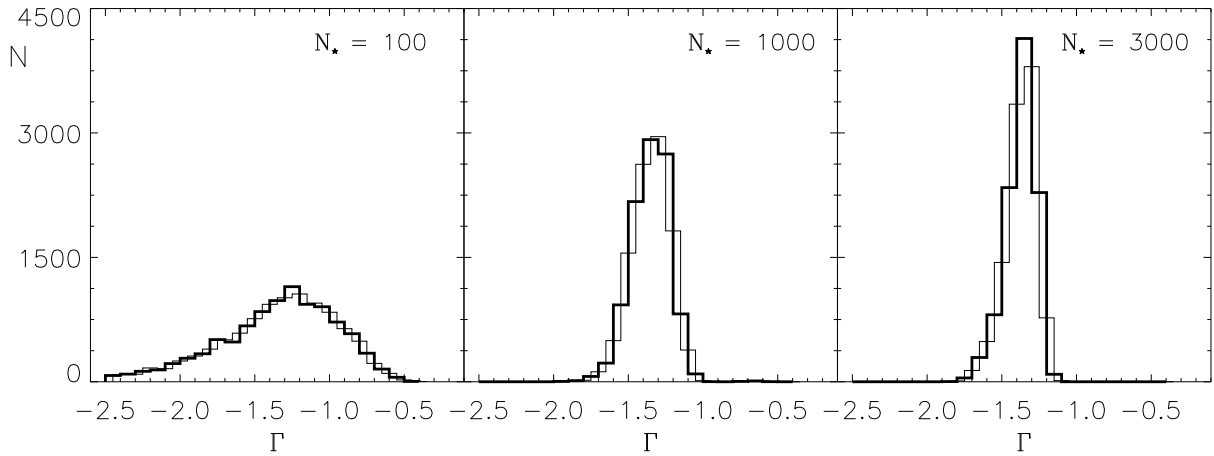}
\caption{Distribution of the IMF slope ($\Gamma$) with the total number of cluster members. A 
Monte-Carlo Simulation was performed 10,000 times for the model clusters consisting of 100, 1,000, 
and 3,000 members, respectively. The histograms are based on different binning of the same artificial 
stars (thick and thin solid lines). }
\label{fig15}
\end{figure}

The total number of members belonging to NGC 1624 and NGC 1931 is so small, that its measured 
IMF is vulnerable to stochastic effects. In order to test for this effect on the slope of 
the IMF, multiple sets of Monte Carlo simulation were conducted for three cases with the total number 
of cluster members ($N_{\star} = 100$, 1,000, and 3,000). The Kroupa IMF 
\citep{K01} was used as the underlying IMF for stars with mass larger 
than 0.5 $M_{\sun}$. The simulation for each case was made 10,000 times. 
The IMF of the model clusters was derived using the same procedure described as above. 
The size of the mass bin was set to $\Delta \log m = 0.4$, while a 
bin size of 1.0 was adopted for the highest mass bin if the number of stars 
with mass larger than 10 $M_{\sun}$ was smaller than 5. Figure~\ref{fig15} displays 
the distribution of $\Gamma$ with the total number of cluster members. The dispersion 
of the distribution appears to be large in model clusters with a small number of members, while 
the IMF of model clusters with a large number of members shows a higher consistency 
with the underlying IMF. However, the possibility that the IMF of a given cluster has a slope 
steeper than $\Gamma = -2.0$ is only about 10 percent in the case of $N_{\star} = 100$, and 
the likelihood decreases with the increase of the total number of cluster members. It 
implies that the possibility that the IMF of the two clusters in the present work exhibits a steeper 
slope than $-2.0$ is about 1 percent. Therefore, the stochastic effect seems not to explain 
our results reasonably. Although the Kroupa/Salpeter IMF provides a representative slope of 
the stellar IMF in a given mass range, it remains to be seen whether the slope is universal 
or not. Future work on the IMF of small, young open clusters will observationally be used 
to examine this issue. 

\section{DISCUSSIONS ON THE AGE SPREAD OF PMS STARS}

\citet{PaSt99,PaSt00,PaSt02} argued that the star formation activity within nearby 
SFRs has persisted for about 10 Myr and that the star formation rate 
may be accelerating in the present epoch, on timescales of 1 -- 2 Myr. Later, 
\citet{PRFP05} found a significant depletion of lithium in four PMS members of 
the Orion Nebula Cluster (ONC), and the result supported their idea 
because it is known that most of lithium at the surface of stars with subsolar 
mass are destroyed on timescales of $\sim$ 10 Myr. On the other hand, 
\citet{H01,H03} refuted the extended star formation timescale invoking the 
influences of binarity, variability, extinction, accretion activities, the inclusion of 
field interlopers, and the birthline effects of intermediate-mass stars. \citet{JONML07} 
found a small age spread of 2.5 Myr for NGC 2169. There are 
still many debates on the age of PMS stars although the age distribution 
is thought to be a key to understanding the processes of cluster formation. 

Only a small number of PMS stars were identified as members of NGC 1624, 
so we were unable to statistically investigate the age spread. The number of 
the PMS members found in NGC 1931 barely allowed us to study their 
age spread, although the membership is still incomplete. The age distribution 
of the PMS members shows a spread of 4.5 Myr. We attempted to interpret 
the observed age distribution. Several observational uncertainties affecting 
the $T_{eff}$ and luminosity of PMS stars were considered, 
e.g. photometric errors, reddening correction, and variability. \citet{H03} pointed 
out that the inclusion of field interlopers can lead the observed age spread to 
be larger. However, we believe that most of the PMS stars identified in this work are 
bona-fide members of NGC 1931 because the membership selection criteria 
were based on the well-known observational properties of such stars together 
with the fundamental parameters of the cluster (see Section 3.2). 
Contamination by field interlopers was excluded as a consideration.

\begin{figure}
\epsscale{1.}
\plotone{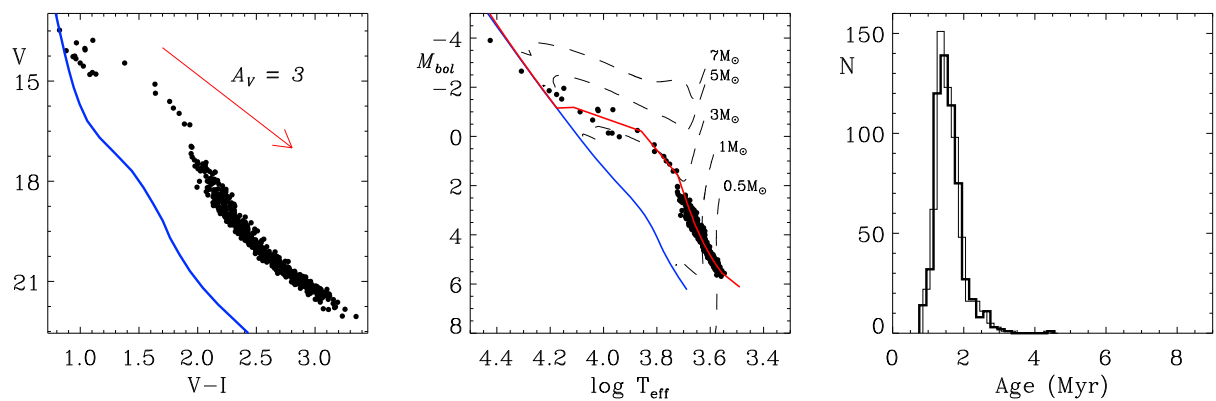}
\caption{The color-magnitude diagram (left), Hertzsprung-Russell diagram (middle), and the age 
distribution (right) of artificial stars. This simulation was conducted to reproduce the age distribution 
of PMS members in NGC 1931. The solid line and arrow in the left-hand panel denote the empirical 
ZAMS relation (Paper 0) and the reddening vector, respectively. In the middle panel, the solid and dashed lines represent 
isochrones (0 and 1.5 Myr) and evolutionary tracks for PMS stars \citep{SDF00,EGE12}. The histograms 
in the right-hand panel are based on different binning of the same stars (thick and thin solid lines). }
\label{fig16}
\end{figure}

We assessed the contribution of the observational uncertainties to the age 
spread of PMS stars using a few sets of simple Monte-Carlo simulations. 
NGC 1931 probably comprises more than 600 stars according to 
the integration of the IMF. We set the number of artificial PMS stars to 
600. The underlying IMF was assumed to be our result ($\Gamma = -2.0$) 
in the mass range of 0.5 to 5 $M_{\sun}$. The age of the stars was set to be 
1.5 Myr (the peak age of NGC 1931) with no intrinsic age spread. 
Based on the PMS evolutionary models of \citet{SDF00} artificial PMS stars 
were generated. $T_{eff}$ and luminosity were converted into $V_0$ magnitude 
and $(V-I)_0$ color using the relations of \citet{B95} and \citet{BCP98}. We 
introduced photometric errors with a similar distribution to the observed one. 
The distribution of reddening was assumed to be a 
normal distribution, where the mean reddening and $3\sigma$ dispersion 
were set to be 0.74 and 0.17 mag, respectively. In order to reproduce 
the extinction in the $V$ band we adopted the segmented reddening law as 
shown in Section 3.4 [$E(B-V)_{\mathrm{fg}} = 0.42$ mag, $R_{V,\mathrm{fg}} = 3.1$, and 
$R_{V,\mathrm{cl}} = 5.2$]. We re-analyzed the photometric data of the artificial stars 
to investigate their age spread after correcting for the mean reddening. 
Figure~\ref{fig16} displays the result of the Monte-Carlo simulation. 
The resulting age spread of the artificial stars was about 0.8 Myr although the age of a few 
PMS stars was overestimated. Hence, the photometric errors and the uncertainty 
in the reddening correction may not be enough to explain the observed age 
spread (4.5 Myr).

\begin{figure}
\epsscale{1.}
\plotone{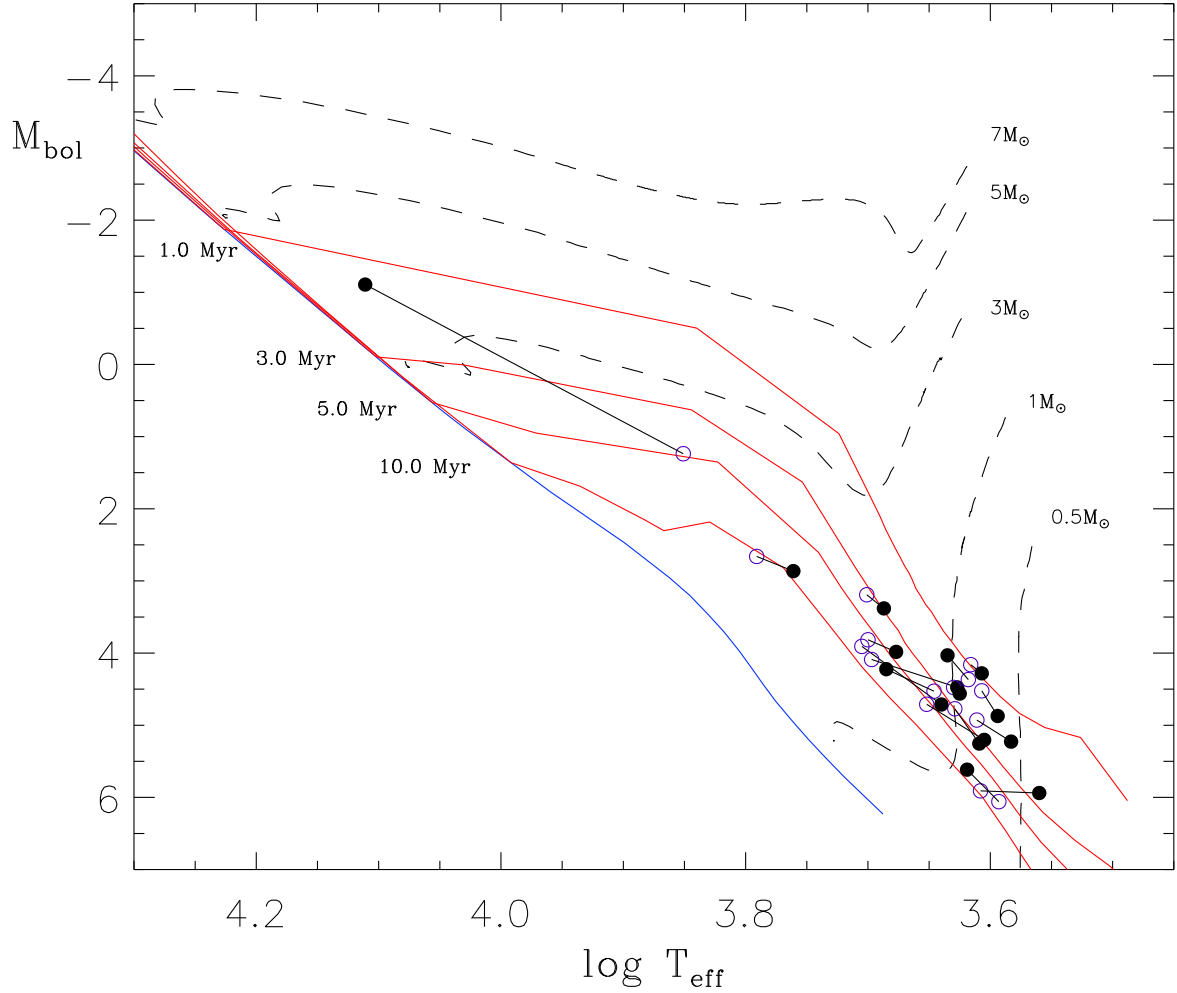}
\caption{Variable stars in the Hertzsprung-Russell diagram of NGC 1931. Open 
and filled circles represent the position of the variable PMS stars observed on 2011 October 29, 
and on 2013 February 5, respectively. Straight solid lines trace the variations. 
Several PMS evolutionary tracks and isochrones for 0, 1, 3, 5, and 10 Myr are indicated by dashed 
and solid lines \citep{SDF00,EGE12}.}
\label{fig17}
\end{figure}

The color variability of PMS stars can also increase the age spread because 
the position of the stars in the HRD is very sensitive to $T_{eff}$. We investigated 
the variation of their physical quantities in the HRD using the data observed at 
two different epochs. The variable PMS stars identified in this work changed by 
-1.2 -- 1.5 mag in $V$ and -0.5 -- 0.5 mag in $V-I$ over the 1.2 years. 
The variation in $V$ appeared larger than that in $I$. This aspect 
may be related to accretion activities, as well as to variable reddening by 
surrounding material \citep{BB90,GMBHS07}. Figure~\ref{fig17} shows the 
HRD of the variable PMS stars. Assuming that the variation of reddening in the 
line of sight is negligible on a timescale of 1 year, we examined the change in 
age. The intermediate-mass PMS star ID 727 (the most luminous star in 
the HRD) varied the most in $T_{eff}$ and luminosity over the 1.2 years. 
As a result of this variation, the apparent age changed by 2 -- 3 Myr. 
Some low-mass stars also revealed as large an age 
variation although the change in the physical quantities was not as large as 
those of the star ID 727. This is because the resolution for distinguishing the 
age of PMS stars becomes low in the low-mass regime of the HRD. Consequently, the 
variability of PMS stars very likely contributes to the observed age spread. 

The fraction of PMS variables is at least 4.6 percent (16/351), where the total number 
of PMS stars (denominator) down to our detection limit ($V \sim  22$ mag) was estimated 
from integration of the IMF. This value is likely a lower limit. We expect more 
PMS variables to exist in the cluster because stars varying on much 
shorter or longer timescales may not have been identified in our observations.
A couple of Monte-Carlo simulations were performed assuming that 
30 and 50 percent of the PMS members are variable. The variation in $V$ 
and $V-I$ was set to have a normal distribution with the same dispersion 
as the observed one. The photometric errors and reddening were treated with 
the same procedure as above. If we assume a variable fraction of 30 percent, the 
resultant age spread was about 1.5 Myr. A spread of 2.6 Myr was found for 
a variable fraction of 50 percent. These simple simulations showed that the variability of PMS stars 
can contribute a portion of the observed age spread. However, additional sources 
may still be required to explain the observed spread of 4.5 Myr. 

\begin{figure}
\epsscale{0.8}
\plotone{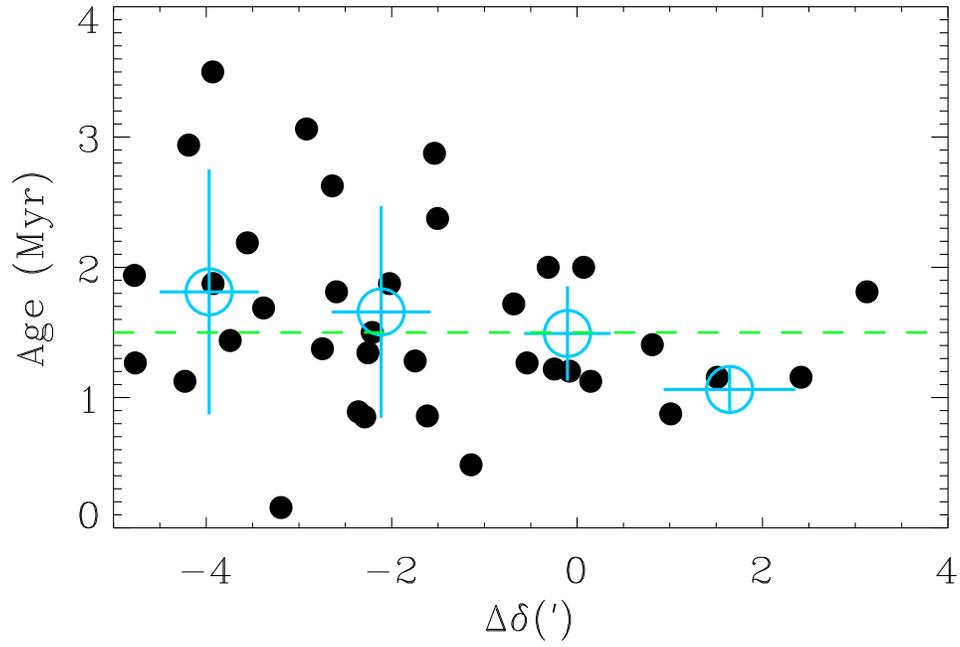}
\caption{The age variation of the PMS members of NGC 1931 along the declination axis. The mean 
age and standard deviation at a given declination bin are shown as an open circle and 
error bar. The dashed line indicates the peak age of 1.5 Myr in the age distribution (Figure~\ref{fig13}).}
\label{fig18}
\end{figure}

We briefly mentioned in Section 3.3 that the structure of NGC 1931 
may have resulted from its star formation history. If star formation propagated 
in a specific direction, one could expect there to be a correlation between the location 
and age of PMS stars. A systematic age variation on a few parsec 
scale has been found in the young open cluster NGC 1893 \citep{SPO07,PSC13,
LSKBP14}. Such a star formation history can influence the age spread of PMS 
stars. We investigated the age variation along the declination axis, as the cluster is elongated 
in the north-south direction (Figure~\ref{fig8}). 
PMS members with masses larger than 1.5 $M_{\sun}$ were excluded from this 
analysis because evolutionary models for intermediate-mass PMS stars are likely 
to overestimate the age of the stars \citep{SBL97,SBC04,H99}. We considered only 
stars younger than 4 Myr to find their age variation clearly. Figure~\ref{fig18} 
shows the age variation of the PMS members with declination. The age appears 
to decline with the increase in declination. It implies that star formation has propagated 
from south to north with a projected velocity of 6.8 km s$^{-1}$ according to the 
slope of the age variation. The maximum age difference between the stars in the 
northern and southern groups is about 0.7 Myr, on average. Hence, the star formation 
history in NGC 1931 seems to contribute a small portion of the observed age spread. 
In conclusion, a genuine age spread of PMS members is required to explain the 
observed age spread, however the size of the spread may not be as large as the 
timescale ($\sim 10$ Myr) argued by \citet{PaSt00,PaSt02}. 

\section{SUMMARY}
NGC 1624 and NGC 1931 are small, young, open clusters in the direction 
of the Galactic anticenter. These clusters are helpful in studying the property 
of star formation in a different environment from the solar 
neighbourhood. We performed $UBVI$ and H$\alpha$ photometry for the clusters 
as part of the SOS project. This work provided homogeneous photometric 
data as well as the comprehensive results for the clusters.

The members of the clusters were selected using various criteria. For NGC 1624, 
we found 14 early-type and 14 PMS members (3 variable, 3 UV excess emission, 
and 9 H$\alpha$ emission stars and candidates) from photometric criteria and H$\alpha$ 
photometry. On the other hand, 14 early-type and 71 PMS stars (16 variable, 4 UV 
excess emission, 54 MIR excess emission, and 27 H$\alpha$ emission stars and candidates) 
were identified as the members of NGC 1931 using photometric criteria, H$\alpha$ 
photometry, and the GLIMPSE360 catalog. 

The reddening of individual early-type members was determined from the 
($U-B, B-V$) TCD. The mean reddening was $\langle E(B-V) \rangle = 
0.92 \pm 0.05$ mag for NGC 1624 and $0.74 \pm 0.17$ mag for NGC 1931. 
The reddening law toward the clusters was examined using 
various color excess ratios from the optical to NIR passbands (to MIR passbands 
for NGC 1931). We confirmed that the reddening law toward and in NGC 1624 
is normal. On the other hand, the early-type members of NGC 1931 exhibited
color excess ratios that far deviated from the normal reddening. It implies that the size 
distribution of dust grains in the NGC 1931 ICM is somewhat different from that found in 
the general diffuse ISM. The ratio of total-to-selective extinction for the ICM of NGC 1931 was 
estimated to be $R_{V,\mathrm{cl}} = 5.2\pm 0.3$, indicating that the evolution of dust grains is 
still ongoing.

We carried out ZAMS fitting to the reddening-independent CMDs 
of the early-type members. As a result, NGC 1624 was found to be $6.0\pm 0.6$ kpc far away from 
the Sun, and probably located in the outer arm. The distance to NGC 1931 was 
about $2.3 \pm 0.2$ kpc, which places the cluster in the Perseus arm. Reddening-corrected 
CMDs were converted to the HRD using the published temperature scales and 
bolometric correction \citep{B95,BCP98,SLB13}. We performed isochrone fitting 
to the HRD of the clusters. A turn-off age of NGC 1624 was estimated to be 4 Myr, 
and the PMS members are likely to be younger than 4 Myr. The star ID 1773 associated 
with the UCHII region turned out to be a very young PMS star with a high mass 
accretion rate of $3.98 \times 10^{-6} M_{\sun} \ \mathrm{yr}^{-1}$. The most 
luminous star in NGC 1931 seems to be a very young star with an accretion disk 
given that the photometric properties were similar to those of Walker 90 
in NGC 2264. Thus, the age of the star is likely to be 1.6 -- 3.0 Myr. We found 
a median age of 2 Myr with a spread of 4.5 Myr for the age distribution of 
the PMS members.

Finally, we derived the IMF of the clusters and determined the slope in the high-mass 
regime ($1 > M_{\sun}$). The slope was $\Gamma = -2.0 \pm 0.2$ for NGC 1624 
and $\Gamma = -2.0 \pm 0.1$ for NGC 1931, respectively. These results were somewhat different 
from the Salpeter/Kroupa IMF. We attributed the steeper slope to a stochastic effect, the small number 
of stars in the highest mass bin ($> 10 \ M_{\sun}$), and the assumed upper limit of 
stellar mass. The IMF of the clusters exhibits a very similar shape to that of the well-studied young 
open cluster NGC 2264 \citep{SB10}. It implies that the property of star formation in 
the clusters may not be far different from that in nearby SFRs.

\acknowledgments
This work was partly supported by a National Research Foundation of Korea grant funded 
by the Korean Government (Grant No. 20120005318) and partly supported by the KASI 
(Korea Astronomy and Space Science Institute) grant 2014-9-710-02.

{\it Facilities:} \facility{AZT-22 1.5 m telescope}, \facility{Kuiper 1.55m telescope}.



\clearpage



\clearpage

\clearpage

\clearpage

\clearpage

\clearpage

\clearpage

\clearpage

\clearpage

\clearpage

\clearpage

\clearpage

\clearpage

\clearpage

\clearpage

\clearpage

\clearpage

\clearpage

\clearpage

\begin{deluxetable}{lccc}
\tabletypesize{\scriptsize}
\tablewidth{0pt}
\tablecaption{Observation Log \label{tab1}}
\tablehead{\colhead{Target} & \colhead{Date} & \colhead{Filter} &\colhead{Exposure time} } 
\startdata
   Maidanak Astronomical Observatory \\
   NGC 1624 & 2006.11.24. & $I$  & 3s, 60s  \\
                     &                       & $V$ & 3s, 180s \\
                     &                       & $B$ & 5s, 300s \\
                     &                       & $U$ & 15s, 600s \\
   Steward Observatory \\ 
   NGC 1624 & 2011.10.29. & $I$ & 5s, 60s \\
                     &                     & $V$& 5s, 180s \\ 
                     &                     & $B$& 7s, 300s \\ 
                     &                     & $U$& 15s, 600s \\
                     &                     & H$\alpha$ & 30s, 600s\\                  
   NGC 1931 & 2011.10.29.  & $I$ & 5s, 60s \\
                     &                     & $V$& 5s, 180s \\ 
                     &                     & $B$& 7s, 300s \\ 
                     &                     & $U$& 15s, 600s \\
                     &                     & H$\alpha$ & 30s, 600s\\    
                     & 2013.02.05.  & $I$ & 5s, 180s \\
                     &                     & $V$& 5s, 180s \\ 
                     &                     & $B$& 7s, 300s \\ 
                     &                     & $U$& 30s, 600s \\  
\enddata
\end{deluxetable}
\clearpage

\begin{deluxetable}{lcccc}
\tabletypesize{\scriptsize}
\tablewidth{0pt}
\tablecaption{Atmospheric Extinction Coefficients and Photometric Zero points \label{tab2}}
\tablehead{
\colhead{Date} & \colhead{filter} & \colhead{$k_1$} & \colhead{$k_2$} & \colhead{$\zeta$ (mag)}}
\startdata
2006.11.24.  & $I$ & $0.056 \pm 0.013$  & \nodata  & $23.229 \pm 0.008$ \\  
                    & $V$ & $0.118 \pm 0.014$  & \nodata  & $23.632 \pm 0.010$ \\
                    & $B$ & $0.231 \pm 0.003$  & $0.023 \pm 0.002$ & $23.498 \pm 0.010$ \\
                    & $U$ & $0.403 \pm 0.002$  & $0.018 \pm 0.001$ & $21.727 \pm 0.008$ \\
\hline
2011.10.29. &  $I$ & $0.046 \pm 0.006$  & \nodata  & $22.196 \pm 0.010$ \\  
                    & $V$ & $0.117 \pm 0.006$  & \nodata  & $23.543 \pm 0.012$ \\
                    & $B$ & $0.250 \pm 0.005$  & $0.026 \pm 0.002$ & $23.572 \pm 0.009$ \\
                    & $U$ & $0.460 \pm 0.007$  & $0.020 \pm 0.003$ & $22.109 \pm 0.011$ \\
                    & H$\alpha$& $0.094 \pm 0.009$ &   \nodata & 19.549 \\
\hline
2013.02.05  & $I$ & $0.045 \pm 0.008$ & \nodata & $22.170 \pm 0.009$ \\  
                    & $V$ & $0.120 \pm 0.008$ & \nodata & $23.560 \pm 0.007$ \\
                    & $B$ & $0.232 \pm 0.008$ & $0.023 \pm 0.002$ & $23.548 \pm 0.006$ \\  
                    & $U$ & $0.444 \pm 0.018$ & $0.031 \pm 0.005$ & $22.069 \pm 0.008$ \\
\enddata
\end{deluxetable}
\clearpage

\begin{deluxetable}{lcccccccc}
\tabletypesize{\tiny}
\tablewidth{0pt}
\tablecaption{Comparison of Photometry\label{tab3}}
\tablehead{\colhead{Paper} & \colhead{$\Delta V$\tablenotemark{1}} & \colhead{N(m)\tablenotemark{2}} &
\colhead{$\Delta$($V-I$)\tablenotemark{1}}  & \colhead{N(m)\tablenotemark{2}} &
\colhead{$\Delta$(\bv)\tablenotemark{1}} & \colhead{N(m)\tablenotemark{2}} &
\colhead{$\Delta$($\ub$)\tablenotemark{1}} & \colhead{N(m)\tablenotemark{2}} }
\startdata
NGC 1624 \\
This (MAO - SO$_{\mathrm{Oct}})$ & $0.003\pm0.013$ & 121(5) & $0.007\pm0.018$&124(2) & $-0.006\pm0.023$ & 123(3) & $0.001 \pm 0.035$ & 120(5) \\
\citet{SuBa06} & $-0.872\pm0.018$& 29(5)&$-0.046\pm0.014$&29(5)& $-0.032\pm0.022$& 31(3)& $-0.073\pm0.064$&31(3)\\
\citet{JPO11} &$0.020\pm0.034$&125(5)&$0.031\pm0.026$&180(10)&$0.017\pm0.041$&129(1)&$-0.063\pm0.077$&106(13)\\
\hline
NGC 1931  \\
This (SO$_{\mathrm{Feb}}$ - SO$_{\mathrm{Oct}}$) & $0.004\pm0.011$&92(15)&$0.013\pm0.017$&102(5)& $-0.011\pm0.018$&96(11)&$0.004\pm0.027$&97(7)\\
\citet{PES13} & $-0.023\pm0.036$&142(12)&$-0.012\pm0.022$&136(20)&$0.014\pm0.022$&141(13)&$0.023\pm0.063$&136(1)\\
\enddata
\tablenotetext{1}{This - Others}
\tablenotetext{2}{N and m indicates the number of compared stars and excluded stars}
\end{deluxetable}
\clearpage

\begin{deluxetable}{rccccccccccccccccc}
\rotate
\tabletypesize{\tiny}
\tablewidth{0pt}
\tablecaption{Photometric Data for NGC 1624 \label{tab4}}
\tablehead{
\colhead{ID} & \colhead{$\alpha_{\mathrm{J2000}}$}  & \colhead{$\delta_{\mathrm{J2000}}$} &
\colhead{$V$} & \colhead{$I$} & \colhead{$V-I$} & \colhead{$B-V$} & \colhead{$U-B$} & 
\colhead{$H-C$}\tablenotemark{a} & \colhead{$\epsilon_V$} & \colhead{$\epsilon_I$} & \colhead{$\epsilon_{V-I}$} & 
\colhead{$\epsilon_{B-V}$} & \colhead{$\epsilon_{U-B}$} & \colhead{$\epsilon_{H-C}$} & \colhead{N$_{\mathrm{obs}}$} & 
\colhead{2MASSID} & \colhead{H$\alpha$}\tablenotemark{b}}
\startdata
 2325 & 04 40 39.03 & +50 34 39.2 & 21.479  & 19.533  &  1.931  &  1.367  & \nodata & \nodata & 0.091   & 0.086   & 0.125   & 0.194   & \nodata & \nodata & 1 1 1 1 0 0 &        -         & - \\
 2326 & 04 40 39.05 & +50 28 24.4 & 18.969  & 17.068  &  1.883  &  1.625  &  1.419  & -0.353  & 0.010   & 0.002   & 0.023   & 0.056   & 0.176   & 0.057   & 3 5 3 2 1 2 & 04403905+5028245 & h \\
 2327 & 04 40 39.07 & +50 27 15.1 & \nodata & 18.484  & \nodata & \nodata & \nodata & \nodata & \nodata & 0.030   & \nodata & \nodata & \nodata & \nodata & 0 2 0 0 0 0 &        -         & - \\
 2328 & 04 40 39.08 & +50 27 13.4 & \nodata & 19.119  & \nodata & \nodata & \nodata & \nodata & \nodata & 0.226   & \nodata & \nodata & \nodata & \nodata & 0 2 0 0 0 0 &        -         & - \\
 2329 & 04 40 39.08 & +50 30 41.7 & 21.374  & 19.444  &  1.858  &  1.138  & \nodata & \nodata & 0.085   & 0.004   & 0.060   & 0.113   & \nodata & \nodata & 2 3 2 1 0 0 &        -         & - \\
 2330 & 04 40 39.10 & +50 25 35.9 & 19.491  & 17.965  &  1.518  &  1.261  &  0.658  & -0.145  & 0.001   & 0.001   & 0.005   & 0.018   & 0.054   & 0.061   & 2 3 2 2 2 1 & 04403910+5025356 & - \\
 2331 & 04 40 39.12 & +50 27 27.6 & 20.972  & 18.660  &  2.256  &  1.732  & \nodata & \nodata & 0.006   & 0.001   & 0.040   & 0.162   & \nodata & \nodata & 2 3 2 2 0 0 & 04403915+5027275 & - \\
 2332 & 04 40 39.19 & +50 27 05.9 & \nodata & 19.340  & \nodata & \nodata & \nodata & \nodata & \nodata & 0.042   & \nodata & \nodata & \nodata & \nodata & 0 2 0 0 0 0 &        -         & - \\
 2333 & 04 40 39.20 & +50 27 07.9 & 21.751  & 18.229  &  3.523  & \nodata & \nodata & \nodata & 0.186   & 0.012   & 0.186   & \nodata & \nodata & \nodata & 1 2 1 0 0 0 & 04403919+5027075 & - \\
 2334 & 04 40 39.22 & +50 27 40.2 & 16.956  & 15.828  &  1.138  &  0.844  &  0.251  & -0.327  & 0.003   & 0.012   & 0.016   & 0.010   & 0.011   & 0.007   & 4 5 4 4 3 2 & 04403923+5027402 & H \\
 2335 & 04 40 39.24 & +50 18 01.8 & 20.933  & 18.638  &  2.294  &  1.798  & \nodata & \nodata & 0.083   & 0.007   & 0.083   & 0.207   & \nodata & \nodata & 1 2 1 1 0 0 & 04403923+5018016 & - \\
 2336 & 04 40 39.24 & +50 27 14.2 & \nodata & 18.519  & \nodata & \nodata & \nodata & \nodata & \nodata & 0.027   & \nodata & \nodata & \nodata & \nodata & 0 2 0 0 0 0 &        -         & - \\
 2337 & 04 40 39.26 & +50 19 06.5 & 21.818  & 20.007  &  1.811  & \nodata & \nodata & \nodata & 0.152   & 0.145   & 0.210   & \nodata & \nodata & \nodata & 1 1 1 0 0 0 &        -         & - \\
 2338 & 04 40 39.30 & +50 27 11.6 & 15.843  & 14.918  &  0.920  &  0.691  & -0.039  &  0.219  & 0.001   & 0.003   & 0.003   & 0.005   & 0.005   & 0.006   & 4 5 4 4 4 2 & 04403929+5027116 & - \\
 2339 & 04 40 39.32 & +50 28 47.4 & 18.428  & 17.077  &  1.350  &  1.059  &  0.428  &  0.156  & 0.013   & 0.023   & 0.011   & 0.008   & 0.020   & 0.041   & 4 5 4 3 2 1 & 04403931+5028473 & - \\
 2340 & 04 40 39.32 & +50 31 39.1 & 21.388  & 18.611  &  2.801  &  1.465  & \nodata & \nodata & 0.063   & 0.006   & 0.045   & 0.244   & \nodata & \nodata & 2 3 2 1 0 0 & 04403932+5031392 & - \\
 2341 & 04 40 39.34 & +50 27 19.6 & 13.204  & 12.338  &  0.859  &  0.623  & -0.298  &  0.147  & 0.005   & 0.008   & 0.001   & 0.001   & 0.001   & 0.018   & 3 5 3 3 3 1 & 04403935+5027196 & - \\
 2342 & 04 40 39.38 & +50 33 01.5 & 16.611  & 15.251  &  1.352  &  1.131  &  0.627  & \nodata & 0.004   & 0.002   & 0.004   & 0.004   & 0.014   & \nodata & 2 3 2 2 1 0 & 04403938+5033015 & - \\
\enddata
\tablenotetext{a}{$H-C$ represents the H$\alpha$ index [$\equiv \mathrm{H}\alpha - (V+I)/2$]}
\tablenotetext{b}{H: H$\alpha$ emission stars; h: H$\alpha$ emission star candidates}
\end{deluxetable}
\clearpage

\begin{deluxetable}{rccccccccccccccccc}
\rotate
\tabletypesize{\tiny}
\tablewidth{0pt}
\tablecaption{Photometric Data for NGC 1931 \label{tab5}}
\tablehead{
\colhead{ID} & \colhead{$\alpha_{\mathrm{J2000}}$}  & \colhead{$\delta_{\mathrm{J2000}}$} &
\colhead{$V$} & \colhead{$I$} & \colhead{$V-I$} & \colhead{$B-V$} & \colhead{$U-B$} & 
\colhead{$H-C$} & \colhead{$\epsilon_V$} & \colhead{$\epsilon_I$} & \colhead{$\epsilon_{V-I}$} & 
\colhead{$\epsilon_{B-V}$} & \colhead{$\epsilon_{U-B}$} & \colhead{$\epsilon_{H-C}$} & \colhead{N$_{\mathrm{obs}}$} & 
\colhead{2MASSID} & \colhead{H$\alpha$}}
\startdata
 705 & 05 31 23.46 & +34 12 27.6 & 21.091  & 18.103  &  3.031  & \nodata & \nodata & -1.609  & 0.007   & 0.031   & 0.058   & \nodata & \nodata & 0.079   & 2 3 2 0 0 1 & 05312347+3412276 & H \\
  706 & 05 31 23.47 & +34 10 47.2 & 21.032  & 18.326  &  2.690  &  1.529  & \nodata & -0.555  & 0.041   & 0.037   & 0.071   & 0.221   & \nodata & 0.099   & 2 3 2 2 0 1 & 05312345+3410471 & H \\
  707 & 05 31 23.47 & +34 13 28.0 & 19.460  & 17.885  &  1.567  &  1.205  &  0.593  &  0.038  & 0.001   & 0.013   & 0.013   & 0.062   & 0.083   & 0.073   & 3 4 3 2 2 1 & 05312347+3413281 & - \\
  708 & 05 31 23.48 & +34 12 18.4 & 21.762  & 18.790  &  2.950  & \nodata & \nodata & \nodata & 0.085   & 0.029   & 0.050   & \nodata & \nodata & \nodata & 2 2 2 0 0 0 & 05312349+3412182 & - \\
  709 & 05 31 23.49 & +34 13 06.1 & 20.549  & 17.648  &  2.938  &  1.704  & \nodata & -1.336  & 0.001   & 0.045   & 0.056   & 0.105   & \nodata & 0.058   & 2 4 2 2 0 1 & 05312347+3413060 & H \\
  710 & 05 31 23.50 & +34 14 02.0 & 21.590  & 18.543  &  3.048  & \nodata & \nodata & -1.852  & 0.084   & 0.054   & 0.100   & \nodata & \nodata & 0.109   & 1 1 1 0 0 1 &        -         & - \\
  711 & 05 31 23.51 & +34 18 16.3 & 20.080  & 18.087  &  1.985  &  1.519  & \nodata &  0.055  & 0.027   & 0.001   & 0.021   & 0.062   & \nodata & 0.058   & 2 3 2 2 0 1 & 05312351+3418161 & - \\
  712 & 05 31 23.53 & +34 14 44.8 & 21.151  & 18.865  &  2.286  & \nodata & \nodata & \nodata & 0.104   & 0.062   & 0.121   & \nodata & \nodata & \nodata & 1 1 1 0 0 0 & 05312356+3414448 & - \\
  713 & 05 31 23.53 & +34 13 41.5 & 17.125  & 15.433  &  1.629  &  1.346  &  0.648  &  0.023  & 0.020   & 0.019   & 0.004   & 0.031   & 0.014   & 0.013   & 4 4 4 4 4 2 & 05312353+3413415 & - \\
  714 & 05 31 23.55 & +34 15 05.5 & 16.541  & 15.444  &  1.091  &  0.782  &  0.479  &  0.313  & 0.008   & 0.002   & 0.010   & 0.006   & 0.009   & 0.012   & 4 4 4 4 4 2 & 05312355+3415056 & - \\
  715 & 05 31 23.57 & +34 13 09.2 & 21.865  & 18.669  &  3.196  & \nodata & \nodata & -1.347  & 0.116   & 0.047   & 0.125   & \nodata & \nodata & 0.154   & 1 1 1 0 0 1 & 05312357+3413097 & - \\
  716 & 05 31 23.57 & +34 13 35.3 & 19.420  & 17.059  &  2.377  &  1.811  & \nodata & \nodata & 0.045   & 0.004   & 0.044   & 0.057   & \nodata & \nodata & 4 4 4 2 0 0 & 05312356+3413353 & - \\
  717 & 05 31 23.60 & +34 10 59.6 & 20.964  & 17.982  &  2.958  &  1.642  & \nodata & -0.206  & 0.047   & 0.019   & 0.023   & 0.131   & \nodata & 0.102   & 2 4 2 1 0 1 &        -         & - \\
  718 & 05 31 23.60 & +34 13 54.2 & 19.331  & 17.506  &  1.826  &  1.543  & \nodata & -0.105  & 0.001   & 0.038   & 0.044   & 0.100   & \nodata & 0.072   & 4 4 4 2 0 1 &        -         & - \\
  719 & 05 31 23.60 & +34 14 10.2 & 21.592  & 18.259  &  3.333  & \nodata & \nodata & \nodata & 0.104   & 0.031   & 0.109   & \nodata & \nodata & \nodata & 1 2 1 0 0 0 &        -         & - \\
  720 & 05 31 23.61 & +34 10 32.2 & 21.686  & 18.904  &  2.787  & \nodata & \nodata & \nodata & 0.017   & 0.013   & 0.031   & \nodata & \nodata & \nodata & 2 2 2 0 0 0 & 05312361+3410320 & - \\
  721 & 05 31 23.64 & +34 11 12.7 & 21.413  & 19.373  &  2.021  & \nodata & \nodata & \nodata & 0.041   & 0.068   & 0.119   & \nodata & \nodata & \nodata & 2 2 2 0 0 0 &        -         & - \\
  722 & 05 31 23.68 & +34 10 40.7 & 15.000  & 14.352  &  0.640  &  0.472  &  0.201  &  0.314  & 0.005   & 0.001   & 0.004   & 0.003   & 0.005   & 0.007   & 4 4 4 4 4 2 & 05312367+3410407 & - \\
\enddata
\end{deluxetable}
\clearpage

\appendix

\counterwithin{figure}{section}
\counterwithin{table}{section}

\section{TRANSFORMATION RELATIONS FOR THE MONT4K CCD CAMERA OF THE KUIPER 61" TELESCOPE}

The Mont4k photometric system of the Kuiper 61" telescope at SO was used in the SOS project for the 
first time. It is necessary to obtain reliable transformation relations of the instrumental system to 
produce homogeneous standardized photometric data for many open clusters. In this appendix 
we describe the derivation of the transformation relations using a large number of the standard stars 
observed in 2011 -- 2013 (281 in Bessell $U$, 119 in SDSS $u'$, 402 in $B$, 436 in $V$, and 530 in $I$). 
The SDSS $u'$ filter was used on several nights in 2012, instead of the Bessell $U$ filter, 
and so the transformation relation was also addressed for the forthcoming papers.

The Mont4k tasks installed in the observing system automatically corrected for variations in the bias level 
and crosstalk for individual images, converting from the raw extended FITS into normal FITS images. In 
order to remove remaining instrumental artifacts we obtained a number of bias, sky flat, and shutter shading 
images during observing runs. The bias level was very small (a few ADU) because the overscan correction 
had already been done. After subtracting the tiny residual bias, all target images were divided by the master flat images in 
the same passbands. The Mont4k CCD camera system is equipped with an iris-type shutter. This type of 
shutter imprints an iris pattern on short-exposure images. In addition, there could be either a
systematic gain or loss in the observed flux depending on the speed of the shutter. 
These behaviours are crucial for obtaining reliable magnitudes in short 
exposure images. We carefully investigated the shutter shading with various exposure 
time (0.5s, 1s, 2s, 3s, 4s, 5s, 7s, 10s, 15s, and 20s) according to the procedures delineated by 
\citet{LSKI08} and found that images exposed for less than 7s showed the iris pattern. Images with 
exposure times less than 2s revealed a systematic gain variation of more than 0.5 percent 
(up to 4.6 percent at 0.5s) in the observed flux. It implies that the shutter operates more slowly for exposure time 
shorter than 2s. Based on these results we corrected the shutter shading effect on short exposure images (0.5 -- 7s). 

\begin{figure}
\epsscale{.80}
\plotone{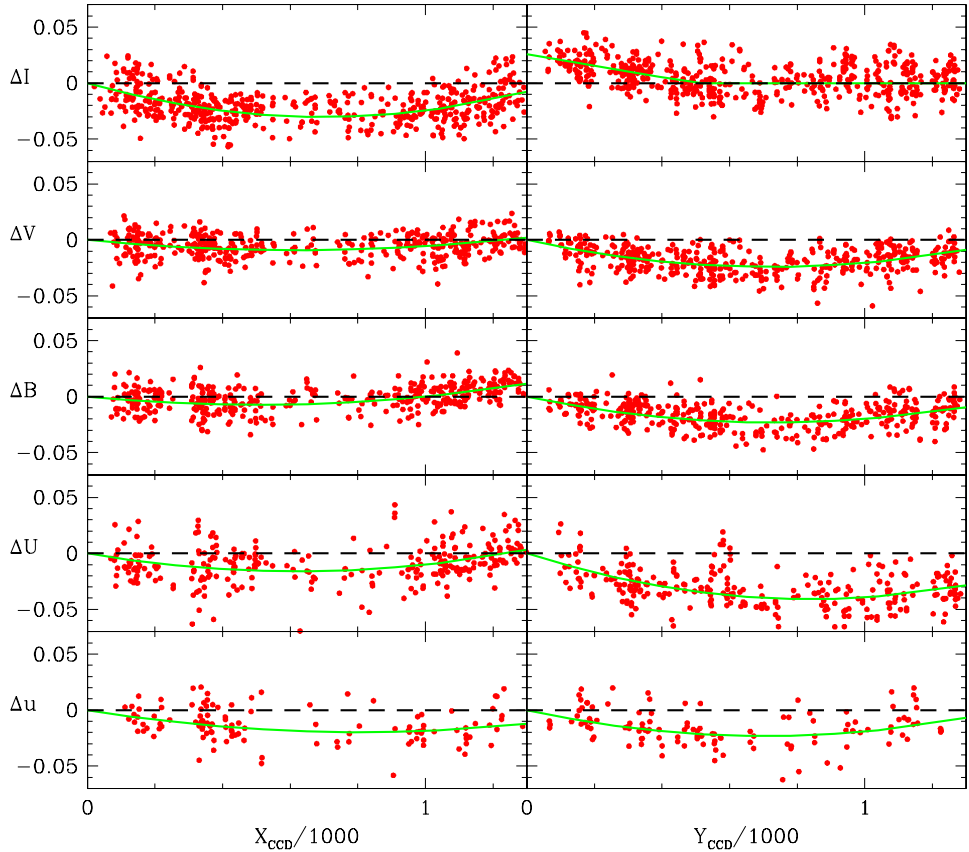}
\caption{Spatial variation of the measured magnitudes with respect to the $X_{CCD}$ (left) or $Y_{CCD}$ 
(right) coordinates of the Mont4k CCD chip. The $X_{CCD}$ coordinates of stars in a positive 
direction from the CCD center ($X_{CCD} = 682$) was increased by 5 pixels because of 
five uncountable columns in the center of the normal FITS images. All the pixel coordinates of the 
standard stars on the CCD chip have been 
divided by 1,000. Most spatial variations appear in the form of a second-order polynomial 
equation. The variation of the $I$ magnitude along the $Y_{CCD}$ can be approximated by a 
combination of two straight lines. The maximum difference between the center and edge of 
the CCD is less than 0.05 mag.  }
\label{a1}
\end{figure}

We performed aperture photometry for the standard stars with an aperture size of 14.0 arcsec (16.3 pixels). 
The enormous amount of photometric data obtained over 11 nights in 2011 -- 2013 allowed us to obtain reliable 
transformation relations. The instrumental magnitudes can be transformed to the standard magnitudes 
and colors using the following equation:

\begin{equation}
M_{\lambda} = m_{\lambda} - (k_{1\lambda} - k_{2\lambda}C_0)\cdot X + \eta _{\lambda} \cdot C_0 
+ \alpha _{\lambda} \cdot \hat{UT} + \beta _{\lambda}\cdot \hat{x}^2_{CCD} 
+ \gamma _{\lambda} \cdot \hat{x}_{CCD} + \delta _{\lambda} \cdot \hat{y}^2_{CCD} 
+ \epsilon \cdot \hat{y}_{CCD} + \zeta _{\lambda}
\end{equation}

\noindent where $\beta _{\lambda}$, $\gamma _{\lambda}$, $\delta _{\lambda}$, and 
$\epsilon _{\lambda}$ represent the coefficients of spatial variation in magnitude. Other symbols are 
the same as in Equation 1. The $X_{CCD}$ and $Y_{CCD}$ coordinates were normalized by 
1,000 pixels ($\hat{x}_{CCD} \equiv X_{CCD}/1,000$, 
$\hat{y}_{CCD} \equiv Y_{CCD}/1,000$). Figure ~\ref{a1} shows the spatial variations in the measured 
magnitudes with respect to $X_{CCD}$ and $Y_{CCD}$ coordinates for each passband. Since there are five 
uncountable columns in the center of the normal FITS images, the $X_{CCD}$ coordinate of stars 
away from the center (682$^{\mathrm{th}}$ pixel) in a positive direction was increased by 5 pixels. The form 
of the spatial variations can be represented by a second-order polynomial. The pattern may 
relate to the shape of the focal plane of the Kuiper 61" telescope (see also \citealt{MC04,SBCKI08,
LCS13}). In most cases, the amount of the variation from the CCD center was less than 0.03 mag. 
The maximum difference was up to 0.05 mag in the Bessell $U$ band. We performed the second-order 
polynomial fitting to the data, presenting the coefficients of the spatial variation in Table~\ref{at1}. 
In addition, a linear variation along the $Y_{CCD}$ coordinate in the $I$ band was found 
between the 0 -- 500$^{\mathrm th}$ pixels, but there was no remarkable spatial variation on the other side. 

\begin{figure}
\epsscale{.80}
\plotone{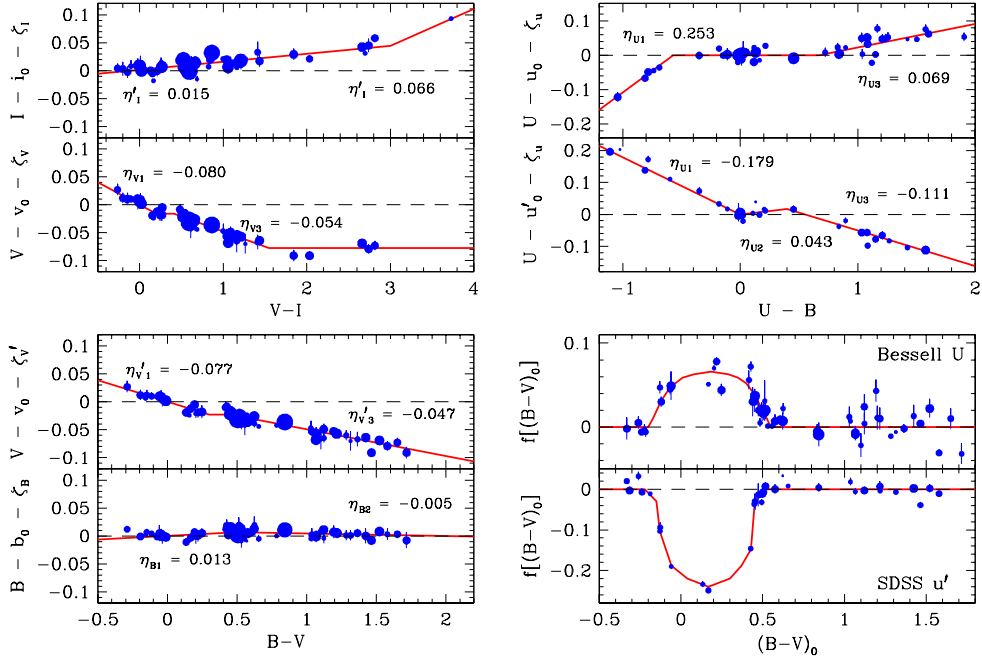}
\caption{Transformation relations of the Mont4k CCD camera system with respect to relevant colors. The size of 
circles is proportional to the number of observations and the degree of precision of 
the standard stars. The error bar is the standard deviation of the observed 
magnitude. The combinations of straight lines in each panel represent 
the transformation relation. The lower-right panels show a non-linear 
correction term against dereddened $B-V$ color in the $U$ band. See the main text for details.  }
\label{a2}
\end{figure}

The atmospheric extinction and transformation coefficients were determined using a weighted least-square 
method after correcting for the spatial variations. Figure~\ref{a2} 
shows the transformation relations for the Mont4k CCD camera system. A combination 
of a few straight lines constitutes the transformation relations. Since there was only one 
extremely red star ($V-I > $ 3 ) in the $I$ transformation with respect to $V-I$ color, the 
transformation relation may be uncertain in that color range. The $V$ transformation with respect 
to $V-I$ shows a plateau for the reddest stars ($V-I > 1.55$). The TiO band spectral features
in late-type stars are related to this plateau as seen in other studies \citep{SB00,
LSBKI09,LCS13}. The $B$ transformation was approximated by a combination of two straight lines, 
and the coefficients are very small. The transformation relations of the Bessell $U$ and SDSS $u'$ filters 
exhibit opposite trends to each other. A non-linear behaviour was found in the $U$ transformation. 
According to \citet{SB00} and \citet{LSBKI09} this aspect is caused by the Balmer discontinuity in 
the fluxes of B--F-type stars and difference between the response functions of the Johnson-Cousins 
standard system and natural photometric systems. Such a non-linearity can be 
corrected using dereddened $B-V$ colors as shown in the lower-right panels of Figure~\ref{a2}. 
The aspect of the non-linear correction term appears as an opposite trend in the 
Bessell $U$ and SDSS $u'$ transformation relations. The transformation coefficients with respect to
given colors are summarized in Table~\ref{at2}.

\begin{figure}
\epsscale{.80}
\plotone{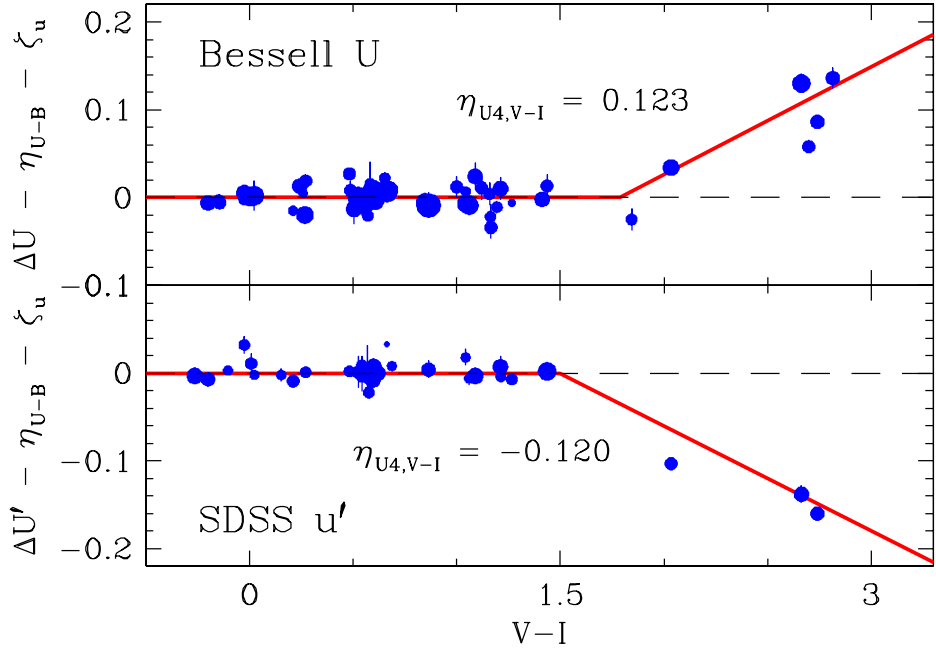}
\caption{Additional correction terms in the $U$ band for the intrinsic red stars ($U-B > 1.0$). 
Upper: Red leak effect found in the Bessell $U$ filter attached to the Mont4k CCD 
camera system. The contribution of the red light can be removed by a linear regression for stars ($V-I > 1.79$). 
Lower: More flux in the $U$ band is required for late-type stars because our SDSS $u'$ filter was 
designed to avoid the Balmer discontinuity toward shorter wavelength. An additional correction 
can be made with a straight line for red stars ($V-I > 1.5$). }
\label{a3}
\end{figure}

On the other hand, we found that standard stars with an intrinsically red color appeared 
systematically brighter than other stars in the Bessell $U$ filter after correcting with all the 
coefficients above (see the upper panel in Figure~\ref{a3}). Unlike the 
Bessell $U$ filter used at MAO, the $U$ filter attached to the Mont4k CCD camera exhibits the 
so-called red leak. Since the contribution of red light drastically increases for $U-B > 1$, a careful correction 
for the red leak effect was carried out with respect to the $V-I$ color as shown in the upper panel of Figure~\ref{a3}. 
The transformation of the SDSS $u'$ filter also required an additional correction term for extremely 
red stars. Because the transmission function of the SDSS $u'$ filter was shifted toward shorter wavelength to 
avoid the Balmer discontinuity, one can expect that the $u'$ magnitude of late-type stars may be fainter than 
the Johnson $U$ magnitude. Thus, a negative correction term for red stars ($V-I > 1.5$) was required 
as shown in the lower panel of Figure~\ref{a3}. We note that all the coefficients in the transformation 
equation were obtained from iteration of the procedures described above.

\clearpage

\clearpage

\clearpage









\begin{deluxetable}{cccccc}
\tabletypesize{\scriptsize}
\tablewidth{0pt}
\tablecaption{Spatial Variation Coefficients of the Mont4k CCD Camera \label{at1}}
\tablehead{
\colhead{Filter} & \colhead{$\beta$} & \colhead{$\gamma$} & \colhead{$\delta$} & \colhead{$\epsilon$} & \colhead{Remark}}
\startdata
  $I$  & 0.062  & -0.086  & \nodata & -0.052 & $Y_{CCD} < 500$ pixels \\  
 $V$ & 0.024 & -0.029 & 0.046 & -0.066 &  \\
 $B$  & 0.029 & -0.029 & 0.043 & -0.063 &  \\
 $U$  & 0.042 & -0.052 & 0.058 & -0.097 & \\
 $u'$   & 0.030 & -0.049 & 0.047 & -0.066 &\\
\enddata
\end{deluxetable}
\clearpage

\begin{deluxetable}{cccccc}
\tabletypesize{\scriptsize}
\tablewidth{0pt}
\tablecaption{Transformation Coefficients of the Mont4k CCD Camera \label{at2}}
\tablehead{
\colhead{Filter} &&  &Color range & \\ 
 & \colhead{$\eta _{\lambda, 1}$} & \colhead{$\eta _{\lambda, 2}$} & 
\colhead{$\eta _{\lambda, 3}$} & \colhead{$\eta _{\lambda, 4}$} & \colhead{Correction term}} 
\startdata
  $I$ &  $V-I \leq 3$ & $V-I > 3$ & & & \\
       & $0.015 \pm 0.001$ & 0.066 & \nodata & \nodata & \nodata \\
\hline
 $V$& $V-I \leq 0.2$ & $0.2 < V-I \leq 0.41$ & $0.41 < V-I \leq 1.55$ & $V-I > 1.55$ & \\
       & $-0.080\pm0.010$ & 0.000 & $-0.054\pm0.002$ & 0.000 & \nodata\\
       & $B-V \leq 0.3$ & $0.3 < B-V \leq 0.45$ & $V-I > 0.45$ & &\\
       & $-0.077\pm0.006$ & 0.000 &$ -0.047\pm 0.001$ & \nodata & \nodata \\
       & $V-R \leq 0.13$ & $0.13 < V-R \leq 0.25$ & $0.25 < V-R \leq 0.45$ & $V-R > 0.9$ &\\
       & $-0.150 \pm 0.013$ & 0.000 & $-0.094 \pm 0.003$ & 0.000 & \nodata\\
\hline
$B$& $B-V \leq 0.5$ & $B-V > 0.5$ & & &\\
       & $0.013\pm0.002$ & $-0.005\pm 0.001$& \nodata & \nodata &\nodata\\
\hline
$U$&$U-B \leq -0.57$& $-0.57 < U-B \leq 0.67$ & $U-B > 0.67$ && $V-I >1.79$\\
       &$0.253\pm0.016$ & 0.000 & $0.069\pm0.007$ & \nodata & $0.123\pm0.019$ \\
\hline
$u'$ & $U-B \leq 0.$ & $0. < U-B \leq 0.4$ & $U-B > 0.4$ & &$V-I > 1.5$\\
       &$-0.179\pm0.004$ & 0.043 & $-0.111\pm0.004$ & \nodata & $-0.120\pm0.007$\\
\enddata
\end{deluxetable}
\clearpage




\begin{thebibliography}{}
\bibitem[Adams et al.(2001)]{ASMSB01} Adams, J. D., Stauffer, J. R., Monet, D. G., Skrutskie, M. F., \& Beichman, C. A. 2001, \aj, 121, 2053
\bibitem[Allison et al.(2010)]{A10} Allison, R. J., Goodwin, S. P., Parker, R. J., Portegies Zwart, S. F., \& de Grijs, R. 2010, \mnras, 407, 1098
\bibitem[Allison et al.(2009)]{A09} Allison, R. J., Goodwin, S. P., Parker, R. J., et al. 2009, \apj, 700, L99
\bibitem[Basri \& Batalha(1990)]{BB90} Basri, G., \& Batalha, C. 1990, \apj, 363, 654
\bibitem[Bastian et al.(2010)]{BCM10} Bastian N., Covey K. R., \& Meyer M. R., 2010, \araa, 48, 339
\bibitem[Bertout et al.(1988)]{BBB88} Bertout, C., Basri, G., \& Bouvier, J. 1988, \apj, 330, 350
\bibitem[Bessell(1990)]{B90} Bessell, M. S. 1990, \pasp, 102, 1181
\bibitem[Bessell(1995)]{B95} Bessell, M. S., 1995, in Proc. ESO Workshop, The Bottom of the Main Sequence and Beyond, ed. C. G. Tinney (Berlin:Springer), 123
\bibitem[Bessell et al.(1998)]{BCP98} Bessell, M. S., Castelli, F., \& Plez B., 1998, \aap, 333, 231
\bibitem[Bhatt et al.(1994)]{BPMP94} Bhatt, B. C., Pandey, A. K., Mahra, H. S., \& Paliwal, D. C. 1994, BASI, 22, 291
\bibitem[Bonatto \& Bica(2009)]{BB09} Bonatto, C., \& Bica, E. 2009, \mnras, 397, 1915
\bibitem[Brott et al.(2011)]{BMC11} Brott, I., de Mink, S. E., Cantiello, M., et al. 2011, \aap, 530, 1115 
\bibitem[Calvet \& Gullbring(1998)]{CG98} Calvet, N., \& Gullbring, E. 1998, \apj, 509, 802
\bibitem[Caramazza et al.(2008)]{CMP08} Caramazza, M., Micela, G., Prisinzano, L., et al. 2008, \aap, 488, 211
\bibitem[Caramazza et al.(2012)]{CMP12} Caramazza M., Micela G., Prisinzano L., et al. 2012, \aap, 539, 74
\bibitem[Chini \& Wink(1984)]{CW84} Chini, R., \& Wink, J. E. 1984, \aap, 139, 5
\bibitem[Clark et al.(2005)]{CNCG05} Clark, J. S., Negueruela, I., Crowther, P. A., \& Goodwin. S. P. 2005, \aap, 434, 949
\bibitem[Deharveng et al.(2008)]{DLK08} Deharveng, L., Lefloch, B., Kurtz, S., et al. 2008, \aap, 482, 585
\bibitem[Dias \& L\'{e}pine(2005)]{DL05} Dias, W. S., \& L\'{e}pine. J. R. D. 2005, \apj, 629, 825
\bibitem[Ekstr\"om et al.(2012)]{EGE12} Ekstr\"om, S., Georgy, C., Eggenberger, P., et al., 2012, \aap, 537, 146
\bibitem[Elmegreen et al.(2000)]{EEPZ00} Elmegreen, B. G., Efremov, Y., Pudritz, R. E., \& Zinnecker, H. 2000, in Protostars and Planets IV, ed. V. Mannings, A. P. Boss, \& S. S. Russell (Tucson: Univ. Arizona Press), 179
\bibitem[Eswaraiah et al.(2011)]{EPM11} Eswaraiah C., Pandey A. K., Maheswar G., et al. 2011, \mnras, 411, 1418
\bibitem[Evans et al.(2010)]{EPG10} Evans, I. N., Primini, F. A., Glotfelty, K. J., et al. 2010, \apjs, 189, 37
\bibitem[Fitzpatrick \& Massa(2007)]{FM07} Fitzpatrick, E. L., \& Massa, D. 2007, \apj, 663, 320
\bibitem[Gennaro et al.(2011)]{GBSH11} Gennaro, M., Brandner, W., Stolte, A., \&  Henning, Th. 2011, \mnras, 412, 2469
\bibitem[Glushkov et al.(1975)]{GDK75} Glushkov, Y. I., Denisyuk, E. K., \& Karyagina, Z. V. 1975, \aap, 39, 481
\bibitem[Goodwin \& Bastian(2006)]{GB06} Goodwin, S. P., \& Bastian, N. 2006, \mnras, 373, 752
\bibitem[Grankin et al.(2007)]{GMBHS07} Grankin, K. N., Melnikov, S. Y., Bouvier, J., Herbst, W., \& Shevchenko, V. S. 2007, \aap, 461, 183
\bibitem[Greve(2010)]{G10} Greve, A. 2010, \aap, 518, 62
\bibitem[Guetter \& Vrba(1989)]{GV89} Guetter, H. H., \& Vrba, F. J. 1989, \aj, 98, 611
\bibitem[Gullbring et al.(1998)]{GHBC98} Gullbring, E., Hartmann, L., Brice\~{n}o, C., \& Calvet, N. 1998, \apj, 492, 323
\bibitem[Gutermuth et al.(2008)]{GMM08} Gutermuth, R. A., Myers, P. C., Megeath, S. T., et al. 2008, \apj, 674, 336
\bibitem[Gutermuth et al.(2009)]{GMM09} Gutermuth, R. A., Megeath, S. T., Myers, P. C., et al. 2009, \apjs, 184, 18
\bibitem[Habibi et al.(2013)]{HSBHM13}  Habibi, M., Stolte, A., Brandner, W., Hu{\ss}mann, B., \& Motohara, K. 2013, \aap, 556, 26
\bibitem[Hartmann(1999)]{H99} Hartmann, L. 1999, NewA Rev., 43, 1
\bibitem[Hartmann(2001)]{H01} Hartmann, L. 2001, \aj, 121, 1030
\bibitem[Hartmann(2003)]{H03} Hartmann, L. 2003, \apj, 585, 398
\bibitem[Hillenbrand \& Hartmann(1998)]{HH98} Hillenbrand, L. A., \& Hartmann, L. W. 1998, \apj, 492, 540
\bibitem[Hur et al.(2014)]{HPS14} Hur, H., Park, B.-G., Sung, H., et al. 2014, \mnras, 446, 3797
\bibitem[Hur et al.(2012)]{HSB12} Hur, H., Sung, H., \& Bessell, M. S. 2012, \aj, 143, 41
\bibitem[Im et al.(2010)]{IKC10} Im, M., Ko J., Cho, Y., Choi, C., Jeon, Y., Lee, I., \& Ibrahimov, M. 2010, J. Korean Astron. Soc., 43, 75 
\bibitem[Jeffries et al.(2007)]{JONML07} Jeffries, R. D., Oliveira, J. M., Naylor, T., Mayne, N. J., \& Littlefair, S. P. 2007, \mnras, 376, 580
\bibitem[Johnson \& Hiltner(1956)]{JH56} Johnson, H. L., \& Hiltner, W. A. 1956, \apj, 123, 267
\bibitem[Jose et al.(2011)]{JPO11} Jose, J., Pandey, A. K., Ogura, K., et al. 2011, \mnras, 411, 2530
\bibitem[Jose et al.(2008)]{JPO08} Jose, J., Pandey, A. K., Ojha, D. K., et al. 2008, \mnras, 384, 1675
\bibitem[Kraus \& Hillenbrand(2007)]{KH07} Kraus, A. L., \& Hillenbrand, L. A. 2007, \aj, 134, 2340
\bibitem[Kilkenny et al.(1998)]{KWRMC98} Kilkenny, D., van Wyk, F., Roberts, G., Marang, F., \& Cooper, D. 1998, \mnras, 294, 93
\bibitem[Koenig et al.(2008)]{KAG08} Koenig, X. P., Allen, L. E., Gutermuth, R. A., Hora, J. L., Brunt, C. M., \& Muzerolle, J. 2008, \apj, 688, 1142
\bibitem[Kook et al.(2010)]{KSB10} Kook, S.-H., Sung, H., \& Bessell, M. S. 2010, J. Korean Astron. Soc., 43, 141
\bibitem[Kroupa(2001)]{K01} Kroupa P., 2001, \mnras, 322, 231
\bibitem[Kroupa(2002)]{K02} Kroupa P., 2002, Science, 295, 82
\bibitem[K\"{o}nigl(1991)]{K91} K\"{o}nigl, A. 1991, \apj, 370, 39
\bibitem[Lada \& Lada(2003)]{LL03} Lada, C., \& Lada, E. 2003, \araa, 41, 57
\bibitem[Landolt(1992)]{L92} Landolt, A. U. 1992, \aj, 104, 340
\bibitem[Lim et al.(2013)]{LCS13} Lim, B., Chun, M.-Y., Sung, H., et al. 2013, \aj, 145, 46
\bibitem[Lim et al. (2009)]{LSBKI09} Lim, B., Sung, H., Bessell, M. S., Karimove, R., \& Ibrahimov, M. 2009, J. Korean Astron. Soc., 42, 161
\bibitem[Lim et al.(2011)]{LSKI11} Lim, B., Sung, H., Karimov, R., \& Ibrahimov, M. 2011, J. Korean Astron. Soc., 44, 39
\bibitem[Lim et al.(2008)]{LSKI08} Lim, B., Sung, H., Karimov, R., \& Ibrahimov, M. 2008, Pub. of the Korean Astron. Soc., 23, 1
\bibitem[Lim et al.(2014a)]{LSKBK14} Lim, B., Sung, H., Kim, J. S., Bessell, M. S., \& Karimov, R., 2014, \mnras, 438, 1451
\bibitem[Lim et al.(2014b)]{LSKBP14} Lim, B., Sung, H., Kim, J. S., Bessell, M. S., \& Park, B.-G. 2014, \mnras, 443, 454
\bibitem[Magnier \& Cuillandre(2004) ]{MC04} Magnier, E. A., \& Cuillandre, J.-C. 2004, \pasp, 116, 449
\bibitem[Martins \& Palacios(2013)]{MP13} Martins, F., \& Palacios, A. 2013, \aap, 560, 16
\bibitem[Massey(2013)]{M13} Massey, P. 2013, \na, 57, 14
\bibitem[McMillan et al. (2007) ]{MVZ07} McMillan, S. L. W., Vesperini, E., \& Portegies Zwart, S. F. 2007, \apj, 655, L45
\bibitem[Menzies et al.(1991)]{MMLCE91} Menzies, J. W., Marang, F., Laing, J. D., Coulson, I. M., Engelbrecht, C. A. 1991, \mnras, 248, 642
\bibitem[Moeckel \& Bonnell(2009)]{MB09} Moeckel, N., \& Bonnell, I. A. 2009, \mnras, 400, 657
\bibitem[Moffat et al.(1979)]{MFJ79} Moffat, A. F. J., Fitzgerald, M. P., \& Jackson, P. D. 1979, \aaps, 38, 197
\bibitem[Offner et al.(2014)]{OCH14} Offner, S. S. R., Clark, P. C., Hennebelle, P., et al. 2013, in Protostars and Planets VI, eds. H. Beuther, et al.(Tucson: University of Arizona Press), in press (arXiv:1312.5326)
\bibitem[Palla et al.(2005)]{PRFP05} Palla F., Randich S., Flaccomio E., \& Pallavicini R. 2005, \apj, 626, L49
\bibitem[Palla \& Stahler(1999)]{PaSt99} Palla, F., \& Stahler, S. W. 1999, \apj, 525, 772
\bibitem[Palla \& Stahler(2000)]{PaSt00} Palla, F., \& Stahler, S. W. 2000, \apj, 540, 255
\bibitem[Palla \& Stahler(2002)]{PaSt02} Palla, F., \& Stahler, S. W. 2002, \apj, 581, 1194
\bibitem[Pandey et al.(2013a)]{PES13} Pandey, A. K., Eswaraiah, C., Sharma, S., et al. 2013a, \apj, 764, 172
\bibitem[Pandey et al.(2013b)]{PSC13} Pandey, A. K., Samal, M. R., Chauhan, N., et al. 2013b, \na, 19, 1
\bibitem[Pandey et al.(2003)]{PUNO03} Pandey, A. K., Upadhyay, K., Nakada, Y., \& Ogura, K. 2003, \aap,  397, 191
\bibitem[Pandey \& Mahra(1986)]{PM86} Pandey, A. K., \& Mahra, H. S. 1986, \apss, 120, 107
\bibitem[Pang et al.(2013)]{PGA13} Pang, X., Grebel, E. K., Allison, R., et al. 2013, \apj, 764, 73
\bibitem[Park et al.(2000)]{PSBK00} Park, B.-G., Sung, H., Bessell, M. S., \& Kang, Y. H. 2000, \aj, 120, 894
\bibitem[Park \& Sung (2002)]{PS02} Park, B.-G., \& Sung, H. 2002, \aj, 123, 892
\bibitem[P\'{e}rez et al.(2008)]{PMAJ08} P\'{e}rez, M. R., McCollum, B., van den Ancker, M. E., \& Joner, M. D. 2008, \aap, 486, 533
\bibitem[Porras et al.(2003)]{PCA03} Porras, A., Christopher, M., Allen, L., et al. 2003, \aj, 126, 1916
\bibitem[Raboud \& Mermilliod(1998)]{RM98} Raboud, D., \& Mermilliod, J.-C. 1998, \aap, 333, 897
\bibitem[Rebull et al.(2000)]{RHS00} Rebull, L. M., Hillenbrand, L. A., Strom, S. E., et al. 2000, \aj, 119, 3026
\bibitem[Robin et al.(2003)]{RRDP03} Robin, A. C., Reyl\'{e} C., Derri\'{e}re, S., \& Picaud, S. 2003, \aap, 409, 523
\bibitem[Salpeter(1955)]{Sp55} Salpeter E. E., 1955, \apj, 121, 161
\bibitem[Sana et al.(2012)]{SMK12} Sana, H., de Mink, S. E., de Koter, A., et al. 2012, Sci, 337, 444
\bibitem[Serkowski et al.(1975)]{SMF75} Serkowski, K., Mathewson, D. S., \& Ford, V. L. 1975, \apj, 196, 261
\bibitem[Sharma et al.(2007)]{SPO07} Sharma S., Pandey A. K., Ojha D. K., et al. 2007, \mnras, 380, 114
\bibitem[Sicilia-Aguilar et al.(2004)]{SHBMC04} Sicilia-Aguilar, A., Hartmann, L. W., Brice\~{n}o, C., Muzerolle, J., \& Calvet, N. 2004, \aj, 128, 805
\bibitem[Siess et al.(2000)]{SDF00} Siess, L., Dufour, E., Forestini, M. 2000, \aap, 358, 5931
\bibitem[Skrutskie et al. (2006)]{2mass} Skrutskie, M. F., et al. 2006, \aj, 131, 1163 
\bibitem[Sota et al.(2011)]{SAW11} Sota A., Apell\'{a}niz J. M., Walborn N. R., et al. 2011, \apjs, 193, 24
\bibitem[Sujatha \& Babu(2006)]{SuBa06} Sujatha, S., \& Babu, G.S.D. 2006, Ap\&SS, 305, 399
\bibitem[Sung \& Bessell (2000)]{SB00} Sung, H., \& Bessell, M. S. 2000, \pasa, 17, 244
\bibitem[Sung \& Bessell (2004)]{SB04} Sung, H., \& Bessell, M. S. 2004, \aj, 127, 1014
\bibitem[Sung \& Bessell (2010)]{SB10} Sung, H., \& Bessell, M. S. 2010, \aj, 140, 2070
\bibitem[Sung \& Bessell(2014)]{SB14} Sung, H., \& Bessell, M. S. 2014, in ASP Conf. Ser. 482, the 10th Pacific Rim Conference on Stellar Astrophysics, ed. H.-W. Lee, Young Woon Kang, and Kam-Ching Leung (San Francisco, CA: ASP), 275
\bibitem[Sung et al.(2004)]{SBC04} Sung, H., Bessell, M. S., \& Chun, M.-Y. 2004, \aj, 128, 1684
\bibitem[Sung et al. (2008)]{SBCKI08} Sung, H., Bessell, M. S., Chun, M.-Y., Karimov, R., \& Ibrahimov, M. 2008, \aj, 135, 441
\bibitem[Sung et al. (1999)]{SBLKL99} Sung, H., Bessell, M. S., Lee, H.-W., Kang, Y. H., \& Lee, S.-W. 1999, \mnras, 310, 982 
\bibitem[Sung et al.(1997)]{SBL97} Sung H., Bessell M. S., \& Lee S.-W. 1997, \aj, 114, 2644
\bibitem[Sung et al. (1998)]{SBL98} Sung, H., Bessell, M. S., \& Lee, S.-W. 1998, \aj, 115, 734
\bibitem[Sung et al. (2000)]{SCB00} Sung, H., Chun, M.-Y., \& Bessell, M. S. 2000, \aj, 120, 333
\bibitem[Sung \& Lee(1995)]{SL95} Sung, H., \& Lee, S.-W. 1995, J. Korean Astron. Soc., 28, 119
\bibitem[Sung et al.(2013a)]{SLB13} Sung, H., Lim, B., Bessell, M. S., et al. 2013a, J. Korean Astron. Soc., 46, 103 (Paper 0)
\bibitem[Sung et al.(2013b)]{SSB13} Sung, H., Sana, H., \& Bessell, M. S. 2013b, \aj, 145, 37
\bibitem[Sung et al.(2009)]{SSB09} Sung, H., Stauffer, J. R., Bessell, M. S. 2009, \aj, 138, 1116
\bibitem[Susa et al.(2014)]{SHT14} Susa, H., Hasegawa, K., \& Tominaga, N. 2014, \apj, 792, 32
\bibitem[Tutukov(1978)]{T78} Tutukov, A. V. 1978, \aap, 70, 57
\bibitem[Uchida \& Shibata (1985)]{US85} Uchida, Y., \& Shibata, K. 1985, \pasj, 37, 515 
\bibitem[Wade et al.(2012)]{WAM12} Wade, G. A., Apell\'{a}niz, J. M., Martins, F., et al. 2012, \mnras, 425, 1278
\bibitem[Walborn et al.(2010)]{WSA10} Walborn, N. R., Sota, A., Ma\'{i}z Apell\'{a}niz, J., et al. 2010, \apj, 711, L143
\bibitem[Whitney et al.(2008)]{WAB08} Whitney, B., Arendt, R., Babler, B., et al. 2008, in Spitzer Proposal ID \#60020
\bibitem[Whitney et al.(2011)]{WBM11} Whitney, B., Benjamin, R., Meade, M., et al. 2011, BAAS, 43, 241.16
\bibitem[Whittet(1977)]{W77} Whittet, D. C. B. 1977, \mnras, 180, 29
\bibitem[Yong et al.(2012)]{YCF12} Yong, D., Carney, B. W., \& Friel, E. D. 2012, \aj, 144, 95


\end{thebibliography}
\end{document}